\documentclass[a4paper,10pt,dvipsnames]{article}
\usepackage[utf8]{inputenc}
\usepackage[english]{babel}
\usepackage{amsthm}
\usepackage{amsfonts}
\usepackage{amssymb}	
\usepackage{amsmath}
\allowdisplaybreaks
\usepackage{mathtools}
\usepackage{slashed}
\usepackage{enumerate}
\usepackage{graphicx}
\usepackage{systeme}
\usepackage{dsfont}
\usepackage{relsize}
\usepackage[margin=0.90in]{geometry}
\usepackage{float}
\usepackage{multirow}
\usepackage{tikz}
\usepackage[labelformat=simple]{subcaption}

\usepackage{graphicx}
\usepackage{bmpsize}
\usepackage{epstopdf}
\usepackage{bbm}
\usepackage{etoolbox}
\usepackage{xcolor}
\usepackage{titling}
\usepackage{authblk}

\usepackage{blindtext,graphicx}
\usepackage[absolute]{textpos}
\usepackage{physics}
\usepackage[hang,flushmargin]{footmisc}

\usepackage{xfrac}
\usepackage{nicefrac}
\usepackage{esvect}
\usepackage{cases}
\usepackage{empheq}
\usepackage{stmaryrd}
\usepackage{cancel}
\usepackage[linguistics]{forest}
\usepackage{url}
\usepackage[font=small]{caption}
\usepackage{array}
\usepackage[super]{nth}
\usepackage[giveninits=true,sortcites=true,date=year,maxbibnames=99,doi=false,isbn=false,url=false,eprint=false]{biblatex}
\renewbibmacro{in:}{}
\DeclareFieldFormat{pages}{#1}
\AtEveryBibitem{%
  \clearlist{language}%
}
\usepackage{csquotes}

\usepackage{setspace}
\makeatletter
\def\hlinewd#1{%
\noalign{\ifnum0=`}\fi\hrule \@height #1 \futurelet
\reserved@a\@xhline}
\makeatother
\theoremstyle{definition}

\makeatletter
\newcommand{\pushright}[1]{\ifmeasuring@#1\else\omit\hfill$\displaystyle#1$\fi\ignorespaces}
\newcommand{\pushleft}[1]{\ifmeasuring@#1\else\omit$\displaystyle#1$\hfill\fi\ignorespaces}
\makeatother
\makeatletter
\newcommand{\thickhline}{%
    \noalign {\ifnum 0=`}\fi \hrule height 1pt
    \futurelet \reserved@a \@xhline
}
\newcolumntype{"}{@{\hskip\tabcolsep\vrule width 1pt\hskip\tabcolsep}}
\makeatother
\title{Unfolding engineering metamaterials design: relaxed micromorphic modeling of large-scale acoustic meta-structures.}
\author[1]{F. Demore}
\author[2]{G. Rizzi}
\author[1]{M. Collet}
\author[3]{P. Neff}
\author[2]{A. Madeo}
\affil[1]{Laboratoire de Tribologie et de Dynamique des Système, Ecole Centrale de Lyon, Ecully 69134, France}
\affil[2]{Institute for Structural Mechanics and Dynamics, Technical University Dortmund, August-Schmidt-Str. 8, 44227 Dortmund, Germany}
\affil[3]{Chair for Nonlinear Analysis and Modeling, Faculty of Mathematics, University of Duisburg-Essen, Thea-Leymann-Str. 9, 45127 Essen, Germany}

\thanksmarkseries{arabic}
\date{\today}

\usepackage{pgfplots}
\pgfplotsset{compat=1.7}
\usepgfplotslibrary{colormaps}

\newcommand{\comsol}{\textit{Comsol Multiphysics}\textsuperscript{\textregistered}}
\newcommand{\matlab}{\textit{MATLAB}\textsuperscript{\textregistered}}

\usepackage{authblk}
\begin{document}
\maketitle
%
%
%
%
%
%%%%%%%%%%%%%%%%%%%%%%%%%%%%%%%%%%%%%%%%%%%%%%%%%%%
\begin{abstract}
In this paper, we present a unit cell showing a band-gap in the lower acoustic domain.
% The corresponding metamaterial is made up of a suitable repetition in space of this unit cell.
The corresponding metamaterial is made up of a periodic arrangement of this unit cell.
We rigorously show that the relaxed micromorphic model can be used for metamaterials’ design at large scales as soon as \textit{sufficiently large} specimens are considered.
We manufacture the metamaterial via metal etching procedures applied to a titanium plate so as to show that its production for realistic applications is viable.
Experimental tests are also carried out confirming that the metamaterials’ response is in good agreement with the theoretical design.
% In order to show that our gold standard model opens unprecedented possibilities in metastructural design, we conceive a large-scale structure that is able to focus elastic energy in a tiny region, thus enabling its possible subsequent re-use.
In order to show that our micromorphic model opens unprecedented possibilities in metastructural design, we conceive a finite-size structure that is able to focus elastic energy in a confined region, thus enabling its possible subsequent re-use.
% Indeed, thanks to the introduction of well-posed micromorphic boundary conditions, we can combine bricks of different metamaterials and of a classical Cauchy material in such a way that the elastic energy produced by a source of vibrations is focused in specific collection points enabling an eventual subsequent reuse.
Indeed, thanks to the introduction of a well-posed set of micromorphic boundary conditions, we can combine different metamaterials and classical Cauchy materials in such a way that the elastic energy produced by a source of vibrations is focused in specific collection points.
% The design of this structure would have not been otherwise possible (via \textit{e.g.}, direct simulations), due to the large dimensions of the considered metamaterials.
The design of this structure would have not been otherwise possible (via \textit{e.g.}, direct simulations), due to the large dimensions of the metastructure, couting hundreds of unit cells.
\end{abstract}
\section{Introduction}
Metamaterials are architectured materials whose mechanical properties go beyond those of classical materials thanks to their heterogeneous microstructure.
This allows them to show exceptional static/dynamic features such as negative Poisson’s ratio \cite{lakes1987foam}, twist in response to being pushed or pulled \cite{frenzel2017three,rizzi2019identification}, band-gaps \cite{liu_broadband_2018,wang_harnessing_2014,bilal,celli_bandgap_2019}, cloaking \cite{buckmann_mechanical_2015,misseroni_cymatics_2016}, focusing \cite{guenneau_acoustic_2007,bacigalupo_second-gradient_2014}, channeling \cite{kaina_slow_2017,tallarico}, negative refraction \cite{zhu_negative_2014,kaina_negative_2015,willis_negative_2016,li2004double,bordiga2019prestress}, etc.

In the last two centuries, the advancement of knowledge on finite-size classical materials modeling has enabled the design of engineering structures (buildings, bridges, airplanes, cars, \textit{etc.}) resisting to static and dynamic loads.
% Today, while the modeling of infinite-size metamaterials is achievable via reliable homogenization techniques \cite{chen2001dispersive,willis2011effective,craster_high-frequency_2010,willis2012construction,boutin2014large,sridhar2018general}, we must acknowledge that these techniques are usually unsuitable for finite-size metamaterials’ modeling.
% In a symbolic sense, homogenization methods cannot provide the right tools to cut finite-size metamaterials’ LEGO bricks out of an infinite block.
Today, while the modeling of infinite-size metamaterials is achievable via reliable homogenization techniques \cite{chen2001dispersive,willis2011effective,craster_high-frequency_2010,willis2012construction,boutin2014large,sridhar2018general}, we must acknowledge that these techniques are usually unsuitable for finite-size metamaterials’ modeling since these homogenization methods cannot provide the right tools to ``cut finite-size metamaterials’ LEGO bricks out'' of an infinite block.
% This conceptual gap prevented us to explore metamaterials/classical-materials structures and optimize them towards efficient wave control and energy recovery.
This gap prevents the exploration of the response of metamaterials/classical-materials structures and their optimization towards efficient wave control and energy recovery.
% In this paper, we show that this gap can be filled by using the relaxed micromorphic model \cite{neff_unifying_2014,neff_relaxed_2015,rizzi_exploring_2021,rizzi2021boundary} as a gold standard to conceive realistic large-scale metamaterials/classical materials structures that can control waves and eventually recover energy.
In this paper, we show that this gap can be filled by using the relaxed micromorphic model \cite{neff_unifying_2014,neff_relaxed_2015,rizzi_exploring_2021,rizzi2021boundary} as a gold standard to conceive realistic finite-size metamaterials/classical-materials structures that can control waves and eventually recover energy.

At present, the response of finite-size metamaterials’ structures is mostly explored via direct Finite Element (FEM) simulations that implement all the details of the involved microstructures (e.g., \cite{krushynska2017coupling}).
% Despite the precise propagation patterns that these direct numerical simulations can provide, they suffer from unsustainable computational costs when considering large metamaterials’ specimens.
Despite the precise propagation patterns that these direct numerical simulations can provide, they suffer from unsustainable computational costs when considering large metamaterials’ specimens or high frequencies.
% Therefore, it is very difficult today to explore large-scale meta-structures combining metamaterials and classical-materials bricks of different type, size and shape. The awareness of this limitation triggered all the recent advances on dynamical homogenization methods \cite{chen2001dispersive,willis2011effective,craster_high-frequency_2010,willis2012construction,boutin2014large,sridhar2018general}.
Therefore, it is very difficult today to explore large meta-structures combining metamaterials and classical-materials bricks of different type, size and shape.
The awareness of this limitation triggered all the recent advances on dynamical homogenization methods \cite{chen2001dispersive,willis2011effective,craster_high-frequency_2010,willis2012construction,boutin2014large,sridhar2018general}.
Such methods share the idea that a periodic infinite-size metamaterial can be replaced by a homogenized continuum, mimicking its response without accounting for all the microstructures’ details.
This leads to an important simplification of metamaterials’ description at the macroscopic scale.
Unfortunately, often homogenization methods cannot describe the response of finite-size metamaterials due to the difficulty of establishing well-posed boundary conditions at the macro-level.
The unsuitability of classical homogenization methods for finite-size metamaterials’ modeling in dynamic regime has been very recently acknowledged by the cutting-edge research groups in dynamical homogenization \cite{srivastava2017evanescent,sridhar2016homogenization}.
Being aware of the homogenization’s limits concerning finite-size metamaterials’ modeling in dynamics, Sridhar et al. \cite{sridhar2016homogenization} recently proposed an alternative \textit{ad hoc} upscaling procedure, only valid for locally resonant metamaterials, leading to a homogenized equation which the authors recognize to be of the micromorphic type.
In a similar spirit, \cite{srivastava2017evanescent} obtained a homogenized continuum with extended kinematics, classifying it as micromorphic, and proposed its use to study a simple 1D boundary value problem for a periodic metamaterial.
It is also common today to find models, alternative to homogenization, that entail frequency-dependent parameters to describe dynamic metamaterials’ response at different frequencies (\textit{e.g.}, \cite{haberman,chen2001dispersive}).
Even if these models can give useful information about metamaterials’ response under particular loading conditions, they cannot provide a comprehensive characterization of metamaterials.

The results presented in this paper show that the relaxed micromorphic model’s structure, coupled with the introduction of well-posed boundary conditions, allowed us to unveil both the static and dynamic response of metamaterials bricks of finite size.
% Playing LEGO with such bricks enabled the design of a surprising structure, combining metamaterials and classical-materials in such a way to focus energy in specific collection points for eventual subsequent re-use.
Playing LEGO with such bricks enables the design of a highly performing structure, combining metamaterials and classical-materials in such a way to focus energy in specific collection points for eventual subsequent re-use.
%
%
%
%
%
%%%%%%%%%%%%%%%%%%%%%%%%%%%%%%%%%%%%%%%%%%%%%%%%%%%
\section{A titanium-based metamaterial for acoustic control: experimental set-up}
\label{sec:set-up}
\begin{figure}[H]
	%% left-hand side: a single subfigure
	\begin{subfigure}{0.49\textwidth}
		\centering
		\includegraphics[width=0.8\textwidth]{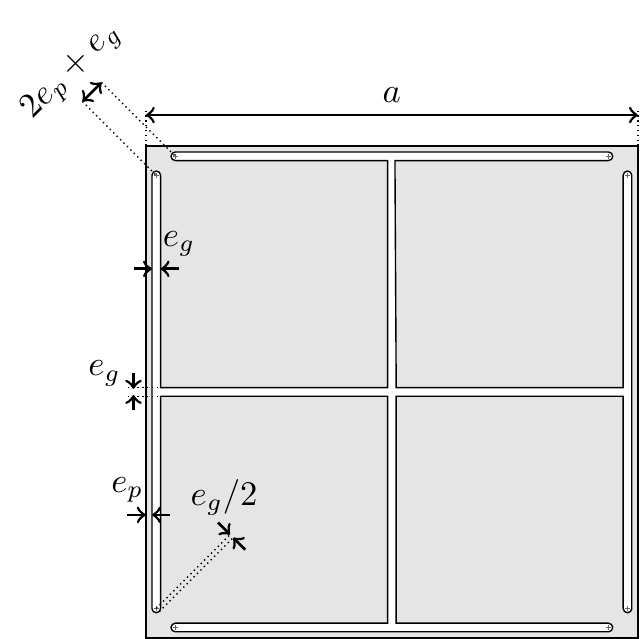}
	\end{subfigure}
	%% horizontal separation between the left and right hand sides
% 	\hspace*{-1.5cm}
	%% right-hand side: a minipage that contains two more subfigures
	\begin{minipage}{0.49\textwidth}
		\begin{subfigure}{\textwidth}
			\centering
			\vspace{0.7cm}
			\renewcommand{\arraystretch}{1.2}
            \begin{tabular}{cccccc}
				\thickhline
				\thickhline
    			$a$ & $e_{g}$ & $e_{p}$      \\ 
				{[mm]} &  {[mm]}   &  {[mm]}   \\
				\hline
				20 & 0.35 & 0.25 \\ 
				\thickhline
				$\rho_{\text{Ti}}$ & $\lambda_{\text{Ti}}$ & $\mu_{\text{Ti}}$          \\
				{[kg/m$^3$]} &   {[GPa]}  &    {[GPa]} \\
				\hline
				4400 & 88.8 & 41.8\\ 
				\thickhline
				\thickhline
				\vspace{0.2cm}
		    \end{tabular}
		\end{subfigure}
	\end{minipage}
	\caption{
	(\textit{left}) unit cell whose periodic repetition in space gives rise to the metamaterial studied in this paper.
	(\textit{right}) geometry and material parameters characterising the unit cell.
	The parameters $\rho_{\tiny \mbox{Ti}}$, $\lambda_{\tiny \mbox{Ti}}$, and $\mu_{\tiny \mbox{Ti}}$ are the density and the Lamé constants of the titanium alloy used, respectively.
	}
	\label{fig:fig_tab_unit_cell}
\end{figure}

%%%%%%%%%%%%%%%%%%%%%%%%%%%%%%%%%%%%%%%%%%%%%%%%%%%
A consistent branch of research on metamaterials focuses on how to engineer the unit cell geometry and material properties to optimize their response with respect to elastic wave manipulation \cite{pham_transient_2013,sridhar_homogenization_2016,sridhar_general_2018,sridhar_frequency_2020,miniaci2016spider,misseroni_cymatics_2016,misseroni2019omnidirectional,willis_effective_2011,willis_negative_2016,nassar_willis_2015,craster_high-frequency_2010,nolde_high_2011,miniaci2016large,krushynska2018accordion}.
In particular, optimizing the size, the mass, and the stiffness distribution within the unit cell can synergize to define the position of the band-gap. Often, in order to obtain band-gaps in the acoustic regime, it is necessary to have unit cells whose size are in the range of tens of centimeters.
The unit cell that we present in this paper (Figure \ref{fig:fig_tab_unit_cell}) has an optimized geometry that allows us to obtain a band-gap in the low acoustic frequency range with a cell size of only 2 centimeters.

\begin{figure}[H]
    \centering
    \includegraphics[height=6.5cm]{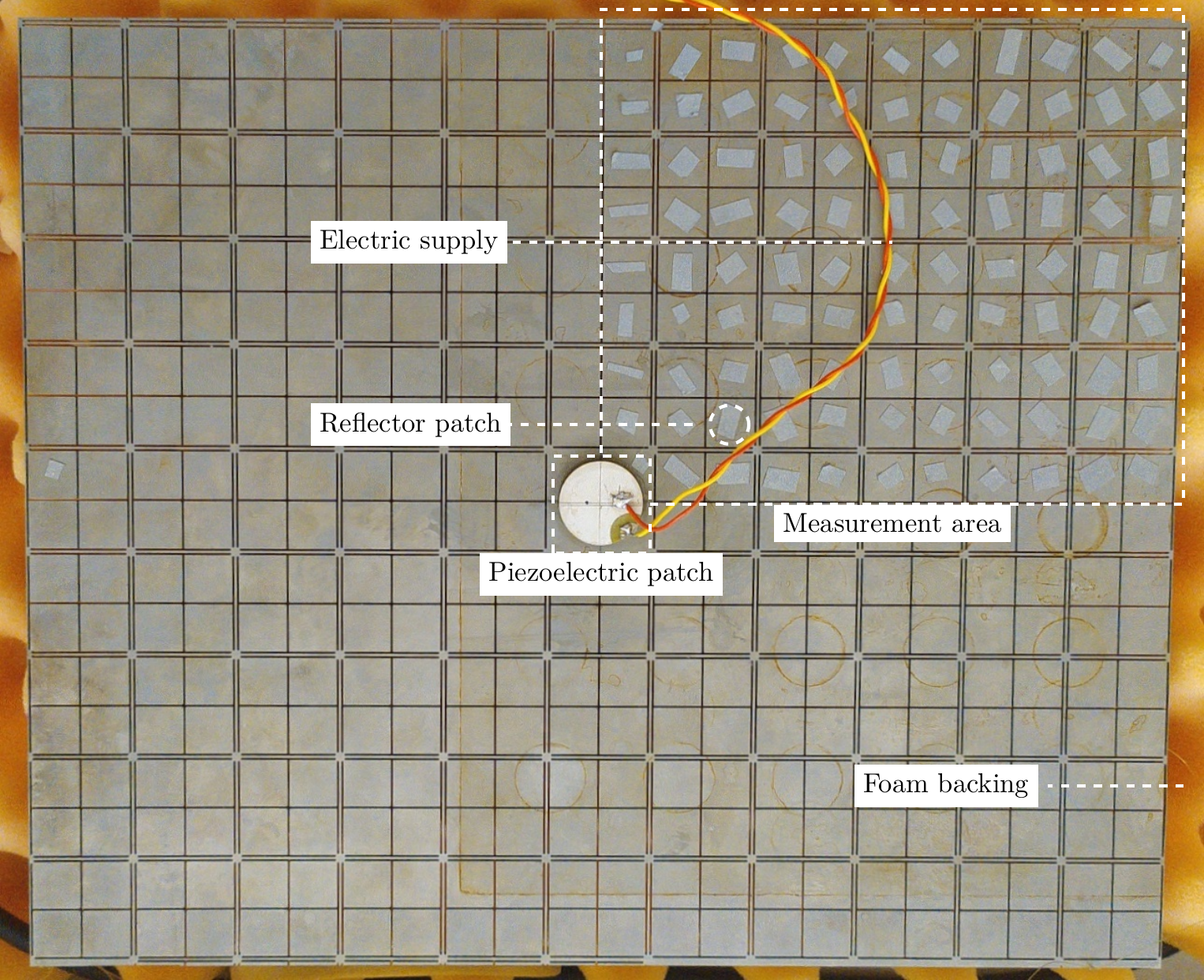}
    \hfill
    \includegraphics[height=6.5cm]{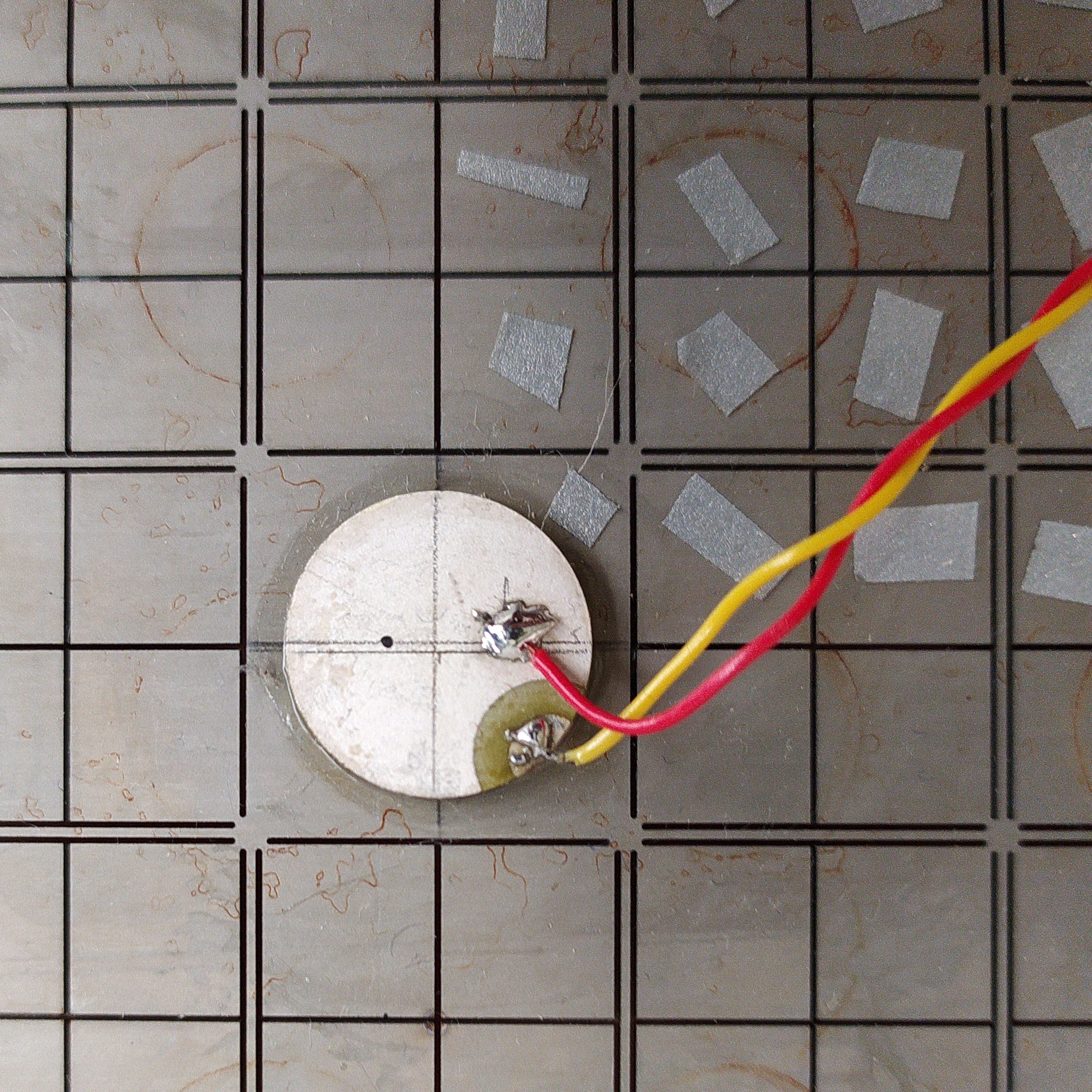}
    \caption{
    (\textit{left}) Experimental set-up: glued in the center of the metamaterial's plate there is the top piezoelectric patch (another one being placed on the other side of the plate) that has been used as an actuator for the external excitation.
    The tapes placed on the top-right quarter of the plate can reflect a laser's beam for speed measurements.
    (\textit{right}) Detail of the upper piezoelectric patch with its electric supply.
    }
    \label{fig:experimental_set_uP^and_2Dplate_2}
\end{figure}

%%%%%%%%%%%%%%%%%%%%%%%%%%%%%%%%%%%%%%%%%%%%%%%%%%%
The four squares in which the unit cell can be divided in act as local resonators (Figure~\ref{fig:fig_tab_unit_cell} \textit{left}) localizing the energy at the microscopic level, thus creating the band-gap effect.
In order to lower the band-gap to the acoustic frequency range, it is required simultaneously to increase the mass of the resonators (\textit{i.e.} maximize the size of such internal squares) and decrease their stiffness (\textit{i.e.} minimize the thickness $e_{\rm p}$ of the stripe on the outline of the cell).
To minimize the unavoidable presence of defects related to the extremely thin holes that must be drilled in the metallic plate, the Electrical Discharge Machining wire erosion was used for metal etching.
Titanium was chosen as base material to maximize the overall strength while minimizing losses due to damping.
The values of $a$, $e_{\rm p}$ and $e_g$ are given in Figure \ref{fig:fig_tab_unit_cell} (\textit{right}) and the out-of-plane cell thickness $e$=1[mm] were thus determined by taking into account:
\begin{itemize}
    \item desired characteristics of the band-gap (acoustic regime);
    \item manufacturing constraints of the chosen process (limitations for the possible values of $e_{\rm p}$ and $e_g$);
    \item static resistance of the structure.
\end{itemize}
Using this unit cell, a 9$\times$11 cells meta-structure has been designed to experimentally explore the band-gap attenuation: the used set-up is presented in Figure \ref{fig:experimental_set_uP^and_2Dplate_2}.
%%%%%%%%%%%%%%%%%%%%%%%%%%%%%%%%%%%%%%%%%%%%%%%%%%%

Two piezoelectric patches (MEGGIT PZ 21, $\varnothing$16[mm] 2[mm]-thick) are used as actuators to generate in-plane extension pulse waves in the plate (see Figure \ref{fig:experimental_set_uP^and_2Dplate_2} \textit{right}).
Excitation signals are generated by a function generator and then amplified to power the piezoelectric patches.
Sine sweeps are chosen to impose the external load and the signal's frequency is swept from 0 to 2500~Hz.
Measured speeds are acquired by a 3-D laser (Polytec CLV-3D).
An interface under \matlab has also been designed, allowing to easily choose the main parameters for each test, namely the required frequency range, the resolution and coordinates of the considered measurement points.

\begin{figure}[H]
\centering
\includegraphics{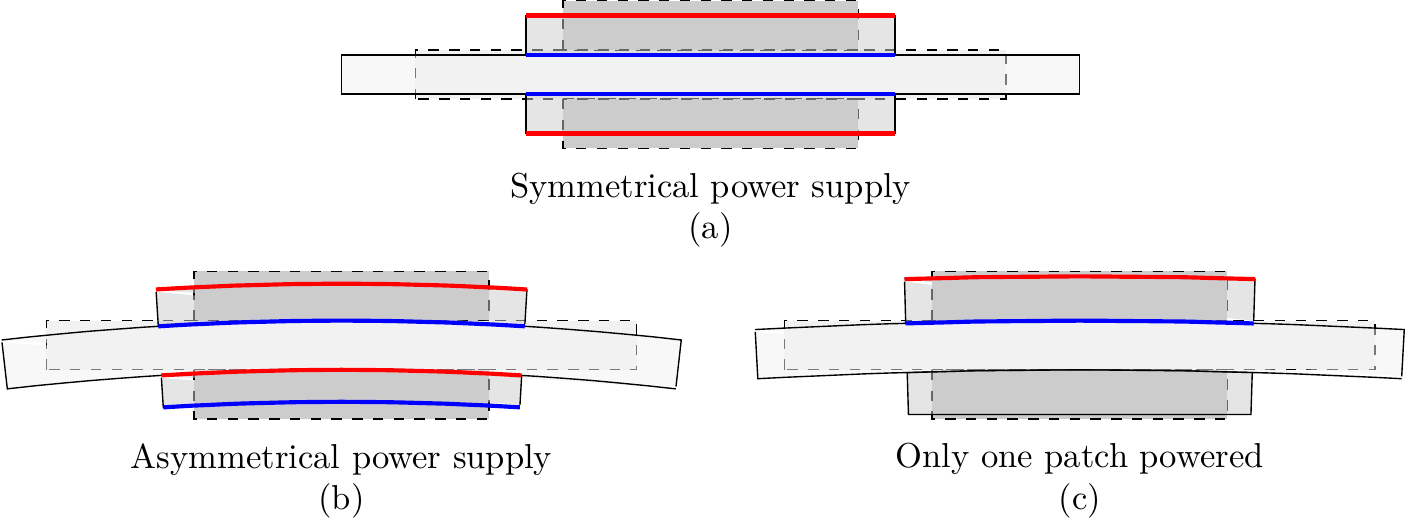}
\caption{
Scheme of possible power supplies of the piezoelectric patches.
The dashed lines outline the non-powered piezoelectric patches and undeformed center of the plate while the solid lines depict the deformed powered piezoelectric patches deforming the microstructured plate.
In particular, blue lines represent the side of the piezoelectric patches connected to the ground while red lines represent the powered side of the piezoelectric patches.
The same electric potential is applied on both red areas.
}
\label{alims}
\end{figure}

The power supply of the piezoelectric patches is designed to avoid flexural vibration modes in the plate at the considered frequencies, so that the applied load is a pure in-plane expansion as shown in Figure~\ref{alims} (a).
In Figures \ref{alims} (b) and \ref{alims} (c) other possible loads are depicted but not used in the experiment.

%%%%%%%%%%%%%%%%%%%%%%%%%%%%%%%%%%%%%%%%%%%%%%%%%%%%%%%%%%%%%%%
\section{Modeling and simulation}
\label{sec:modelling_simulation}
%%%%%%%%%%%%%%%%%%%%%%%%%%%%%%%%%%%%%%%%%%%%%%%%%%%%%%%%%%%%%%%
In this section, we give an outlook on the possible modeling tools allowing to catch the dynamical response of the metamaterial presented in Section \ref{sec:set-up}.
In particular, we simulate the experimental set-up presented in Figure \ref{fig:experimental_set_uP^and_2Dplate_2} both by using a detailed finite element model and a novel micromorphic homogenized approach. The model's features will be presented in detail except the loading conditions that will be introduced in section 4 (pulse load) and 5 (piezoelectric load).
%%%%%%%%%%%%%%%%%%%%%%%%%%%%%%%%%%%%%%%%%%%%%%%%%%%%%%%%%%%%%%%
\subsection{Detailed direct element simulations}
\label{sec:micro_mate_characteristic}
\begin{figure}[H]
\centering
\raisebox{-.75\height}{\includegraphics[width=0.4725\textwidth]{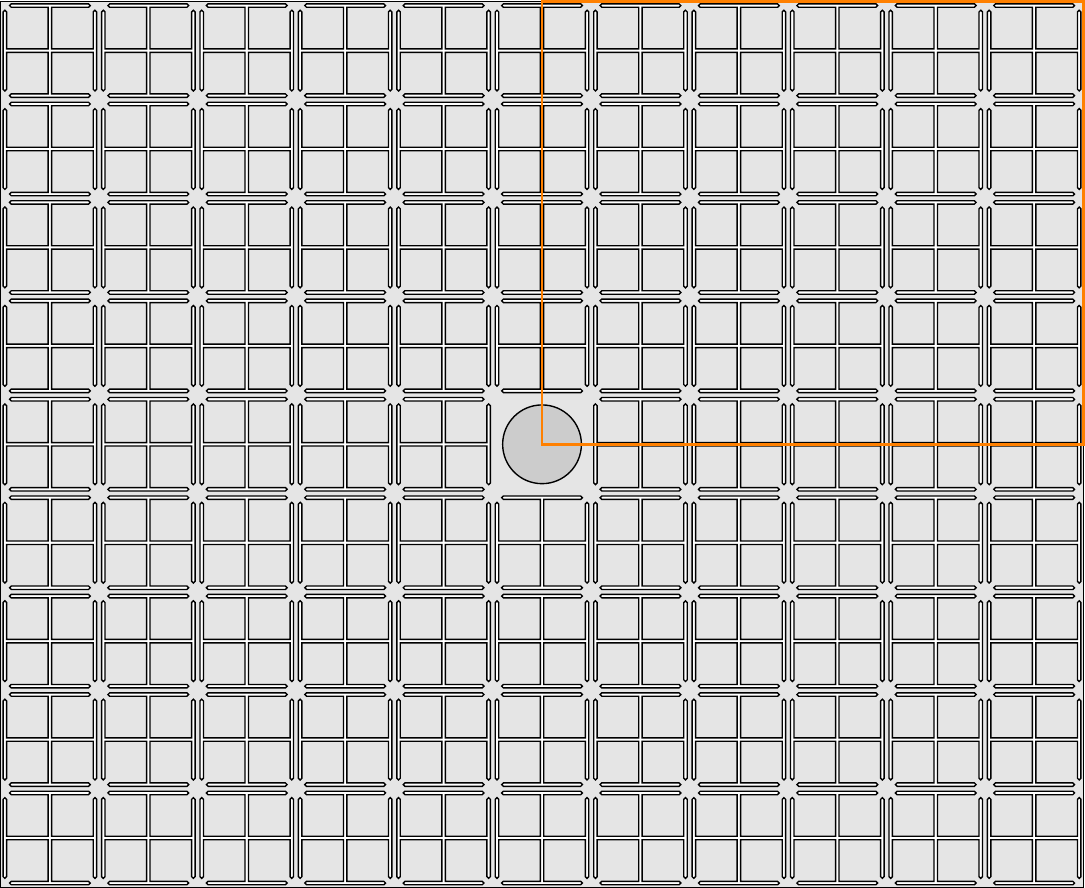}}
$\longrightarrow$
\raisebox{-.75\height}{\includegraphics[width=0.4725\textwidth]{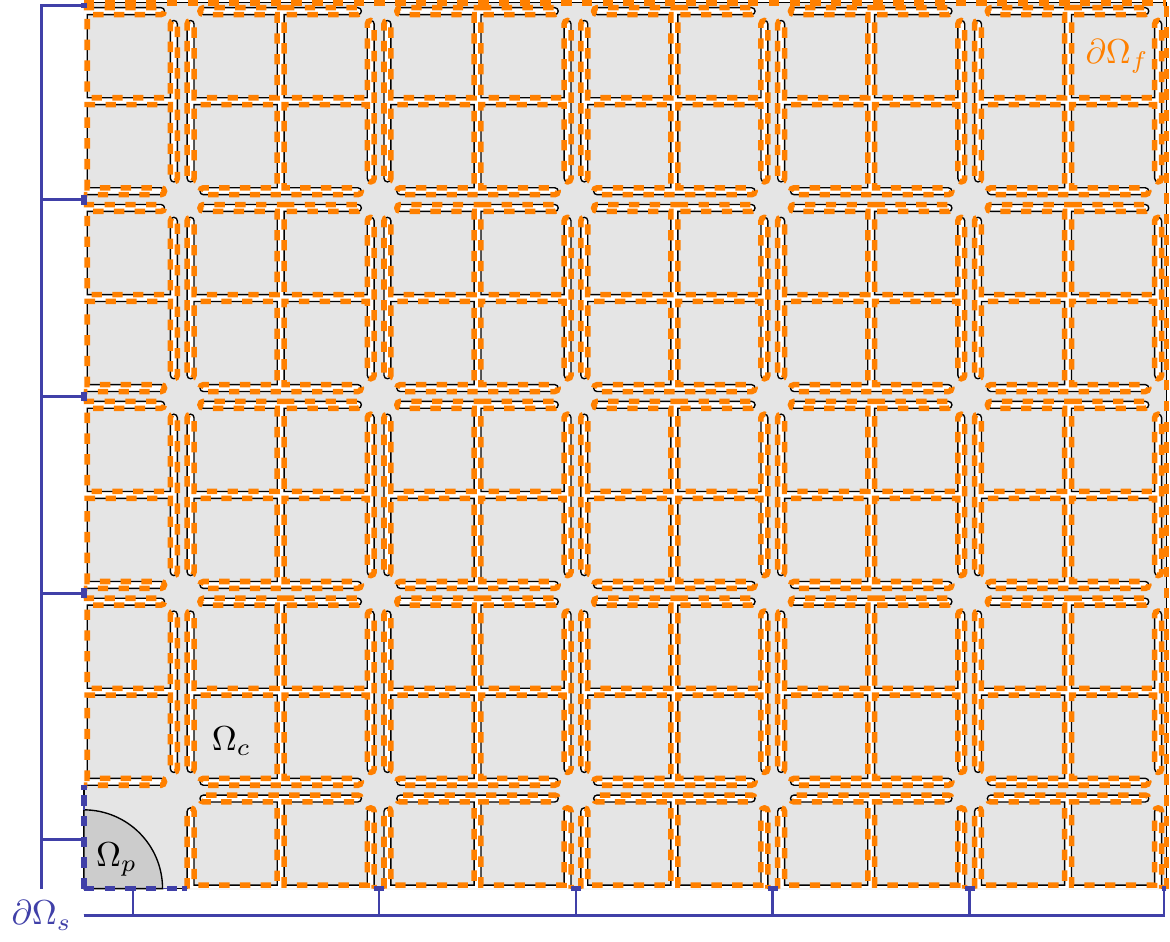}}
\caption{
(\textit{left}) Full microstructured plate with its top piezoelectric patch.
(\textit{right}) Reduced microstructured symmetrical plate: $\Omega_{\rm p}$ is the region occupied by the piezoelectric patch, $\Omega_{\rm c}$ is the region made up of titanium, and $\partial \Omega_s$ is the union of the boundaries on the plane of symmetry.
The thickness of the reduced plate is 0.5 mm instead of 1 mm.
}
\label{fig:m3dsym}
\end{figure}

%%%%%%%%%%%%%%%%%%%%%%%%%%%%%%%%%%%%%%%%%%%%%%%%%%%%%%%%%%%%%%%
We present there the detailed finite element simulations of the structure presented in Section \ref{sec:set-up}, where the material composing the unit cell is modelled as a classical isotropic Cauchy continuum.
The structure and load symmetry allow us to consider one eighth of the system as presented in Figure \ref{fig:m3dsym} (the plate's thickness implemented in the simulation is 0.5 mm instead of 1 mm).

The action functional of the considered reduced system is\footnote{Here and in the sequel, we consider the plane-strain hypothesis which allows to sort out the thickness $e$ out of the intergration with respect to $x_3$.}
\begin{equation}
    \mathcal{A} \left ( u,V \right ) = \int^T_0 \left [ \iiint_{\Omega_{\rm p}} (K_{\rm p}-W_{\rm p}-Q) \mathrm{d}x_1\mathrm{d}x_2\mathrm{d}x_3
    + \frac{e}{2} \iint_{\Omega_{\rm c}} (K_{\rm t}-W_{\rm t})\mathrm{d}x_1\mathrm{d}x_2 \right ] \mathrm{d}t
    \label{action}
\end{equation}
where $K_{\rm p}$, $W_{\rm p}$, and $Q$ being respectively the kinetic, strain and electric potential energy density of the piezoelectric patch while $K_{\rm t}$ and $W_{\rm t}$ are the kinetic and potential energy density of the isotropic Cauchy material (titanium) constituting the domain $\Omega_{\rm c}$, respectively.
Finally $e$ is the plate's thickness and $[0,T]$ is the time interval during which the system's response is observed. Given the symmetry of the system, only one piezoelectric patch can be considered.

The strain and kinetic energy density, and the electric potential expression are respectively given by
\begin{gather}
    W_{\rm p} =
    \frac{1}{2} \langle \mathbb{C}_{\rm p} \, \text{sym} \, \nabla u + \xi^T \, E , \text{sym} \, \nabla u \rangle
    \, ,
    \qquad
    K_{\rm p} =
    \frac{1}{2}  \rho_{\rm p} \, \langle \dot{u} ,\dot{u} \rangle
    \, ,
    \qquad
    Q =
    \frac{1}{2} \langle\epsilon_0  \varepsilon E + \xi \, \text{sym} \, \nabla u, E \rangle \, ,
    \\
    W_{\rm t} =
    \frac{1}{2} \langle \mathbb{C}_{\rm t} \, \text{sym} \, \nabla u , \text{sym} \, \nabla u \rangle
    \, ,
    \qquad
    K_{\rm t} =
    \frac{1}{2}  \rho_{\rm t} \, \langle \dot{u} ,\dot{u} \rangle \, ,
    \,
    \notag
\end{gather}
where $\mathbb{C}_{\rm p}$ and $\mathbb{C}_{\rm t}$ are 4th order elasticity tensors, $\text{sym} \,\nabla u$ is the symmetric part of the gradient of the displacement field, $\xi$ is the 3rd order piezoelectric coupling tensor (in C/m$^2$), $E$ is the electric vector field (in V/m), $\varepsilon$ is the relative permittivity tensor, and ${\epsilon_0 = 8.86 \times 10^{-12}}$[F/m] is the vacuum permittivity. Using the approximation of electrostatic and the Maxwell-Faraday equation, $E$ derives from the potential $V$, \textit{i.e.}
\begin{equation}
    E = - \nabla V
\end{equation}
Requiring the first variation of the total energy with respect $u$ and $V$ to be zero gives the following equilibrium equations
\begin{align}
    \rho_{i} \, \ddot{u} = \text{Div} \, \sigma
    \quad
    \text{(Cauchy equilibrium)}
    \, ,
    \qquad\qquad
    \text{Div} \, D = 0
    \quad
    \text{(Maxwell-Gauss law)}
    \label{eq:equiequa}
\end{align}
where the 2nd order Cauchy stress tensor $\sigma$ and the electric induction vector $D\in\mathbb{R}^{6}$ are given by
\begin{align}
\sigma \coloneqq \mathbb{C}_{i} \, \text{sym} \, \nabla u - \xi^T \, E
\qquad\qquad\qquad
D \coloneqq \varepsilon_0 \, \varepsilon E + \xi \, \text{sym} \, \nabla u
\, .
\label{eq:sigma_induction_def}
\end{align}
with $i=\{{\rm p,t}\}$.
In the domain $\Omega_{\rm c}$, only equation (\ref{eq:equiequa})$_{1}$ is required and the definition (\ref{eq:sigma_induction_def})$_{1}$ becomes $\sigma \coloneqq \mathbb{C}_{\rm t} \, \text{sym} \, \nabla u $.
Given the cylindrical symmetry of the piezoelectric patches, the Voigt representation of elastic tensor $\mathbb{C}_{\rm p}$, the piezoelectric coupling tensor $\xi$, and the relative permittivity tensor $\varepsilon$ is
\begin{align}
\mathbb{C}_{\rm p} = 
\begin{pmatrix}
C_{11} & C_{12} & C_{13} & 0 & 0 & 0 \\ 
C_{12} & C_{11} & C_{13} & 0 & 0 & 0 \\ 
C_{13} & C_{13} & C_{33} & 0 & 0 & 0 \\ 
0 & 0 & 0 & C_{44} & 0 & 0 \\ 
0 & 0 & 0 & 0 & C_{44} & 0 \\ 
0 & 0 & 0 & 0 & 0 & C_{66}
\end{pmatrix}
, \,
\xi = \begin{pmatrix}
0 & 0 & 0 & 0 & \xi_{15} & 0 \\
0 & 0 & 0 & \xi_{15} & 0 & 0 \\
\xi_{31} & \xi_{31} & \xi_{33} & 0 & 0 & 0 \\
\end{pmatrix}
, \, 
\varepsilon = 
\begin{pmatrix}
\varepsilon_{11} & 0 & 0 \\ 
0 & \varepsilon_{11} & 0 \\ 
0 & 0 & \varepsilon_{33} \\ 
\end{pmatrix}
.
\end{align}
The matrix $\mathbb{C}_{\rm t}$ has the structure of the classical elasticity isotropic tensor and its coefficients are reported in Table \ref{fig:fig_tab_unit_cell} in terms of Lamé constants together with the density $\rho_{\rm t}$.
The values of the parameters in $\mathbb{C}_{\rm p}$, $\xi$, and $\varepsilon$, together with the density $\rho_{\rm p}$ are reported in Table \ref{table:piezo12}.
\renewcommand{\arraystretch}{1.2}
\begin{table}[H]
    \centering
    \begin{tabular}{cccccccccc}
    \thickhline
    \thickhline
    & $\rho_{\rm p}$ & $C_{11}$ & $C_{12}$ & $C_{13}$ & $C_{33}$ & $C_{44}$ & $C_{66}$ & \\
    & [kg /m$^{3}$] & [GPa] & [GPa] & [GPa] & [GPa] & [GPa] & [GPa] & \\
    \hline
    & 7780 & 1140 & 757 & 724 & 1110 & 263 & 403 & \\
    \thickhline
    & $\varepsilon_{11}$ & $\varepsilon_{33}$ & $\xi_{31}$ & $\xi_{33}$ & $\xi_{15}$ &\\
    & [-] & [-] & [C/m$^{2}$] & [C/m$^{2}$] & [C/m$^{2}$] &  \\
    \hline
    & $3.24\cdot 10^3$ & $3.98 \cdot 10^3$ & -2.92 & 23.4 & 16.2 & \\
    \thickhline
    \thickhline
    \end{tabular}
    \caption{Mechanical parameters of the piezoelectric patches, coupling and electrical parameters.}
    \label{table:piezo12}
\end{table}

\begin{figure}[H]
    \centering
    \includegraphics{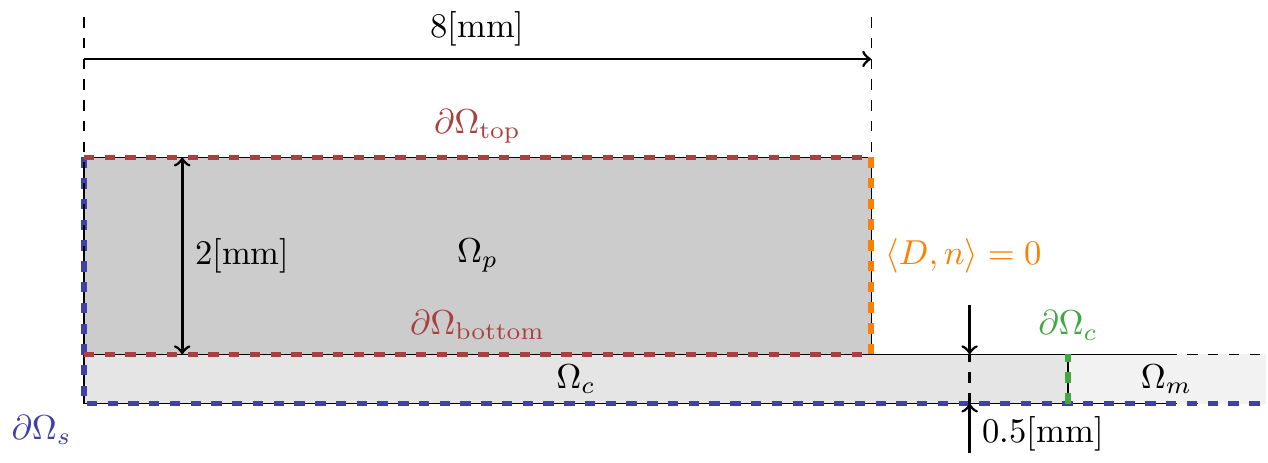}
    \caption{Section of the symmetrized microstructured plate with the piezoelectric patch (one eight on the whole system) and the boundaries' denomination $\partial \Omega_\text{top}$ and $\partial \Omega_\text{bottom}$ where the electric potential is imposed.}
    \label{fig:front_sym}
\end{figure}
To simulate the experimental setup of Section \ref{sec:set-up} \textit{via} the reduced problem of Figure \ref{fig:m3dsym}, proper boundary conditions need to be imposed on the boundaries defined in Figure~\ref{fig:front_sym}.
The first interface conditions represent the imposition of the electric potential while the second ones are associated to the symmetry conditions of the reduced problem
\begin{align}
\begin{cases}
        V = 0
        & \text{on } \partial\Omega_\text{bottom}
        \\
        V = V_0
        &  \text{on } \partial\Omega_\text{top}
\end{cases}, \qquad
\begin{cases}
    \langle u , n \rangle = 0 \\
    \langle D , n \rangle = 0
\end{cases} \text{on } \partial \Omega_s
\label{eq:interf_cond_piezo}
\end{align}
where $V_0=100$[V]. No condition have to be imposed on the free boundaries $\partial \Omega_f$. The microstructured plate is studied under the plain strain hypothesis while the piezoelectric region is kept as a full 3D medium:
\begin{equation}
    u^{\rm c} =
    \left(
    u^{\rm c}_{1},
    u^{\rm c}_{2},
    0
    \right)^{\rm T}
    \quad
    \text{in}
    \quad
    \Omega_{\rm c}
    \qquad\qquad
    \text{and}
    \qquad\qquad
    u^{\rm p} =
    \left(
    u^{\rm p}_1,
    u^{\rm p}_2,
    u^{\rm p}_3
    \right)^{\rm T}
    \quad
    \text{in}
    \quad
    \Omega_{\rm p}.
\end{equation}

\begin{figure}[H]
\centering
\includegraphics[height=6.75cm]{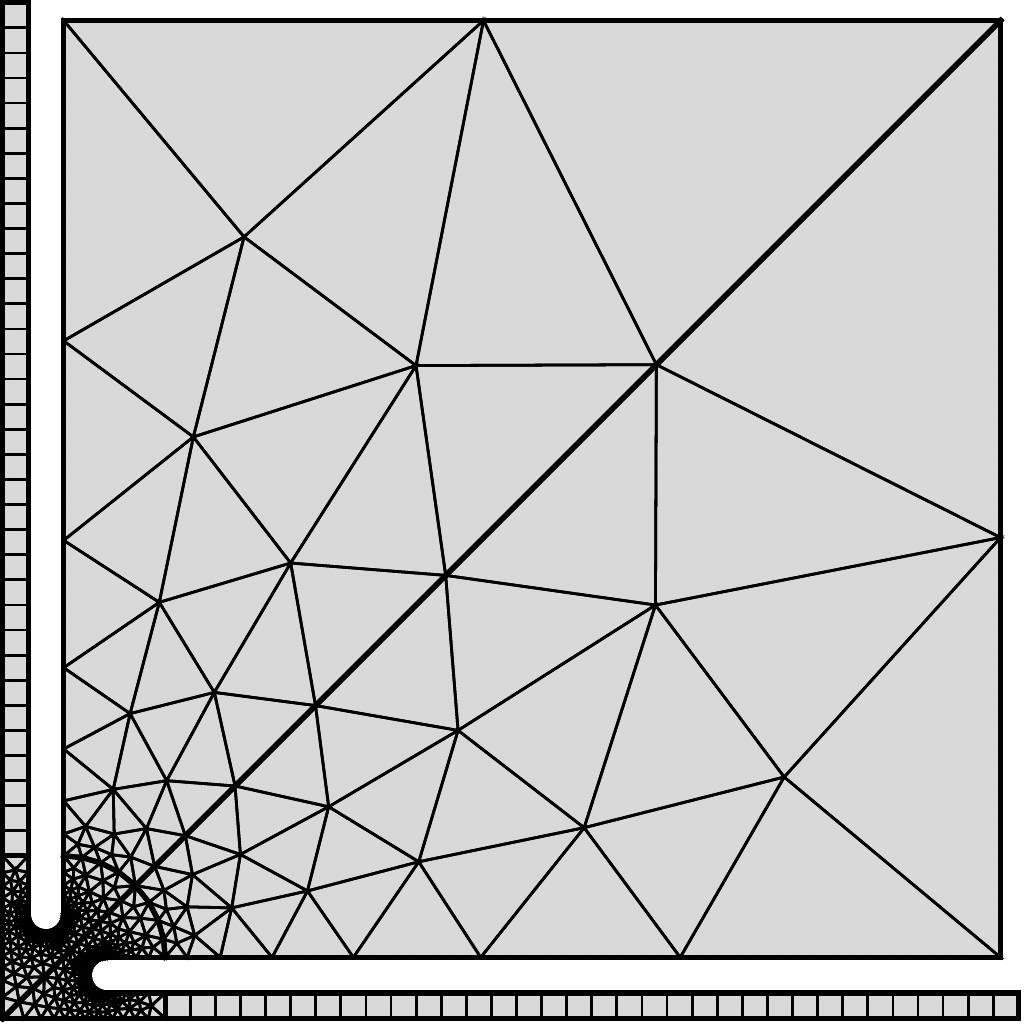}
   \caption{
   Mesh for a quarter unit cell of the microstructured model: the portions of domain with coarse mesh allow a reduction of the total number of degrees of freedom, while a finer mesh is needed in the slender portions of the domain in order to be able to properly describe the behaviour of the microstructure in the band-gap frequency range.
   }
   \label{microstructured-mesh}
\end{figure}

Since a thin plate is considered here, a plane stress hypothesis could have been used instead, but since the difference between the response of the system in terms of displacement for the two hypothesis is always smaller than $5 \%$, the plane strain hypothesis is kept for the rest of the study for the sake of simplicity. This plane strain hypothesis in $\Omega_{\rm c}$ together with the perfect contact condition interface on $\partial \Omega_\text{bottom}$ requires the following interface conditions
\begin{align}
\begin{cases}
    u^{\rm c}_1 = u^{\rm p}_1 \\
    u^{\rm c}_2 = u^{\rm p}_2 \\
    u^{\rm p}_3 = 0
    \end{cases}
     \text{on}
     \quad
     \partial \Omega_\text{bottom}
\end{align}
The differential problem in equations (4), (7) and (9), to which we refer in the following sections as ``the microstructured simulation'' or ``direct simulation'' has been implemented and numerically solved under the finite element multiphysics software \comsol using the ``Solid Mechanics'' and ``Electrostatics'' modules.
Special attention has been brought to the domain meshing as presented in Figure~\ref{microstructured-mesh}.
The small dimension of the quadrangular and triangular elements used to mesh the slender portion of the domain of the unit cell is necessary for the description of the band-gap through the local resonance, and this significantly increases the number of degrees of freedom of the finite element problem that has to be solved.
To reduce the computational burden, we chose a coarser mesh on the remaining domain that allows us to reduce the total number of degrees of freedom while still guaranteeing the reliability of the results.
Since there is a considerable number of modes in the frequency range that have been studied (0 to 2500~Hz), numerical damping has been implemented for numerical stability reasons, modifying the strain energy density of the system by
\begin{equation}
W^\text{non-conservative}_{\rm t}
=
(1+i \times 0.002)W^\text{conservative}_{\rm t}
\end{equation}
where \textit{i} is the imaginary unit.
%%%%%%%%%%%%%%%%%%%%%%%%%%%%%%%%%%%%%%%%%%%%%%%%%%%%%%%%%%%%%%%
\subsection{Relaxed micromorphic simulations}
%%%%%%%%%%%%%%%%%%%%%%%%%%%%%%%%%%%%%%%%%%%%%%%%%%%%%%%%%%%%%%%
We introduce here the effective model that we will use to describe the metamaterial's response at the macroscopic scale and which is known as relaxed micromorphic medium. The strain and kinetic energy density expressions for the relaxed micromorphic model are \cite{rizzi_exploring_2021,rizzi2021boundary} \footnote{The presence of curvature terms associated to higher space derivatives of $P$ is not accounted for in the present paper since their effect is not predominant in the dynamic regime.}
\begin{align}
W \left(\nabla u, P\right)
&= 
\dfrac{1}{2} \langle \mathbb{C}_{\rm e} \, \mbox{sym}\left(\nabla u -  \, P \right), \mbox{sym}\left(\nabla u -  \, P \right) \rangle
+ \dfrac{1}{2} \langle \mathbb{C}_{\rm c} \, \mbox{skew}\left(\nabla u -  \, P \right), \mbox{skew}\left(\nabla u -  \, P \right) \rangle
\notag
\\*[5pt]
& \quad
+ \dfrac{1}{2} \langle \mathbb{C}_{\rm micro} \, \mbox{sym}  \, P,\mbox{sym}  \, P \rangle
\, ,
\label{eq:ene_relax}
\\*[10pt]
K \left(\dot{u},\nabla \dot{u}, \dot{P}\right)
&=
\dfrac{1}{2}\rho \, \langle \dot{u},\dot{u} \rangle + 
\dfrac{1}{2} \langle \mathbb{J}_{\rm m}  \, \mbox{sym} \, \dot{P}, \mbox{sym} \, \dot{P} \rangle 
+ \dfrac{1}{2} \langle \mathbb{J}_{\rm c} \, \mbox{skew} \, \dot{P}, \mbox{skew} \, \dot{P} \rangle
+ \dfrac{1}{2} \langle \mathbb{T}_{\rm e} \, \mbox{sym}\nabla \dot{u}, \mbox{sym}\nabla \dot{u} \rangle
\notag
\\*[5pt]
& \quad 
+ \dfrac{1}{2} \langle \mathbb{T}_{\rm c} \, \mbox{skew}\nabla \dot{u}, \mbox{skew}\nabla \dot{u} \rangle
\, ,
\notag
\end{align}
where $u \in \mathbb{R}^{3}$ is the macroscopic displacement field, $P \in \mathbb{R}^{3\times3}$ is the non-symmetric micro-distortion tensor, $\mathbb{C}_{\rm e}$, $\mathbb{C}_{\rm micro}$ and $\mathbb{C}_{\rm c}$ are \nth{4} order elastic tensors, $\rho$ is the macroscopic apparent density, and $\mathbb{J}_{\rm m}$, $\mathbb{J}_{\rm c}$, $\mathbb{T}_{\rm e}$ and $\mathbb{T}_{\rm c}$ are \nth{4} order micro-inertia tensors. The action functional for the micromorphic medium is defined as :

\begin{equation}
    \mathcal{A} \left [ u,P \right ] = \int^T_0 \iint_{\Omega} (K-W)\mathrm{d}\Omega  \mathrm{d}t
\end{equation}

\begin{figure}[H]
    \centering
    \includegraphics{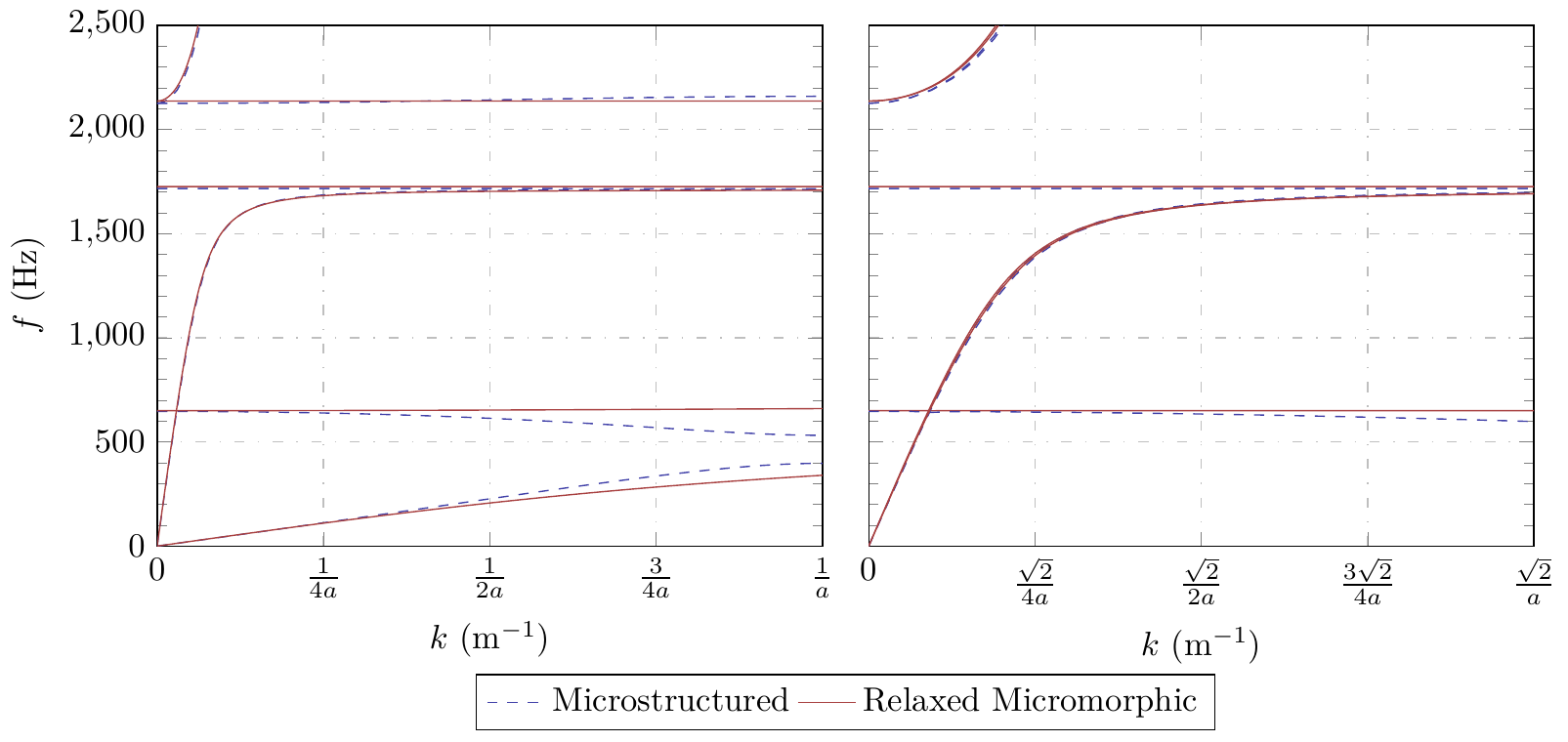}
    \caption{(\textit{left}) Dispersion curves of the microstructured and the relaxed micromorphic systems along $\Gamma$X (propagation at 0°). (\textit{right}) Dispersion curves of the microstructured and the relaxed micromorphic systems along $\Gamma$M (propagation at 45°).}
    \label{fig:dispersion}
\end{figure}

Where $\Omega$ is the domain occupied by the relaxed micromorphic medium. Requiring the first variation of the action functional with respect to $u$ and $P$ to be zero gives the following two sets of equilibrium equations, as well as the associated boundary conditions \cite{rizzi_exploring_2021,rizzi2021boundary,dagostino_effective_2020}
\begin{align}
\begin{cases}
    \rho \, \ddot{u} - \text{Div} \, \widehat{\sigma} = \text{Div} \, \widetilde{\sigma}
    \, ,
    \\
    \mathbb{J}_{\rm m} \, \text{sym } \ddot{P}
    +
    \mathbb{J}_{\rm c} \, \text{skew } \ddot{P} 
    =
    \widetilde{\sigma} - s
    \, ,
\end{cases}
\qquad
\text{in $\Omega$} \, ,
\qquad\qquad
t_{m}
\coloneqq
\left(\widetilde{\sigma} + \widehat{\sigma} \right) n
=t_{\rm m}^{\rm ext}
\qquad
\text{on $\partial \Omega$}
\, ,
\end{align}
where $n$ is the normal to the boundary $\partial \Omega$, $t_m$ is the generalized traction vector, $t_{\rm m}^{\rm ext}$ represent the external traction load, and
\begin{align}
    \widehat{\sigma} = \mathbb{T}_{\rm e} \,  \text{sym} \, \nabla \ddot{u} + \mathbb{T}_{\rm c} \, \text{skew} \, \nabla \ddot{u}
     \, ,
    \qquad
    \widetilde{\sigma} = \mathbb{C}_{\rm e} \, \text{sym} \,  ( \nabla u - P ) + \mathbb{C}_{\rm c} \, \text{skew} ( \nabla u - P )
    \, ,
    \qquad
    s = \mathbb{C}_{\rm m} \, \text{sym} \, P
    \, .
\end{align}
All the previous results hold for a generic class of material symmetry and in the following the elastic and the micro-inertia tensors will be presented in the Voigt notation for the tetragonal class of symmetry
\begin{equation}
\begin{array}{rlrl}
	&
    \mathbb{C}_{\rm e} = 
    \begin{pmatrix}
    \lambda_{\rm e} + 2\mu_{\rm e}	& \lambda_{\rm e}				& \dots		& \bullet\\ 
    \lambda_{\rm e}				& \lambda_{\rm e} + 2\mu_{\rm e}	& \dots		& \bullet\\ 
    \vdots					& \vdots					& \ddots	& 		 \\ 
    \bullet					& \bullet					& 			& \mu_{\rm e}^{*}\\ 
    \end{pmatrix} \, ,
    \quad\qquad\qquad
    \mathbb{C}_{\rm micro} = 
    \begin{pmatrix}
    \lambda_{\rm m} + 2\mu_{\rm m}	& \lambda_{\rm m}				& \dots		& \bullet\\ 
    \lambda_{\rm m}				& \lambda_{\rm m} + 2\mu_{\rm m}	& \dots		& \bullet\\ 
    \vdots					& \vdots					& \ddots	&  \\ 
    \bullet					& \bullet					& 			& \mu_{\rm m}^{*}\\ 
    \end{pmatrix} \, ,
    \\[1cm]
    &
    \mathbb{J}_{\rm m} =
	\begin{pmatrix}
	\eta_{3} + 2\eta_{1} & \eta_{3}              & \dots 			& \bullet \\ 
	\eta_{3}            & \eta_{3} + 2\eta_{1} & \dots 			& \bullet \\ 
	\vdots               & \vdots                 & \ddots 			& \bullet \\ 
	\bullet              & \bullet                & \bullet 		& \eta^{*}_{1} \\ 
	\end{pmatrix}
	\, ,
	\qquad\qquad\quad\;\;
	\mathbb{T}_{\rm e} =
	\begin{pmatrix}
	\overline{\eta}_{3} + 2\overline{\eta}_{1}	& \overline{\eta}_{3}        			   	& \dots		& \bullet\\ 
	\overline{\eta}_{3}         				    & \overline{\eta}_{3} + 2\overline{\eta}_{1} 	& \dots		& \bullet\\ 
	\vdots                    					& \vdots                 				   	& \ddots 	&		 \\ 
	\bullet                   					& \bullet									& 	 		& \overline{\eta}^{*}_{1}
	\end{pmatrix} \, ,
    \\[1cm]
	&\;
	\mathbb{C}_{\rm c} = 
    \begin{pmatrix}
    \bullet & 			& \bullet\\ 
    		& \ddots 	& \vdots\\ 
    \bullet & \dots		& 4\mu_{\rm c}
    \end{pmatrix} \, ,
    \qquad\qquad
	\mathbb{J}_{\rm c} =
	\begin{pmatrix}
	\bullet & 			& \bullet\\ 
	& \ddots 	& \vdots\\ 
	\bullet & \dots & 4\eta_{2}
	\end{pmatrix}
    \, ,
    \qquad\qquad
    \mathbb{T}_{\rm c} =
	\begin{pmatrix}
	\bullet & 			& \bullet\\ 
	& \ddots 	& \vdots\\ 
	\bullet & \dots & 4\overline{\eta}_{2}
	\end{pmatrix}
	\, ,
    \end{array}
\label{eq:micro_ine_1}
\end{equation}
Given the plane strain hypothesis, the values of the "dotted" coefficients are not necessary to our study. The assumption regarding the class of symmetry is driven by considerations regarding the symmetry of the unit cell in Figure~\ref{fig:fig_tab_unit_cell}.
In previous papers \cite{rizzi_exploring_2021,rizzi2021boundary,aivaliotis_frequency-_2020}, we showed that a specific calibration procedure can be applied to compute the values of the relaxed micromorphic parameters for the metamaterial issued from the unit cell in Figure~\ref{fig:fig_tab_unit_cell}.
This procedure is based on the comparison of the relaxed micromorphic dispersion curves with those obtained \textit{via} a classic Bloch-Floquet analysis done on the unit cell.
The comparison of the dispersion curves of the relaxed micromorphic medium with those obtained \textit{via} Bloch-Floquet analysis is shown in Figure~\ref{fig:dispersion}.

The values of the relaxed micromorphic parameters relative to the metamaterial in Figure~\ref{fig:fig_tab_unit_cell} are presented in Table \ref{table:microp} (\textit{left}), while in Table \ref{table:microp} (\textit{right}) are reported also the coefficients of the Cauchy material resulting from the long-wave limit of the relaxed micromorphic one.
\begin{table}[H]
    \renewcommand{\arraystretch}{1.2}
    \centering
    \begin{subtable}[t]{.55\textwidth}
    \centering
    \begin{tabular}{cccccc}
        \\
        \thickhline
        \thickhline
        & $\rho$ & $\mu_{\rm e}$ & $\lambda_{\rm e}$ & $\mu_{\rm e}^\star$ &
        \\
        & [\text{kg/m}$^3$] & [Pa] & [Pa] & [Pa] &
        \\
        \hline
        & 3841 & 2.53$\cross 10^9$ & 1.01$\cross 10^8$ & 1.26 $\cross 10^{6}$ &
        \\
        \thickhline
        & $\mu_{\rm m}$ & $\lambda_{\rm m}$ & $\mu_{\rm m}^\star$ & $\mu{\rm c}$ &
        \\
        & [Pa] & [Pa] & [Pa] & [Pa] &
        \\
        \hline
        & 4.51 $\cross 10^{9}$ & 1.83 $\cross 10^{8}$ & 2.70 $\cross 10^{8}$ & $10^{5}$ &
        \\
        \thickhline
        & $\eta_1$ & $\eta_2$ & $\eta_3$ & $\eta_1^\star$ &
        \\
        & [\text{kg/m}] & [\text{kg/m}] & [\text{kg/m}] & [\text{kg/m}] &
        \\
        \hline
        & 38.99 & 5.99$\cross 10^{-3}$ & 1.58 & 2.31 &
        \\
        \thickhline
        & $\overline{\eta}_1$ & $\overline{\eta}_2$ & $\overline{\eta}_3$ & $\overline{\eta}^\star_1$ &
        \\
        & [\text{kg/m}] & [\text{kg/m}] & [\text{kg/m}] & [\text{kg/m}] &
        \\
        \hline
        & 8$\cross 10^{-4}$ & 0.02 & 0.008 & 0.09 &
        \\
        \thickhline
        \thickhline
    \end{tabular}
    \end{subtable}
    \hfill
	\centering
	\begin{subtable}[t]{.44\textwidth}
    \centering
    \begin{tabular}{ccccc}
        \thickhline
        \thickhline
    	& $\lambda_{\tiny \rm macro}$ & $\mu_{\tiny \rm macro}$ & $\mu^{*}_{\tiny \rm macro}$ &
    	\\
    	& [Pa] & [Pa] & [Pa] &
    	\\
    	\hline
    	& $6.51\times10^7$ & $1.62\times10^9$ & $1.25\times10^6$ &
    	\\
    	\thickhline
    	\thickhline
    \end{tabular}
	\end{subtable}
    \caption{
    (\textit{left})
    Values of the elastic and micro-inertia relaxed micromorphic parameters calibrated on the metamaterial whose unit cell is reported in Fig.~\ref{fig:fig_tab_unit_cell}, and 
    (\textit{right})
    the corresponding long-wave limit Cauchy material $\mathbb{C}_{\tiny \rm macro}$.}
     \label{table:microp}
\end{table}

\begin{figure}[H]
\centering
\raisebox{-.5\height}{\includegraphics[width=0.4725\textwidth]{drawings/sym_3d_microstructured_plate.pdf}}
\hfill
\raisebox{-.5\height}{\includegraphics[width=0.4725\textwidth]{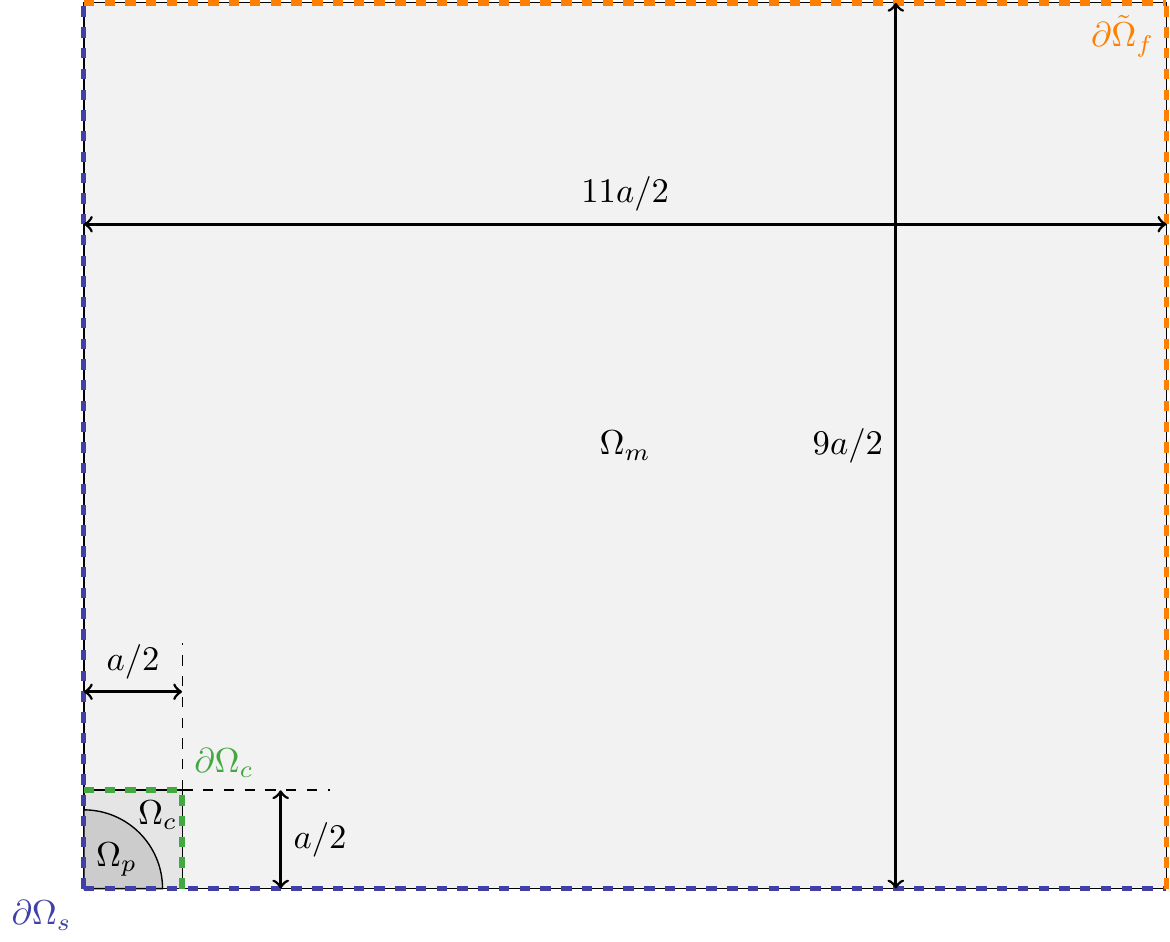}}
\includegraphics{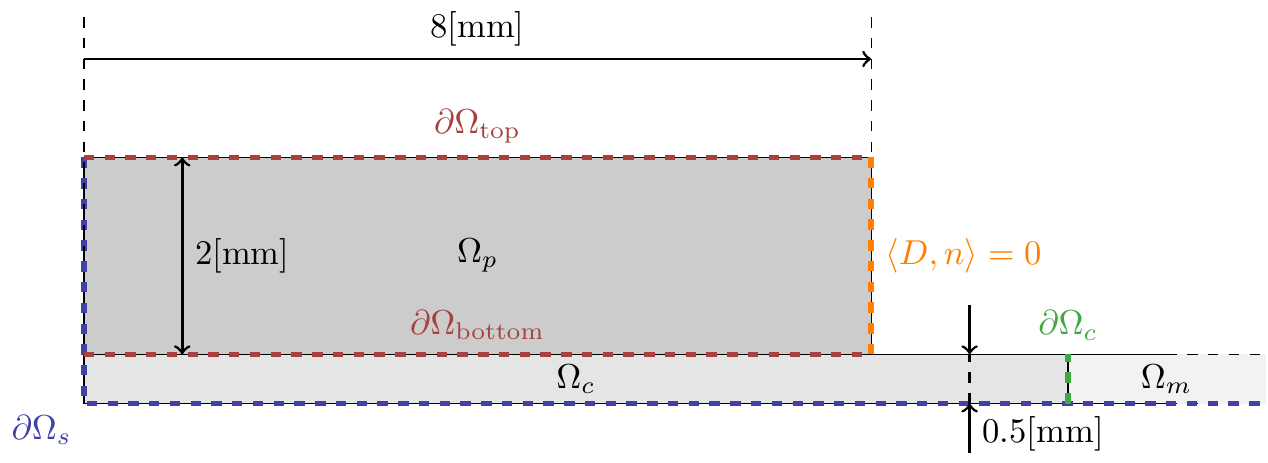}  
\caption{
(\textit{left, top}) One eighth of the microstructured plate with its top piezoelectric patch. 
(\textit{right, top}) Equivalent relaxed micromorphic plate with its top piezoelectric patch: $\Omega_{\rm m}$ is the domain of the relaxed micromorphic medium, $\partial \tilde{\Omega}_{\rm f}$ is the generalized traction-free border, $\partial \Omega_{\rm c}$ is the interface between the Cauchy and relaxed micromorphic domains, and $\partial \Omega_{\rm s}$ is the union of the boundaries on the plane of symmetry.
We refer to this configuration as ``reduced micromorphic problem''.
(\textit{bottom}) Front view of the reduced relaxed micromorphic plate.
}
\label{fig:micromorphic3d}
\end{figure}

In Figure \ref{fig:micromorphic3d} we present a reduced problem also for the relaxed micromorphic model consisting of an equivalent plate mimicking the considered microstructured plate under piezoelectric excitation.

The piezoelectric patch and the central isotropic Cauchy medium description presented in Section \ref{sec:micro_mate_characteristic} remain unchanged, while it is worth to focus on the plain strain hypothesis for the relaxed micromorphic model, which implies :
\begin{equation}
u^{\rm c} =
    \begin{pmatrix}
    u^{\rm c}_{1} \\
    u^{\rm c}_{2} \\
    u^{\rm c}_{3}
    \end{pmatrix}
    \quad
    \text{in}
    \quad
    \Omega_{\rm p},
    \quad
    u^{\rm c} =
    \begin{pmatrix}
    u^{\rm c}_{1} \\
    u^{\rm c}_{2} \\
    0
    \end{pmatrix}
    \quad
    \text{in}
    \quad
    \Omega_{\rm c},
    \quad
    u^{\rm m} =
    \begin{pmatrix}
    u^{\rm m}_{1} \\
    u^{\rm m}_{2} \\
    0
    \end{pmatrix}
    \,\,\,
    \text{and}
    \,\,\,
    P =
    \begin{pmatrix}
    P_{11} & P_{12} & 0 \\
    P_{21} & P_{22} & 0 \\
    0 & 0 & 0
    \end{pmatrix}
    \quad
    \text{in}
    \quad
    \Omega_{\rm m}.
\end{equation}

The perfect contact conditions between the Cauchy material and the relaxed micromorphic material at the interfaces $\partial \Omega_{\rm c}$ and the traction-free conditions on $\partial \tilde{\Omega}_{\rm f}$ are
\begin{equation}
    \begin{cases}
        u^{c} = u^{m} \\
        (\widehat{\sigma}+\widetilde{\sigma}) \cdot n = \sigma \cdot n
    \end{cases}
    \quad
    \text{on}
    \quad
    \partial\Omega_{\rm c},
    \qquad
    (\widehat{\sigma}+\widetilde{\sigma}) \cdot n = 0 \quad \text{on} \quad \partial \tilde{\Omega}_{\rm f}
    \label{continuity}
\end{equation}

On the boundary of symmetry $\partial \Omega_{\rm s}$ we have to impose the following boundary conditions\footnote{See Appendix A for a derivation of these conditions.}
\begin{equation}
\begin{aligned}
    &
    \begin{cases}
    u_i n_i= 0 \\
    (\delta_{ki}-n_k n_i) (P_{ij} n_j) = 0
    \end{cases}
\end{aligned}
\label{eq:sym3d}
\end{equation}
where, $n_i$ are the components of the unit normal to each surface and $\delta_{ij}$ is the Kronecker delta operator. The relaxed micromorphic model not being implemented under Comsol by default, the ``Weak form PDE'' module has been used, requiring to write explicitly energy densities and imposing manually boundary conditions presented above. The associated Lagrangian for the plate consisting of Cauchy and relaxed micromorphic media, under the plane-strain hypothesis, is

\begin{equation}
    \mathcal{A} \left [ u,P \right ] = \frac{e}{2} \int^T_0 \left [ \iint_{\Omega_{\rm c}} (K_{\rm c}-W_{\rm c})\mathrm{d}x_1\mathrm{d}x_2 + \iint_{\Omega_m} (K_m-W_m)\mathrm{d}x_1\mathrm{d}x_2 \right ] \mathrm{d}t
\end{equation}

\begin{figure}[H]
\centering
\includegraphics[height=6.75cm]{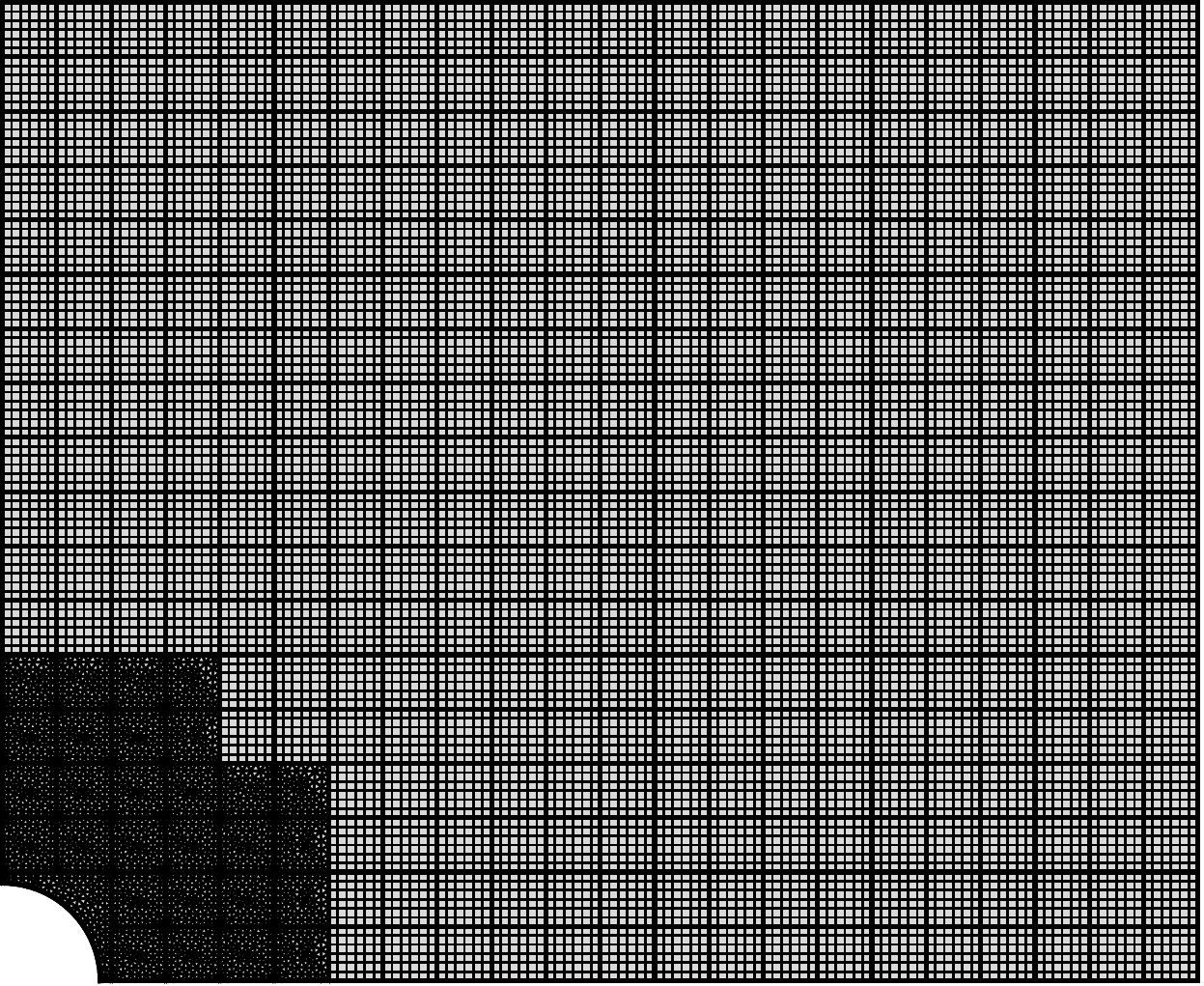}
   \caption{Chosen mesh for the reduced relaxed micromophic plate with its central Cauchy medium inclusion.}
   \label{micromesh}
\end{figure}

Using a unique field $u$ which is equal to $u^{\rm c}$ when restricted to $\Omega_{\rm c}$ and to $u^{\rm m}$ when restricted to $\Omega_m$ allows to simplify the numerical implementation of the problem, automatically guaranteeing the continuity of displacement at the interface $\partial \Omega_{\rm c}$. The mesh used for the relaxed micrmorphic relaxed problem is given in Figure~\ref{micromesh}.

Quadratic Lagrange elements are used for the discretization of $u$, which allows us to recover the continuity of $\nabla u$ and therefore of generalized tractions. No gradient being applied upon $P$, we discretize $P$ through form functions an order below (here, linear Lagrange elements), in terms of regularity, of the one used for $u$. Despite the introduction of an additional 2nd order tensor to describe the response of the plate and the tight mesh around the excitation domain required to assure the slow convergence of $P$, the relaxed micromorphic model allows a considerable reduction in the number of degrees of freedom with respect to the microstructured model, as shown in Table \ref{dof}.

\begin{table}[H]
    \renewcommand{\arraystretch}{1.2}
    \centering
\begin{tabular}{ccccc}
    \thickhline
        \thickhline
    	& Degrees of freedom & Microstructured plate & Relaxed micromorphic plate & \\
    	\hline
    	& $9\times 11$ plate & 493,674 & 67,048 & \\
    	& $49\times 51$ plate & 9,726,194 & 1,300,302 & \\
    	\thickhline
    	\thickhline
\end{tabular}
\caption{Numbers of degrees of freedom for the symmetrized microstructured and the relaxed micromorphic plates.}
\label{dof}
\end{table}

As justified in section 3.1, the considered structure requires to be damped for numerical stability. As before, a hysteretic damping is introduced in the plate, modifying the strain energy density in the bidimensionnal plate as
\begin{align}
    \begin{cases}
        W^{\rm c}_\text{non-conservative}=(1+i \eta)W^{\rm c}_\text{conservative} \\
        W^{\rm m}_\text{non-conservative}=(1+i \eta)W^{\rm m}_\text{conservative}
    \end{cases}\text{ where } \eta = 0.002
\end{align}
% Given the small damping considered here, the precise value of $\eta$ chosen for the simulations does not have much importance : using Basile hypothesis, one could prove that the only effect of such damping is to clip resonance peaks at the resonance frequencies.
%%%%%%%%%%%%%%%%%%%%%%%%%%%%%%%%%%%%%%%%%%%%%%%%%%%%%%%%%%%%%%%
\section{Metamaterial's plate under a pulse excitation : microstructured versus relaxed micromorphic simulations}
%%%%%%%%%%%%%%%%%%%%%%%%%%%%%%%%%%%%%%%%%%%%%%%%%%%%%%%%%%%%%%%
In this section, we start simulating the experimental set-up presented in section 2 both \textit{via} the microstructured simulations and the relaxed micromorphic ones when considering a pulse excitation to give a first description of the piezoelectric load.

\begin{figure}[H]
\centering
\includegraphics[width=0.495\textwidth]{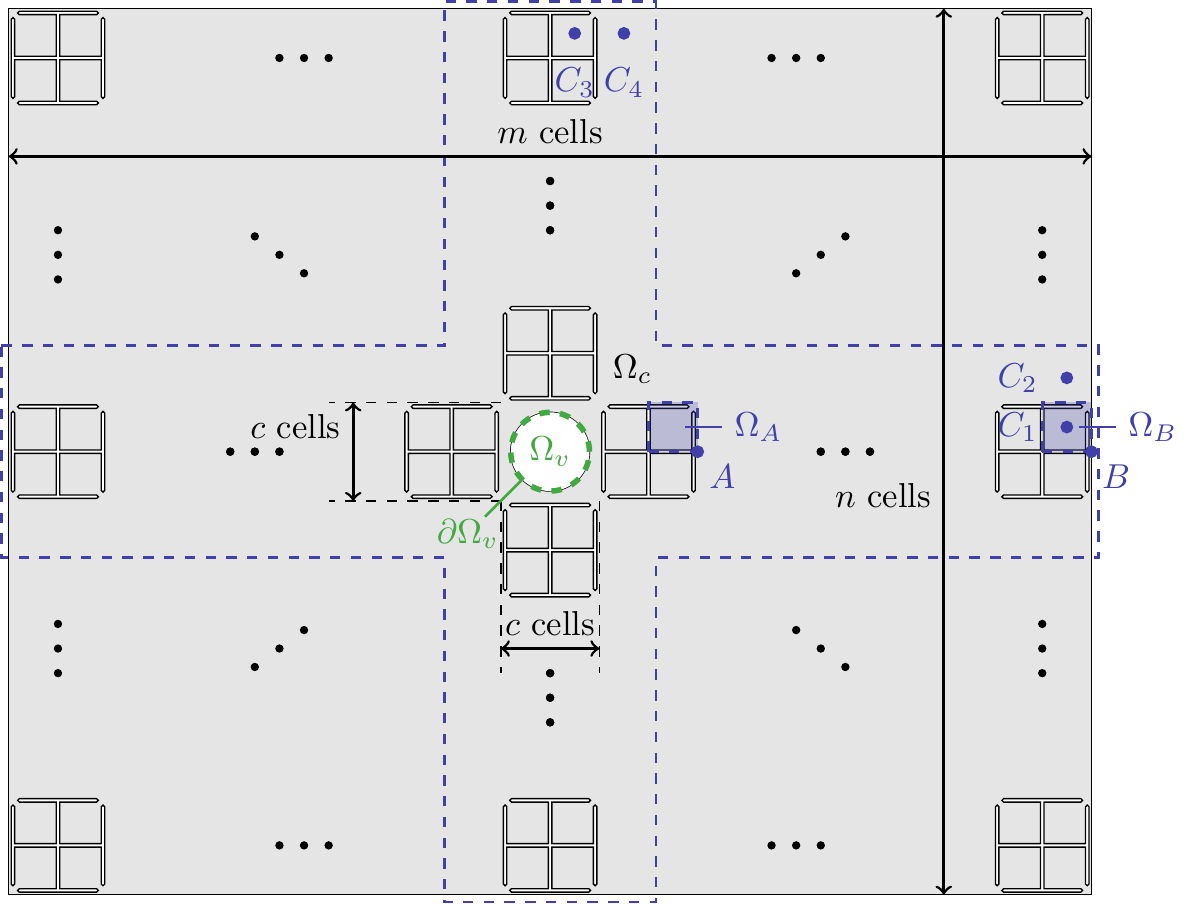}
\includegraphics[width=0.495\textwidth]{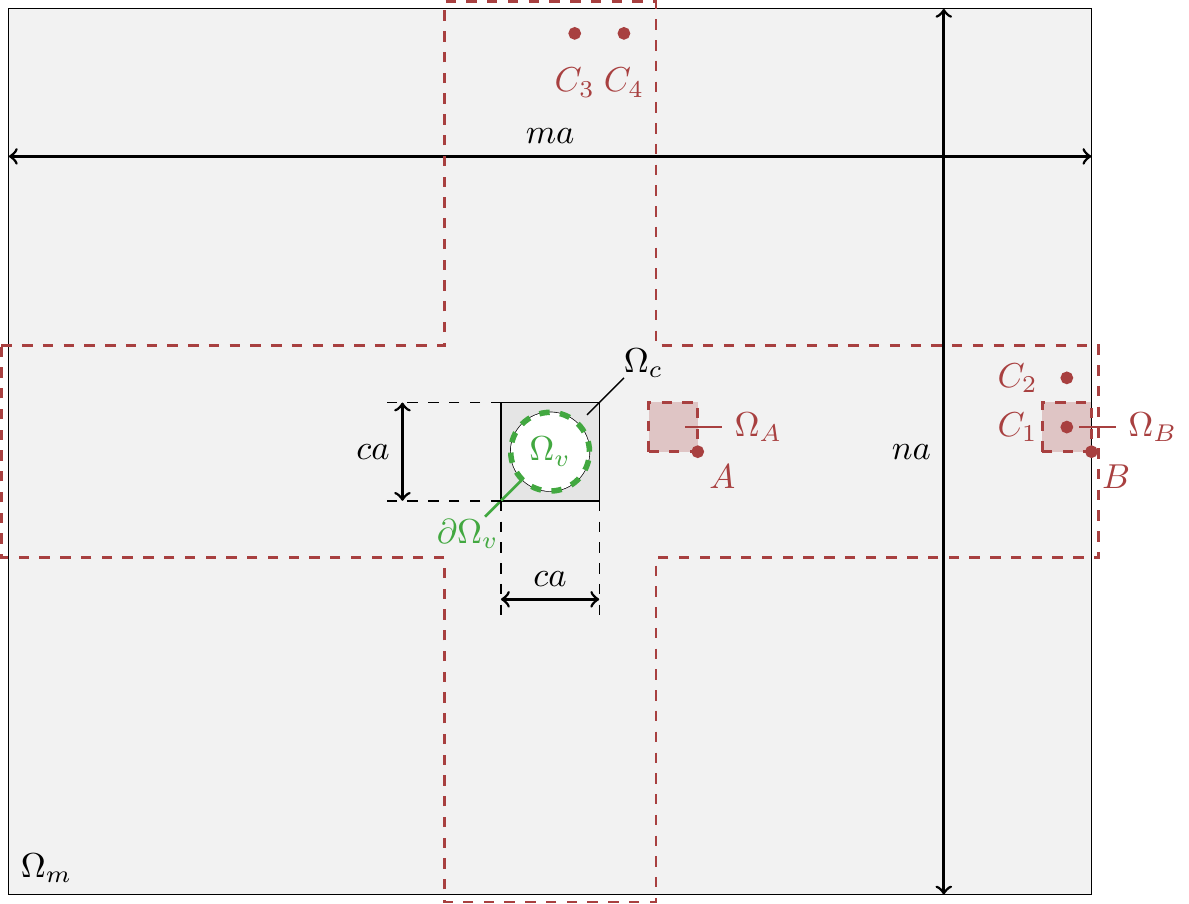}
\caption{(\textit{left}) Top view of the full microstructured plate and identification of points A and B. (\textit{right}) Top view of the full relaxed micromorphic plate  and identification of the corresponding volumes $\Omega_A$ and $\Omega_B$. Given the strong directionality of the plate, we do not consider other points outside the dashed domain to show the simulation's results. Points $C_1$, $C_2$, $C_3$ and $C_4$ are also located since they are used to present further results.}
\label{fig:interestingzones}
\end{figure}

More precisely, to mimick the application of the piezoelectric load, we start implementing a simplified framework in which the piezoelectric region $\Omega_{\rm p}$ is modeled as a void region $\Omega_v$.
The applied load is given as an imposed displacement on the boundary $\partial \Omega_v$ (see Figure~\ref{fig:interestingzones}) in the form :
\begin{equation}
        u = \psi \, n \text{ (expansion load)}
\end{equation}
where $n$ is the unit normal to the each surface and $\psi=10^{-3}$ [m]. The other interface conditions are given in equations (\ref{continuity}) and (\ref{eq:sym3d}).
Since the relaxed micromorphic model is a macroscopic model it is not always worth comparing the solution displacement field pointwise with the one issued \textit{via} the microstructured simulations. A consistent difference between these pointwise fields may be expected. To provide a more stable comparison an average displacement field over a representative portion of the unit cell can be considered. To this aim we start identifying the points $A$ and $B$ in the considered structure as (see also Figure~\ref{fig:interestingzones}):
\begin{align}
\begin{cases}
A & = \Big( 3a/2,0 \Big)^T \\
B & = \Big( ma/2,0 \Big) ^T %\text{ (respecting Saint's Venant Principle)}
\end{cases}
\text{ and surfaces }
\begin{cases}
\Omega_A & = \Big[a,3a/2 \Big] \times \Big[0,a/2 \Big] \\
\Omega_B & = \Big[ (m-1)a/2,ma/2 \Big] \times \Big[ 0,a/2 \Big] %\text{ (respecting Saint's Venant Principle)}
\end{cases}
\end{align}
where $m$ is the number of cells of the plate on its main axis (see Figure~\ref{fig:interestingzones}). We then introduce a pointwise measure of displacement $p$ and a mean measure of displacement $m$ as :
\begin{equation}
\begin{aligned}
    p_X = \frac{1}{\psi} \sqrt{\langle u(X) , \bar{u}(X) \rangle} 
    \text{ and }
    m_X = \frac{4}{\psi a^2} \iint_{\Omega_X} \sqrt{\langle u(x_1,x_2),\bar{u}(x_1,x_2) \rangle} \mathrm{d}x_1\mathrm{d}x_2 \text{ for } X = \{ A,B \} \\
\end{aligned}
\end{equation}
where a superposed bar indicates the complex conjugate operation and $\psi$ the amplitude of the expansion load introduced before. The hermitian norm used here, not necessary for the static response of the system, where the displacement stays real despite the hysteretic damping, finds its use computing the dynamic response of the plate. For the subsequent purposes of comparison with the experiment, we also introduce four points $C_i$ $i \in \{ 1,4 \}$ far from the excitation. As these points are only considered in comparison with the experimental setup, we directly give their coordinates for $m=11$, $n=9$ and $c=1$ in Table \ref{mesp}.

\begin{table}[H]
    \renewcommand{\arraystretch}{1.2}
    \centering
\begin{tabular}{ccccccc}
    \thickhline
        \thickhline
    	& Point & $C_1$ & $C_2$ & $C_3$ & $C_4$ & \\
    	\hline
    	& $x_1$ coordinate [mm] & 105 & 105 & 5 & 15 & \\
    	& $x_2$ coordinate [mm] & 5 & 15 & 85 & 85 & \\
    	\thickhline
    	\thickhline
\end{tabular}
\caption{Coordinates of the measurement points.}
\label{mesp}
\end{table}

%%%%%%%%%%%%%%%%%%%%%%%%%%%%%%%%%%%%%%%%%%%%%%%%%%%%%%%%%%%%%%%
\subsection{The long-wave limit : statics}
%%%%%%%%%%%%%%%%%%%%%%%%%%%%%%%%%%%%%%%%%%%%%%%%%%%%%%%%%%%%%%%
The main challenge for metamaterials' modelling consists in the description of their broadband mechanical response. More particularly, a suitable model must be able to describe metamaterials' response for the larger possible set of frequencies. We will show in the following sections that the RMM is able to correctly describe the metamaterial's response for a very wide side set of frequencies going well beyond the first band-gap.

Nevertheless, specific attention must be payed to the so-called long-wave or static limit which can be recovered from the dynamic model when considering very small frequencies, in the limit $\omega \rightarrow 0$.

\begin{figure}[H]
    \centering
    \includegraphics{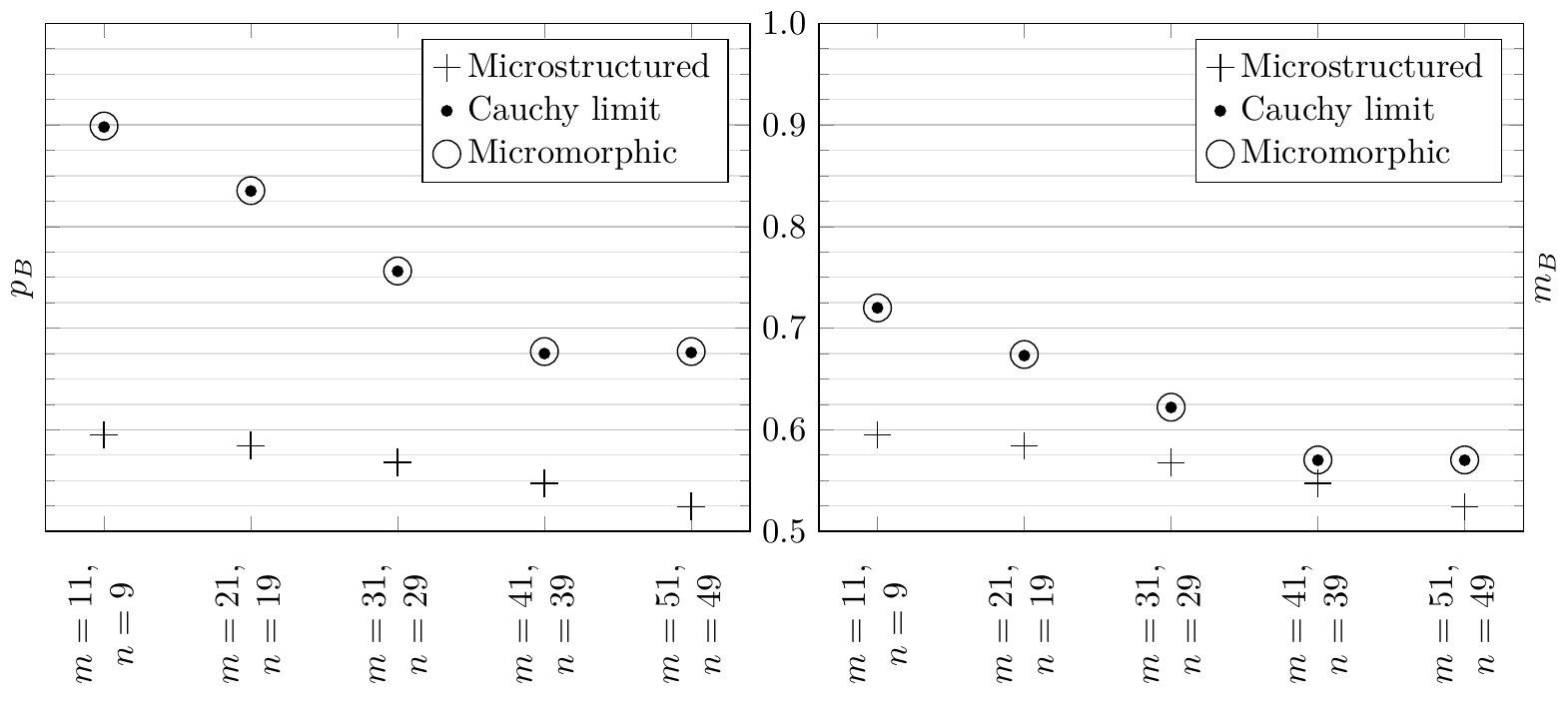}
    \caption{Pointwise (\textit{left}) and mean (\textit{right}) displacement for the static responses of the microstructured, the homogenized and the relaxed micromorphic models at point B for $c=1$. Similar results are valid for point A.}
    \label{fig:mn}
\end{figure}

In this section, we explicitely point out this static limit both for the microstructured and the relaxed micromorphic model. We show that, since internal lengths are neglected, the relaxed micromorphic static limit coincides with an equivalent Cauchy medium. We remark that for the experimental metamaterial's specimen's size (9 $\times$ 11 cells), this equivalent Cauchy medium slightly deviates from the static response of the full microstructured metamaterial. However, this difference remains smaller than 10 $\%$ (see Figure~\ref{fig:mn}) and becomes even smaller as soon as higher frequencies are considered.

To improve the relaxed micromorphic response of these small samples for the static limit, internal lengths should be introduced. This would lead, on the other hand, to a more complex identification procedure for the dynamic regime. We thus limit ourselves to the case of negligible internal lengths, knowing that this leads to a controlled inaccuracy in the static limit for small specimens.

It is well known that a given metamaterial can be modeled as a Cauchy continuum in the static regime as soon as a specimen of \textit{suitable large} size is considered. By direct inspection of Figure~\ref{fig:mn} we can infer that the metamaterial considered in this paper behaves as a Cauchy continuum in the static regime when considering specimens that are greater than 30$\times$30 cells, for the pulse excitation shown in Figure \ref{fig:interestingzones}.

\begin{figure}[H]
    \centering
    \includegraphics{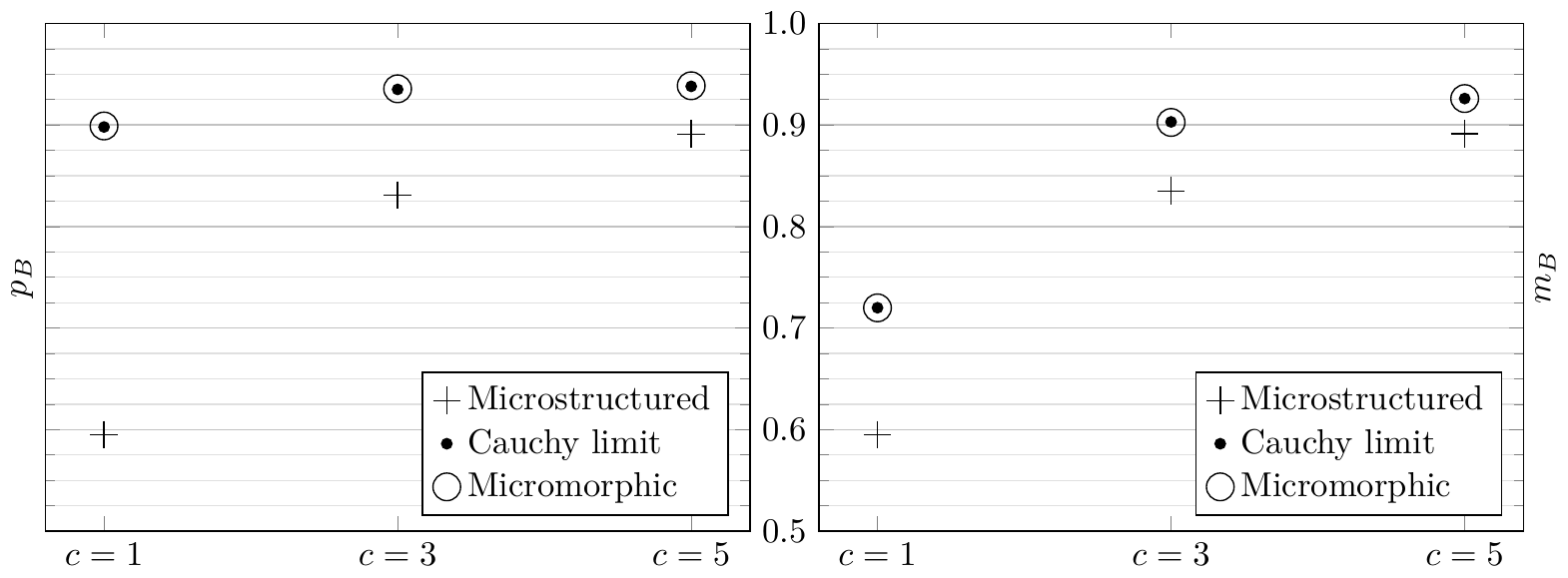}
    \caption{Pointwise (\textit{left}) and mean (\textit{right}) displacement for the static responses of the microstructured, the homogenized and the relaxed micromorphic models at point B for ($m=11$,$n=9$). Similar results are valid for point A.}
\label{fig:csmall}
\end{figure}

\begin{figure}[H]
    \centering
    \includegraphics{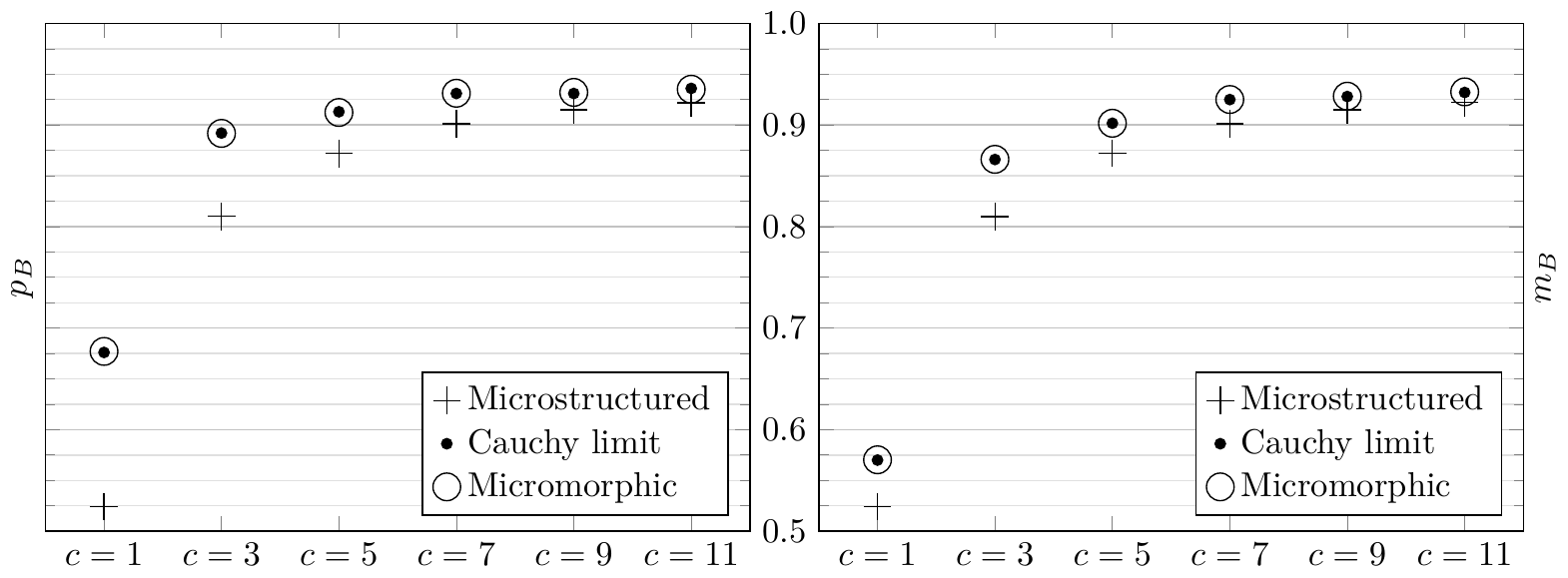}
    \caption{Pointwise (\textit{left}) and mean (\textit{right}) displacement for the static responses of the microstructured, the homogenized and the relaxed micromorphic models at point B for ($m=51$,$n=49$). Similar results are valid for point A.}
\label{fig:cbig}
\end{figure}

The convergence of the considered metamaterial towards a proper Cauchy material in the static regime can be also achieved on smaller specimens by suitably changing the applied load (\textit{e.g.}, increasing the size of the region where the load is applied). This can be seen from Figures \ref{fig:csmall} and \ref{fig:cbig} ; when considering a load applied on a square whose side is 3 unit cells, the obtained static response can be modeled as a Cauchy continuum even for small specimen sizes (9$\times$11 cells).

In summary, we have shown that our hypothesis of neglecting static internal lengths may produce a small and controlled inaccuracy for the static limiting case when considering a specimen of the size considered in our experiment (9$\times$11 cells) and an external load applied on a unique unit cell ($c=1$)\footnote{The parameter $c$ indicates, as presented in Figure \ref{fig:interestingzones}, the extension of the region where the applied load is applied : $c=1$ represents a $1\times1$ cells central region, $c=2$ a $2\times2$ cells central region, etc. We will show in section 4.2 that the extension of this region is necessary to the correct description of the microstructured plate by the relaxed micromorphic model.}.

Since, on the other hand, this hypothesis drastically simplifies the characterization procedure for the dynamic regime, we decide to keep it in the remainder of the paper.

%%%%%%%%%%%%%%%%%%%%%%%%%%%%%%%%%%%%%%%%%%%%%%%%%%%%%%%%%%%%%%%
\subsection{Broadband dynamics of the metamaterial's plate}
%%%%%%%%%%%%%%%%%%%%%%%%%%%%%%%%%%%%%%%%%%%%%%%%%%%%%%%%%%%%%%%

In this section, we present the broadband response for both the microstructured and relaxed micromorphic plate when considering a plate size of 51$\times$49 cells and a load applied on a square whose side is 11 unit cells. Based on the results of the previous section this choice allows to precisely recover the long-wave limit and shows the capability of the relaxed micromorphic model to correctly reproduce the dynamical response of the considered metamaterial for a wide set of frequencies (from zero to beyond the first band-gap).

\begin{figure}[H]
\centering
\includegraphics{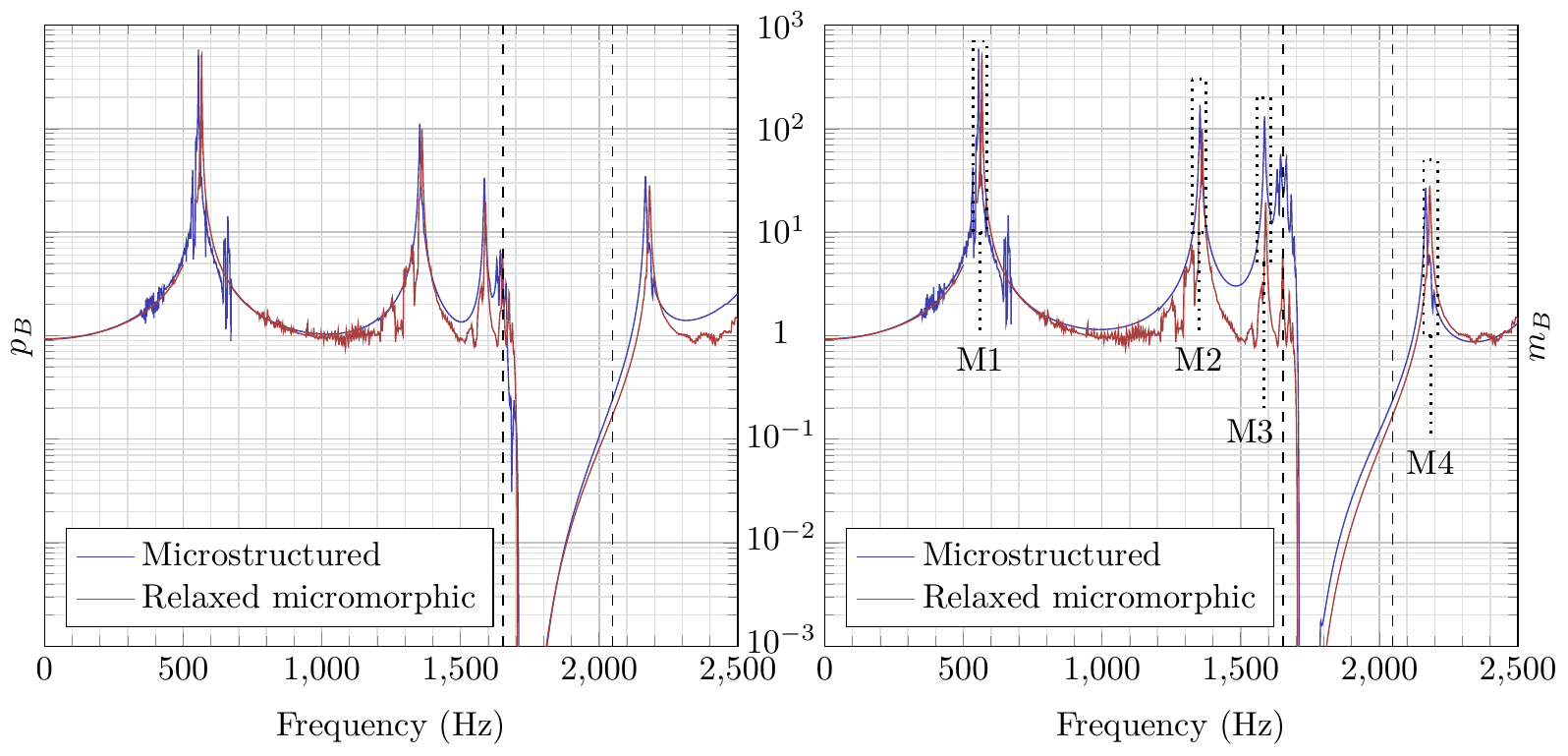}
\caption{(\textit{left}) Pointwise displacement $p_B$ of the microstructured and relaxed micromorphic models at point $B$ with the theoretical band-gap (dashed) for $m=51$, $n=49$ and $c=11$. (\textit{right}) Mean displacement $m_B$ of the microstructured and relaxed micromorphic models with the theoretical band-gap (dashed) for $m=51$, $n=49$ and $c=11$.}
\label{fig:frfconverge}
\end{figure}

Figure \ref{fig:frfconverge} shows this broadband response for the considered metamaterial plate : it is apparent that the relaxed micromorphic model describes well the plate's behavior for the whole considered frequency range.
The size of the considered plate (49$\times$51) was still allowing a direct comparison of the relaxed micromorphic model with the microstructured simulations. However, the computational time was considerably higher for the microstructured plate. Explicit comparison for larger plates would be out of reach with standard computational tools. This calls for the importance of our model in view of its use for the design of large-scale engineering metastructures.

\begin{figure}[H]
\centering
\includegraphics{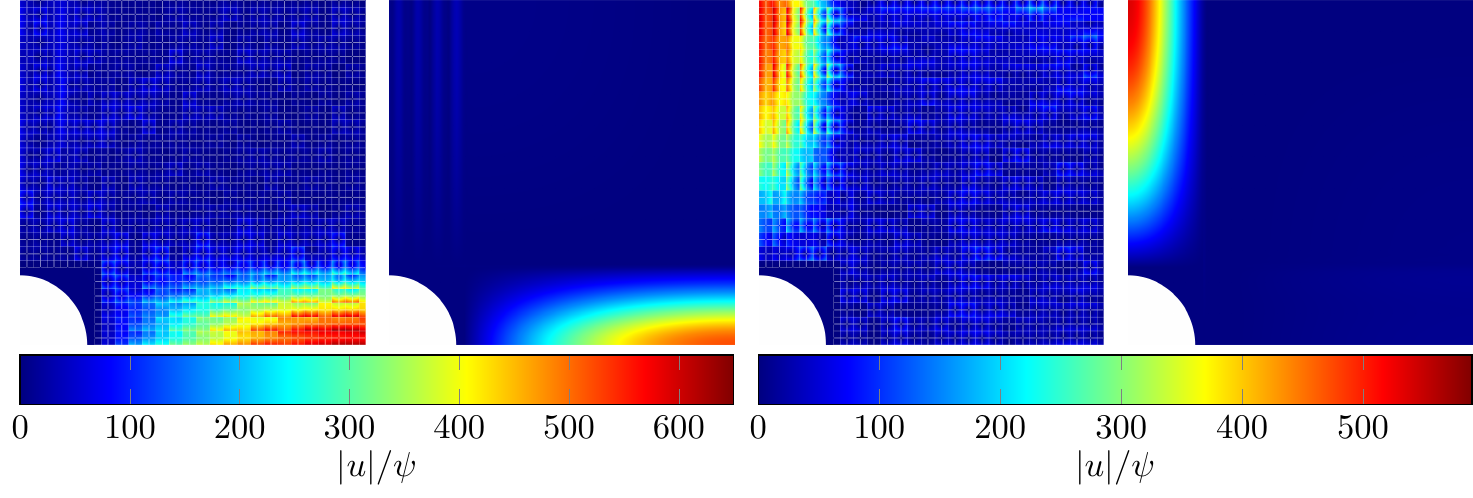}
\caption{$|u|/\psi$ at frequency $M_1$ for the microstructured model and the relaxed micromorphic model, the first two figures correspond to $M_1$ (left) and the last two figures to $M_1$ (right) (see Figure~\ref{fig:frfconverge} for the definition of these points).}
\label{m1}
\end{figure}

\begin{figure}[H]
\centering
\includegraphics{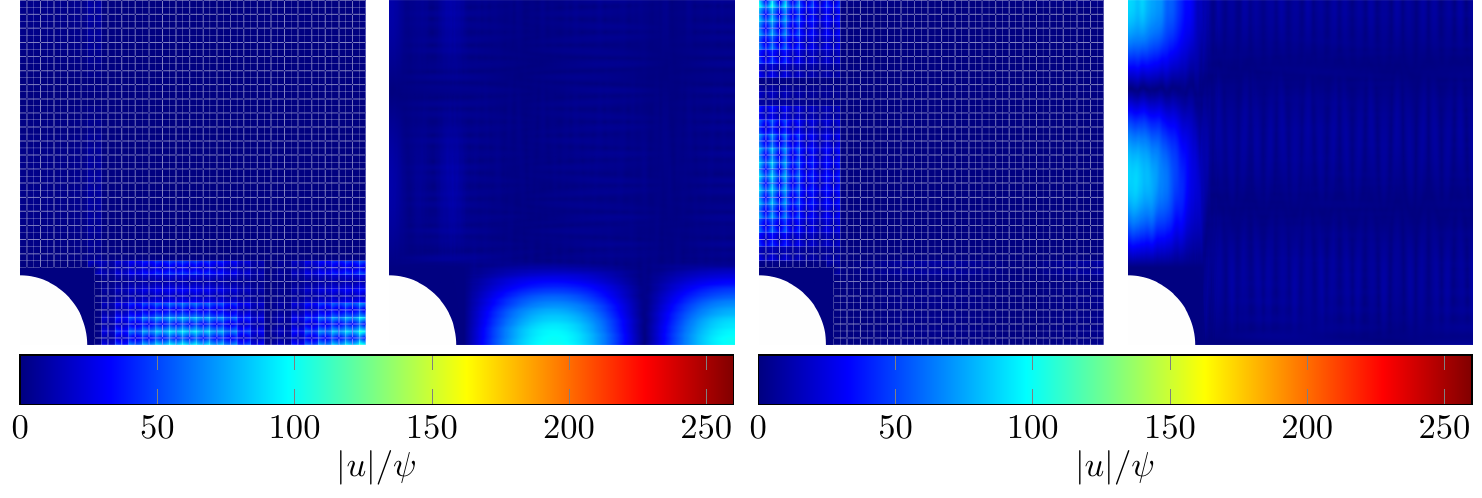}
\caption{$|u|/\psi$ at frequency $M_2$ for the microstructured model and the relaxed micromorphic model, the first two figures correspond to $M_2$ (left) and the last two figures to $M_2$ (right) (see Figure~\ref{fig:frfconverge} for the definition of these points).}
\label{m2}
\end{figure}

\begin{figure}[H]
\centering
\includegraphics{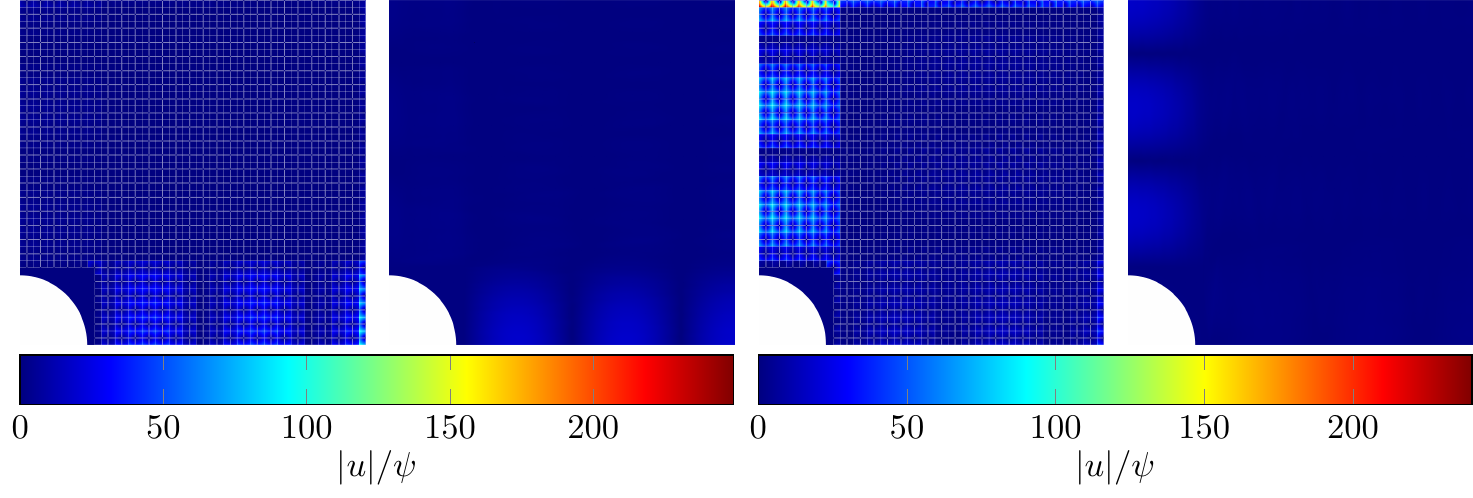}
\caption{$|u|/\psi$ at frequency $M_3$ for the microstructured model and the relaxed micromorphic model, the first two figures correspond to $M_3$ (left) and the last two figures to $M_3$ (right) (see Figure~\ref{fig:frfconverge} for the definition of these points).}
\label{m3}
\end{figure}

\begin{figure}[H]
\centering
\includegraphics{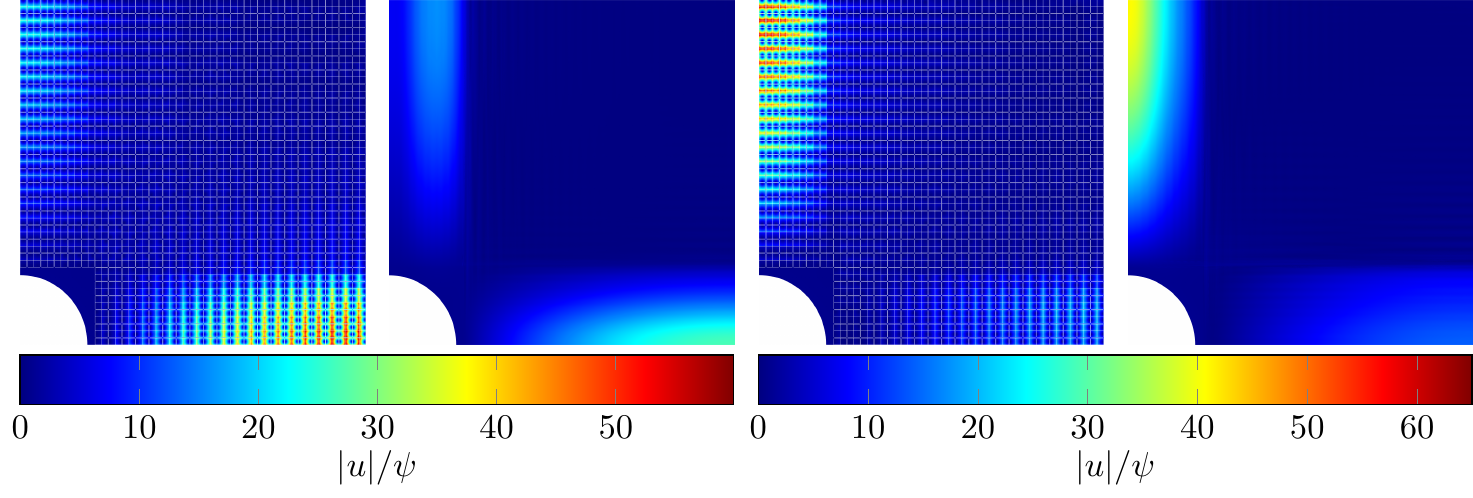}
\caption{$|u|/\psi$ at frequency $M_4$ for the microstructured model and the relaxed micromorphic model, the first two figures correspond to $M_4$ (left) and the last two figures to $M_4$ (right) (see Figure~\ref{fig:frfconverge} for the definition of these points).}
\label{m4}
\end{figure}

To give an outlook of the efficacity of the relaxed micromorphic model, we plot in Figs. \ref{m1}-\ref{m4} the solution for the displacement field at frequencies $M_1$, $M_2$, $M_3$ as defined in Figure~\ref{fig:frfconverge} (right). For each point $M_i$, we actually consider two adjacent values of the frequency to compute the solution (see Figure~\ref{fig:frfconverge} right). It can be clearly inferred that the relaxed micromorphic model encodes all the main features of the metamaterial's reponse at a fraction of the computationnal cost.
%Figure \ref{bg} presents the response of the microstructured and the relaxed micromorphic models in the band gap. In this picture we chose to use a different scale for the two pictures so as to show the load concentration around the Cauchy region that occurs for both the relaxed micromorphic and the microstructured model. Indeed, due to the lack of higher space derivatives of the microdistortion $P$, the relaxed micromorphic medium cannot catch the highly concentrated peaks of displacement occuring in few isolated cells. However, apart from these pointwise difference the solution is well reproduced and, above all, the band-gap behavior is correctly described.
%\begin{figure}[H]
%\centering
%\includegraphics{pics/converge/bg.pdf}
%\caption{$|u|/\psi$ in the band-gap for the microstructured model and the relaxed micromorphic model at 1717~Hz.}
%\label{bg}
%\end{figure}
%%%%%%%%%%%%%%%%%%%%%%%%%%%%%%%%%%%%%%%%%%%%%%%%%%%%%%%%%%%%%%%
\section{Comparison of the relaxed micromorphic and microstructured simulations with the experimental results}
\begin{figure}[H]
    \centering
    \includegraphics{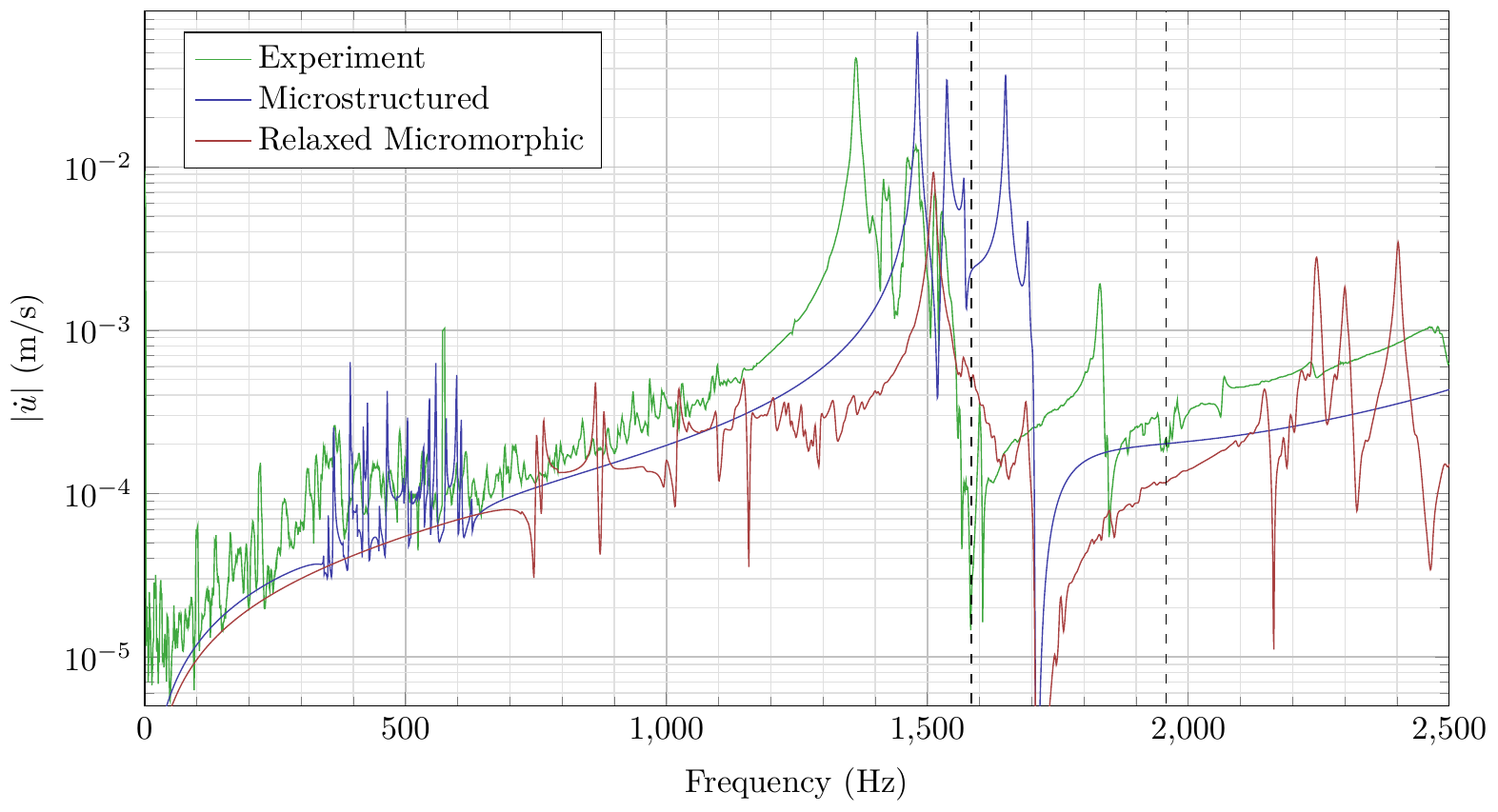}
    \caption{Amplitude of $|\dot{u}|$ at $C_1$ (see Figure~\ref{fig:interestingzones} for the definition of this point) for the experiment, the microstructured and relaxed micromorphic models (the dashed black lines locate the theoretical position of the band-gap for an infinite specimen).}
    \label{fig:r0}
\end{figure}

%%%%%%%%%%%%%%%%%%%%%%%%%%%%%%%%%%%%%%%%%%%%%%%%%%%%%%%%%%%%%%%
To make a direct comparison with the experimental results obtained with the set-up of section 2, we introduce here a more precise modeling of the applied piezoelectric load. The details and micromorphic simulations presented in sections 3.1 and 3.2 are implemented and solved in Comsol. We show that both approaches give similar results which can be directly superposed to the experimental ones as soon as manufacturing defects are taken into account.
While a direct comparison of the results obtained via the geometry detailed finite element simulation and the relaxed micromorphic approach has been possible given the reduced size of the metamaterial's sample, finite element simulations become computationally too expensive for samples of larger size.
On the other hand, the relaxed micromorphic simulations provide a powerful tool for investigating the behaviour of larger samples with a modest increase of the computational costs.
As we will show in section 6, this feature of the relaxed micromorphic model is of major importance to simulate realistic, larger-scale meta-structures that can have a true impact in engineering science. Figure \ref{fig:r0} shows a broadband comparison of the relaxed micromorphic and microstructured response with the experimental one.
It can be seen that both models reproduce well, on average, the experimentally observed response, except for a frequency shift. This frequency shift can be related to defects in the experimental procedure, more particularly to defects that have occurred during the manufacturing process. To account for the presence of these defects, a recalibration procedure is presented in section 5.1.
As we will discuss more in section 5.1.3, the relaxed micromorphic material slightly underestimates the resonance peak before the band-gap. This is due, to a big extent, to the fact that Figs. \ref{fig:r0} and \ref{fig:r3} are relative to pointwise displacements (not to mean displacements) because of the need of a direct comparison with the experiment.
%%%%%%%%%%%%%%%%%%%%%%%%%%%%%%%%%%%%%%%%%%%%%%%%%%%%%%%%%%%%%%%
\subsection{Recalibration procedure accounting for manufacturing defects}
%%%%%%%%%%%%%%%%%%%%%%%%%%%%%%%%%%%%%%%%%%%%%%%%%%%%%%%%%%%%%%%
In this subsection we present a recalibration procedure for both the microstructured and the relaxed micromorphic model allowing to account for the presence of defects in the fabrication process or, to a smaller extent, to measurement inaccuracy. Given the change of nature of the excitation of the system, we redefine $p_X$ and $m_X$ by :
\begin{equation}
\begin{aligned}
    p_X = \sqrt{\langle u(X) , \bar{u}(X) \rangle} 
    \text{ and }
    m_X = \frac{4}{a^2} \iint_{\Omega_X} \sqrt{\langle u(x_1,x_2),\bar{u}(x_1,x_2) \rangle} \mathrm{d}x_1\mathrm{d}x_2.
\end{aligned}
\end{equation}
\subsubsection{Recalibration for the microstructured system}
%%%%%%%%%%%%%%%%%%%%%%%%%%%%%%%%%%%%%%%%%%%%%%%%%%%%%%%%%%%%%%%
To account for the presence of defects and get closer to the experimental results, the system can be modified, taking into account several potential differences between the analytical model and experimental system. To simulate the likely presence of defects in the plate, we chose not to alter geometry and to modify the mechanical parameters. In formulas :
\begin{equation*}
\begin{cases}
    \lambda^\text{recalibration}_\text{Ti} = (1+\kappa_E) \lambda_\text{Ti} \\
    \mu^\text{recalibration}_\text{Ti} = (1+\kappa_E) \mu_\text{Ti}
\end{cases}
\text{and }
\rho^\text{recalibration}_\text{Ti} = (1+\kappa_\rho) \rho_\text{Ti}
\end{equation*}
Leading, for the local energy densities, to
\begin{equation*}
    W_\text{recalibration} = (1+\kappa_E) W_\text{Ti}
    \text{ and }
    K_\text{recalibration} = (1+\kappa_\rho) K_\text{Ti}
\end{equation*}
Confronting the theoretical results to the experiments eventually led to
\begin{align}
\kappa_\rho
= 
+0.05
\, ,
\qquad\qquad\qquad
\kappa_E
=
-0.10875
\, ,
\end{align}
which, considering the uncertainties due to, among other, the manufacturing process, the experimental boundary conditions, the gluing of the piezoelectric patch to the microstructured plate and more particularly the plain-strain hypothesis in the plate, is quite acceptable.
%%%%%%%%%%%%%%%%%%%%%%%%%%%%%%%%%%%%%%%%%%%%%%%%%%%%%%%%%%%%%%%
\subsubsection{Recalibration for the micromorphic system}
\begin{figure}[H]
    \centering
    \includegraphics{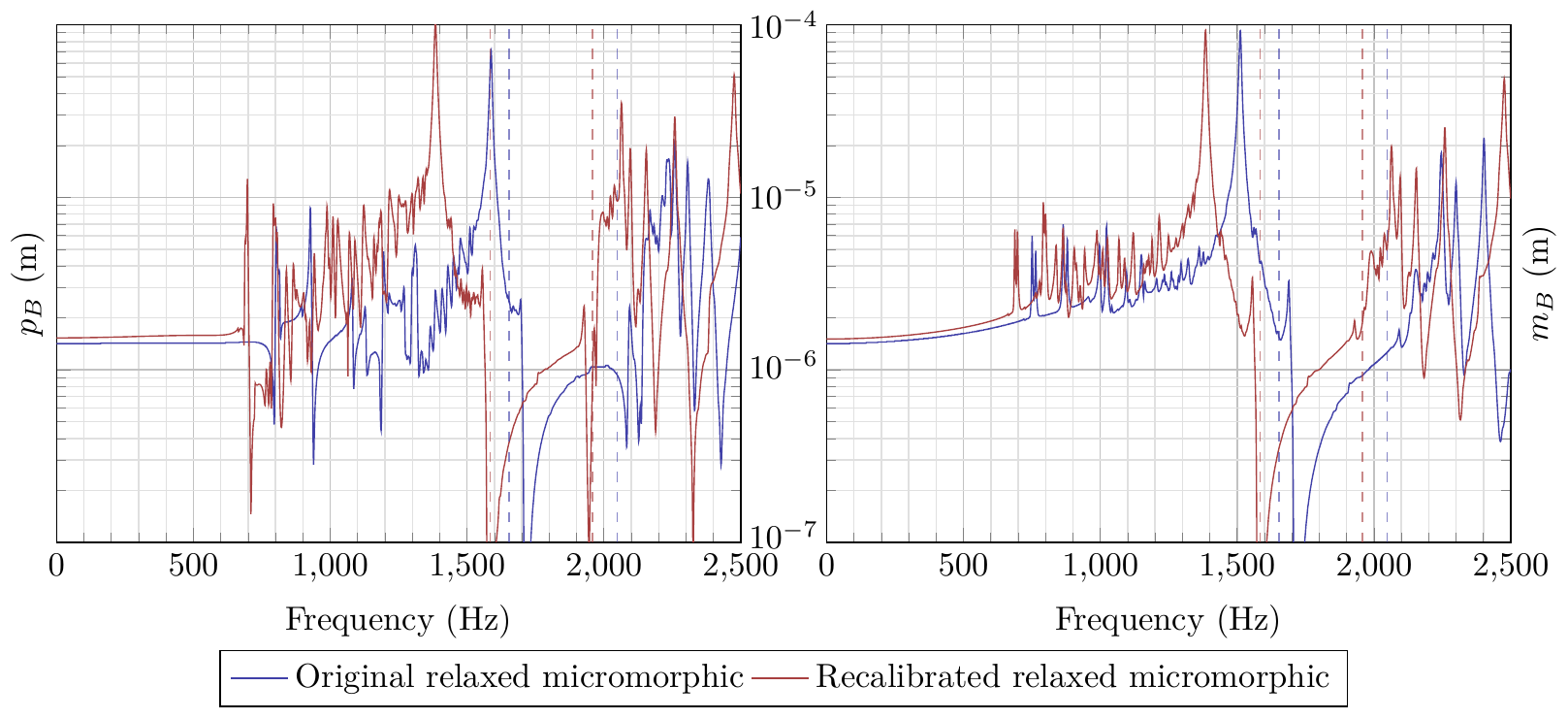}
    \caption{(\textit{left}) Pointwise displacement $p_B$ for the original and recalibrated relaxed micromorphic models at point $B$. (\textit{right}) Mean displacement $m_B$ for the original and recalibrated relaxed micromorphic models.}
    \label{fig:r1}
\end{figure}

\begin{figure}[H]
    \centering
    \includegraphics{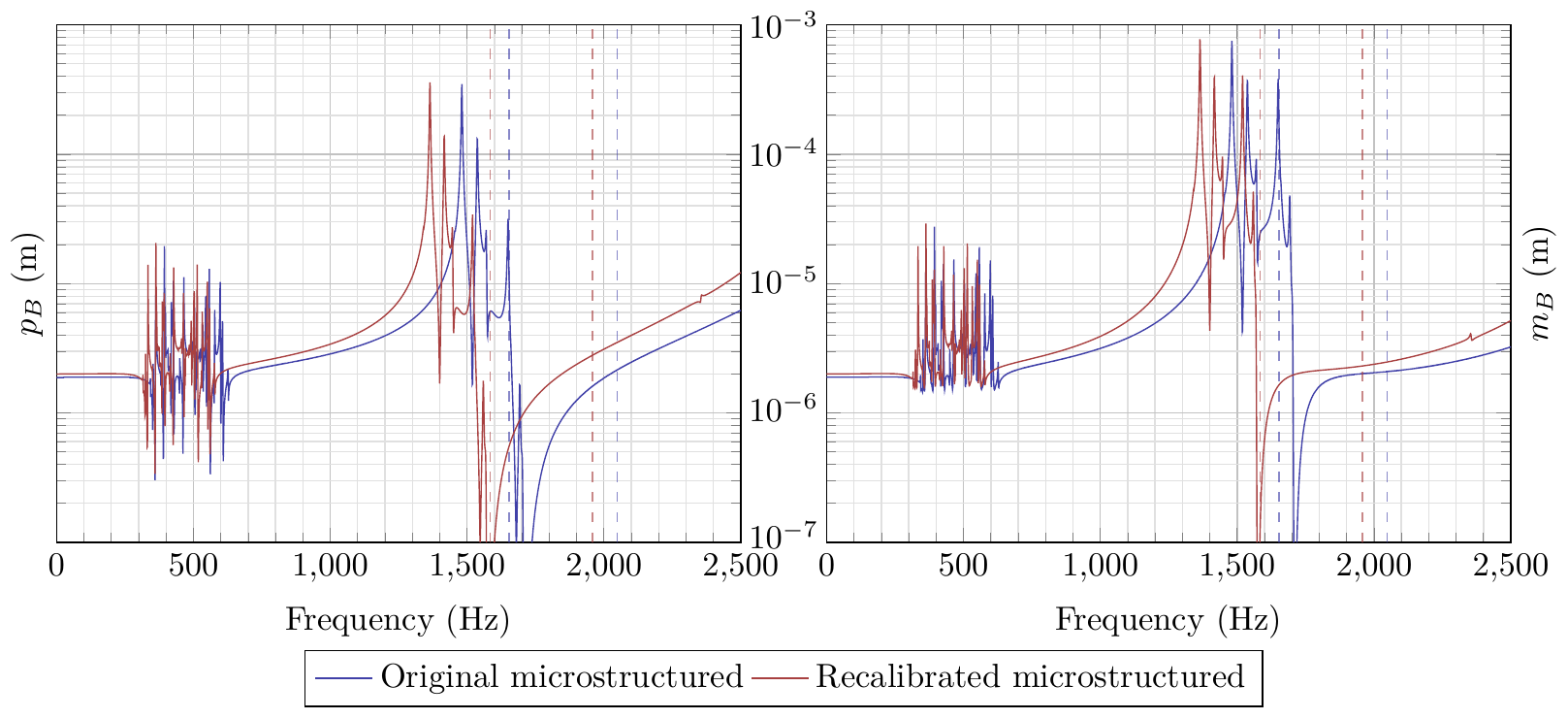}
    \caption{(\textit{left}) Pointwise displacement $p_B$ for the original and recalibrated microstructured models at point $B$. (\textit{right}) Mean displacement $m_B$ for the original and recalibrated microstructured models.}
    \label{fig:r2}
\end{figure}

%%%%%%%%%%%%%%%%%%%%%%%%%%%%%%%%%%%%%%%%%%%%%%%%%%%%%%%%%%%%%%%
In the same way, the kinetic and potential energy density of the micromorphic medium are modified as
\begin{equation*}
    K_\text{recalibration} = (1+\kappa_K) K_0 \text{ and } W_\text{recalibration} = (1+\kappa_W) W_0 
\end{equation*}
To recalibrate the relaxed micromorphic model on the microstructured model, we set
\begin{equation*}
    \kappa_K = \kappa_\rho \text{ and } \kappa_W = \kappa_E
\end{equation*}
The relaxed micromorphic model, less expensive in computing time, was recalibrated on the experiment and the parameters of the microstructured model were then updated by the following rule :
\begin{align*}
\begin{cases}
\eta_i^\text{recalibration} = (1+\alpha)\eta_i \text{, } i \in \{ 1,3 \} \\
\eta_1^{* \text{ recalibration}} = (1+\alpha)\overline{\eta}_1^* \\
\overline{\eta}_i^\text{recalibration} = (1+\alpha)\overline{\eta}_i \text{, } i \in \{ 1,3 \} \\
\overline{\eta}_1^{* \text{ recalibration}} = (1+\alpha)\overline{\eta}_1^* \\
\lambda_i^\text{recalibration} = (1+\beta)\lambda_i \text{, } i \in \{e,m \} \\
\mu_i^\text{recalibration} = (1+\beta)\mu_i \text{, } i \in \{e,m \} \\
\mu_i^{* \text{ recalibration}} = (1+\beta)\mu_i^* \text{, } i \in \{e,m \} \\
\mu_{\rm c}^\text{ recalibration} = (1+\beta)\mu_{\rm c}
\end{cases} \longleftrightarrow
\begin{cases}\rho^\text{recalibration} = (1+\alpha) \rho \\
\mu^\text{recalibration} = (1+\beta)\mu \\
\lambda^\text{recalibration} = (1+\beta)\lambda \\
\end{cases} \text{where }
\begin{cases} \alpha &= 0.05 \\ \beta &= -0.10875 \end{cases}.
\end{align*}
Figures \ref{fig:r1} and \ref{fig:r2} present $p_B$ and $m_B$ for the microstructured and the relaxed micromorphic models with the recalibrated parameters. We remark that, once calibrated, both the pointwise and the mean displacement describe well the local resonance occuring at the lower band-gap limit. As expected, the mean displacements for the relaxed micromorphic and microstructured model show better agreement than the pointwise displacement.


\subsubsection{Discussion of results}

\begin{figure}[H]
    \centering
    \includegraphics{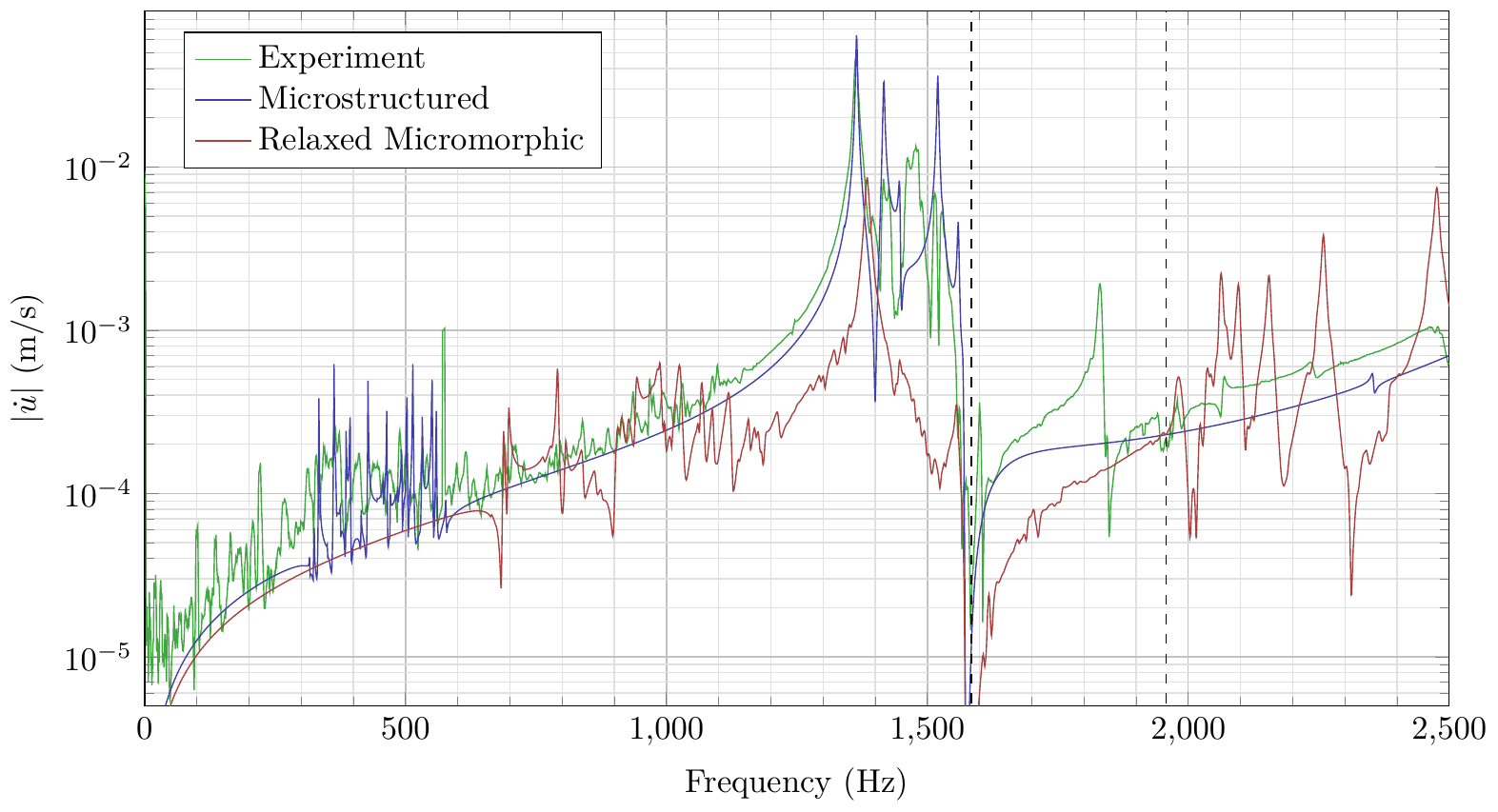}
    \caption{Amplitude of $|\dot{u}|$ at $C_1$ (see Figure \ref{fig:interestingzones} for the definition of this point) for the experiment, the microstructured and relaxed micromorphic models.}
    \label{fig:r3}
\end{figure}

Figure \ref{fig:r3} shows the comparison of the velocity spectrum at point $B$ for the experiment, the microstructured and the relaxed micromorphic model. A generally good agreement is found for both models, the relaxed micromorphic model showing slightly reduced performances due to the reduced size of the plate. This result is to be considered satisfying, since all the main response characteristics are well described.

When increasing the size of the plate the relaxed micromorphic model becomes more and more accurate with no significant increase of the computational cost (see Figure~\ref{fig:frfconverge}). On the other hand the microstructured simulations become very costly in terms of computational time. This opens the way to the effective design of large-scale metastructures that would not be possible without using the relaxed micromorphic model. To prove this statement, we present in the next section the design of a new large-scale engineering meta-structure that would not have been otherwise possible.

\section{Enabling meta-structural engineering design}
\begin{figure}[H]
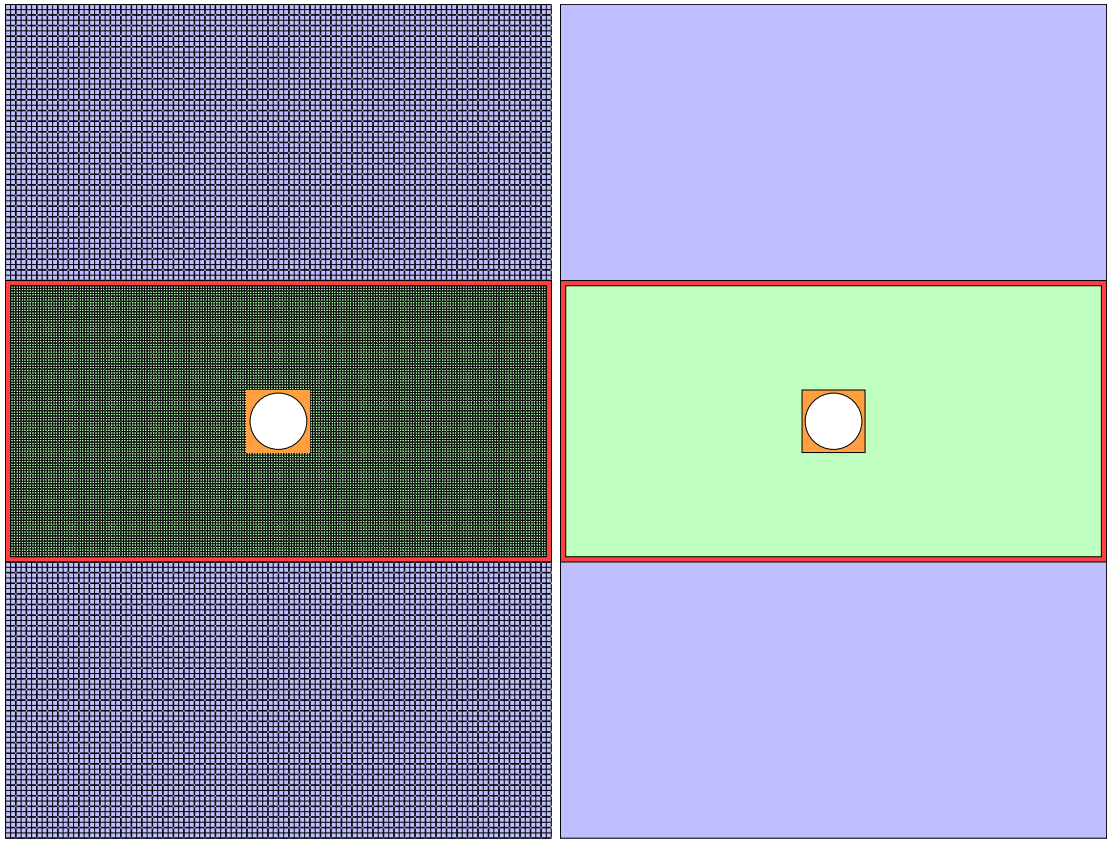

    \centering
    \includegraphics[width=0.4375\textwidth]{drawings/new_model/cauchy_unsym.pdf}
    \includegraphics[width=0.4375\textwidth]{drawings/new_model/rm_unsym.pdf}
    \caption{(\textit{left}) Top view of the full microstructured plate with the two different cells. (\textit{right}) Top view of the equivalent micromorphic plate with the boundaries and medium denominations.}
    \label{newunsym}
\end{figure}

As we pointed out so far, the relaxed micromorphic model can be considered as an appropriate
homogenized limit for mechanical metamaterials as soon as \textit{sufficiently large} specimens are considered. In particular, for the specific unit cell presented in this paper (see Fig. \ref{fig:fig_tab_unit_cell}), we showed that the relaxed micromorphic model starts giving very good results for specimens that are larger than 30x30 unit cells. We also showed that the results ulteriorly improve when considering a central pulse excitation distributed over more than one unit cell. For example, we showed in Fig. \ref{fig:frfconverge} that the metamaterial’s behavior is perfectly reproduced for the whole frequency range when considering a specimen of $51 \times 49$ unit cells and a pulse excitation distributed over a region whose side is 11 cells.
The results presented in this paper open unprecedented opportunities of exploring metamaterial structures at large scales, thus electing the relaxed micromorphic model as a gold standard for engineering design. To support our claiming, we present in this section the design of a complex metastructure that is able to concentrate energy in specific points, so that the eventual use of converters can be eased for the subsequent conversion of elastic energy into, e.g., heat or electricity.
We consider a structure whose geometry is given in Fig. \ref{newunsym}: the central domain $\Omega_m^1$ is made up of the metamaterial studied in this paper (Fig. \ref{fig:fig_tab_unit_cell}), while the outer domain $\Omega_m^2$ is made of a metamaterial with the same geometry whose unit cell is doubled with respect to the one presented in Fig. \ref{fig:fig_tab_unit_cell} (see Appendix C for the details concerning this larger unit cell). Both metamaterials’ domains are very large ($101 \times 51$ unit cells in $\Omega_m^1$) and $51 \times 51$ unit cells in $\Omega_m^2$), so that the use of the homogenized model is certainly pertinent. This large-scale structure points towards realistic structural engineering design (think, for example, that the domain $\Omega_m^1$ is located around a railway truck and that the domain $\Omega_m^2$ are the lateral banks).
The two metamaterials’ domains are separated by a classical Cauchy material occupying the annular domain $\Omega_c$. The elastic properties of such soft Cauchy material are given in Table \ref{tab:cauchy2}. The metastructure setup is given in Figure \ref{newunsym}. The functioning mechanism of this structure can be summarized as follows: we send a pulse signal in the center of the metamaterial plate $\Omega_m^1$ at a frequency that falls in the band-gap of the outer metamaterial (see Fig. \ref{fig:dispersionboth} for the relative position of the dispersion curves for the two metamaterials). When reaching the outer domain the wave cannot propagate due to the presence of the outer metamaterial. The annular Cauchy material is chosen so that a diode effect is triggered (the wave coming from $\Omega_m^1$ can pass, but cannot go back) \cite{rizzi2020towards}. Thanks to this design, the proposed meta-structure can focus an important part of the elastic energy in the annular Cauchy region (see Tab. \ref{tab:energy}). It can be noted that the energy concentration in the annular Cauchy material is evident, especially considering the very narrow area in which it occurs (see Fig. \ref{fig:newdisp}). A structure of this type could be used to locate energy converters in the annular Cauchy region for subsequent energy conversion and re-use. It will be the object of forthcoming works to effectively optimize and realize a structure of this type.
\begin{table}[H]
\centering
\begin{tabular}{cccccccc}
\thickhline
\thickhline
& $\rho_2$ & $\lambda_2$ & $\mu_2$ &
\\
& [kg.m$^{-3}$] & [Pa] & [Pa] & \\
\hline
& $3000$ & $9.74 \cdot 10^8$ & $5.88 \cdot 10^5 $ & \\
\thickhline
\thickhline
\end{tabular}
    \caption{Mechanical parameters of the second isotropic Cauchy medium between the two relaxed micromorphic mediums (red annular region in Fig. \ref{newunsym} and \ref{newsym}).}
\end{table}
\label{tab:cauchy2}
\begin{figure}[H]
    \centering
    \includegraphics[width=0.4375\textwidth]{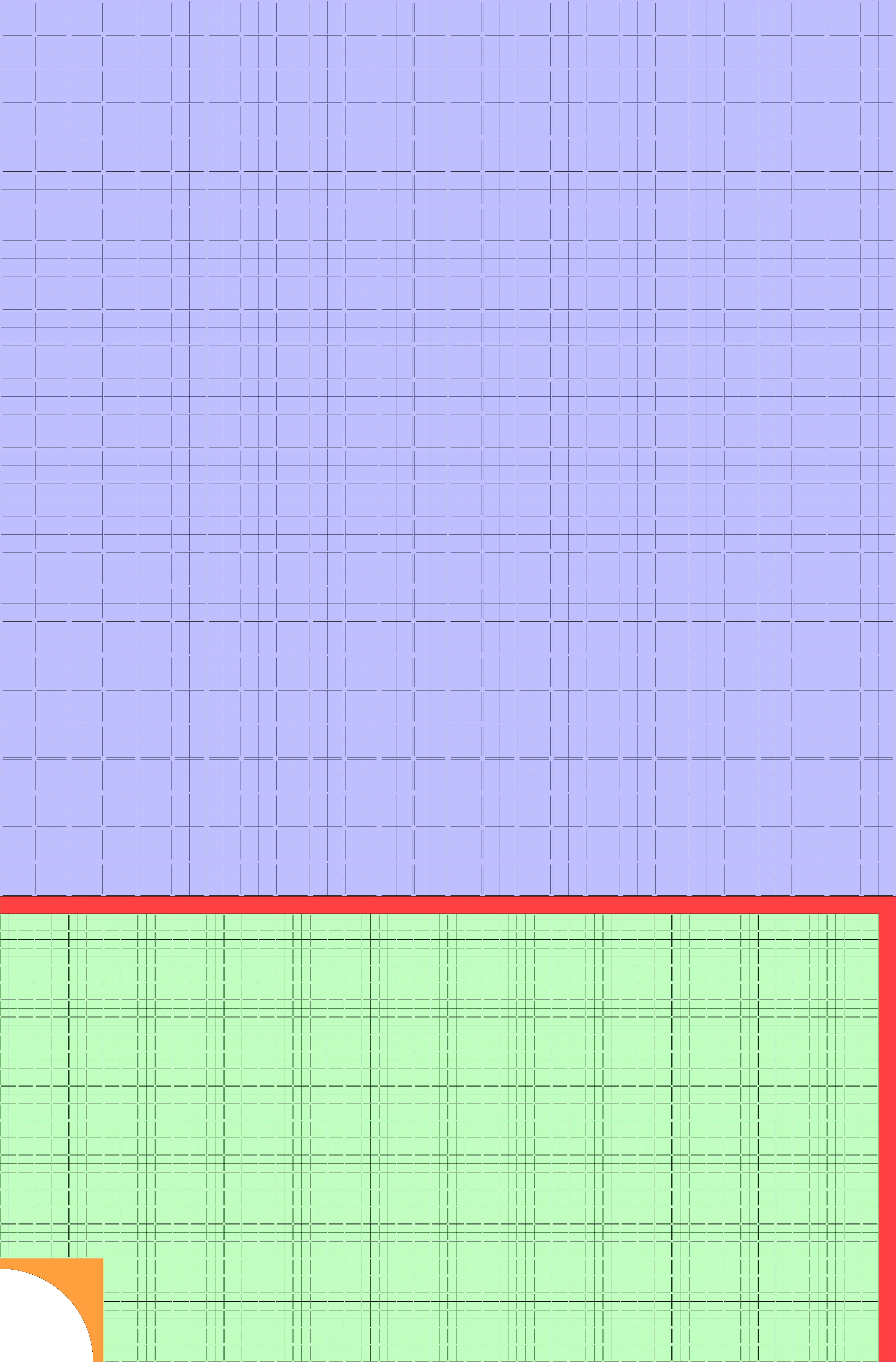}
    \includegraphics[width=0.4375\textwidth]{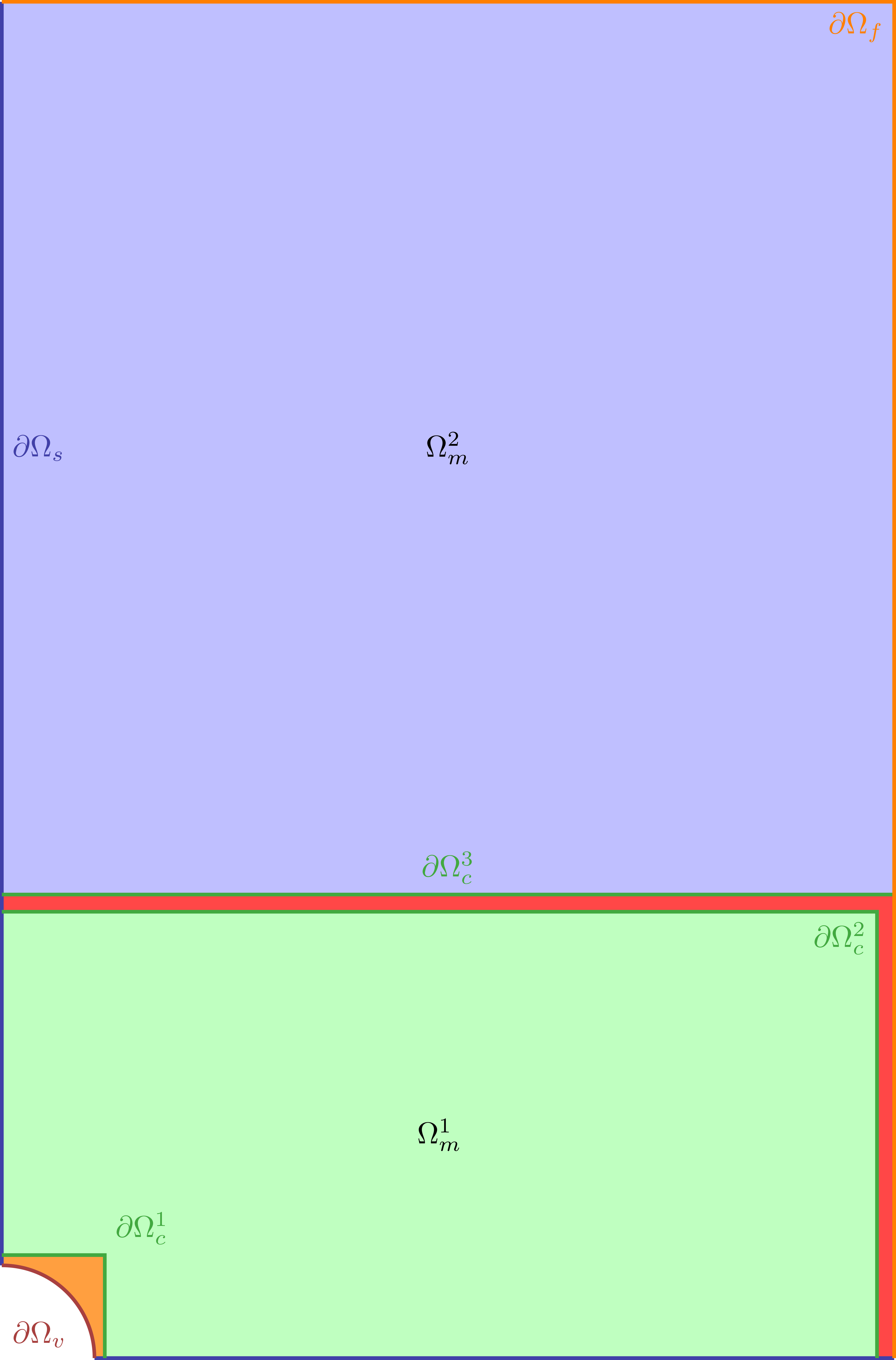}
    \caption{(\textit{left}) Top view of the symmetrized microstructured structure with the two different cells. (\textit{right}) Top view of the symmetrized equivalent micromorphic structure with the boundaries and domains denominations. Only the micromorphic meta-structure has been implemented in the FEM environnement.}
    \label{newsym}
\end{figure}
\begin{table}[H]
\centering
\begin{tabular}{cccccccc}
\thickhline
\thickhline
& \multirow{2}{*}{857.5 Hz} & $\Omega^1_{c}$ & $\Omega^1_m$ & $\Omega^2_{c}$ & $\Omega^2_m$ &
\\
& & [J.m$^{-3}$] & [J.m$^{-3}$] & [J.m$^{-3}$] & [J.m$^{-3}$] & \\
\hline
& W & $55.1 \cdot 10^3$ & $45.5 \cdot 10^3$ & $29.5 \cdot 10^3$ & $7.74 \cdot 10^{3}$ & \\
& K & $3.42 \cdot 10^6$ & $43.6 \cdot 10^3$ & $26.3 \cdot 10^3$ & $7.84 \cdot 10^{3}$ & \\
\thickhline
\thickhline
\end{tabular}
    \caption{Values of the average total energy for each domain of the structure at 857.5~Hz.}
    \label{tab:energy}
\end{table}

\begin{figure}[H]
    \centering
    \includegraphics[scale=1]{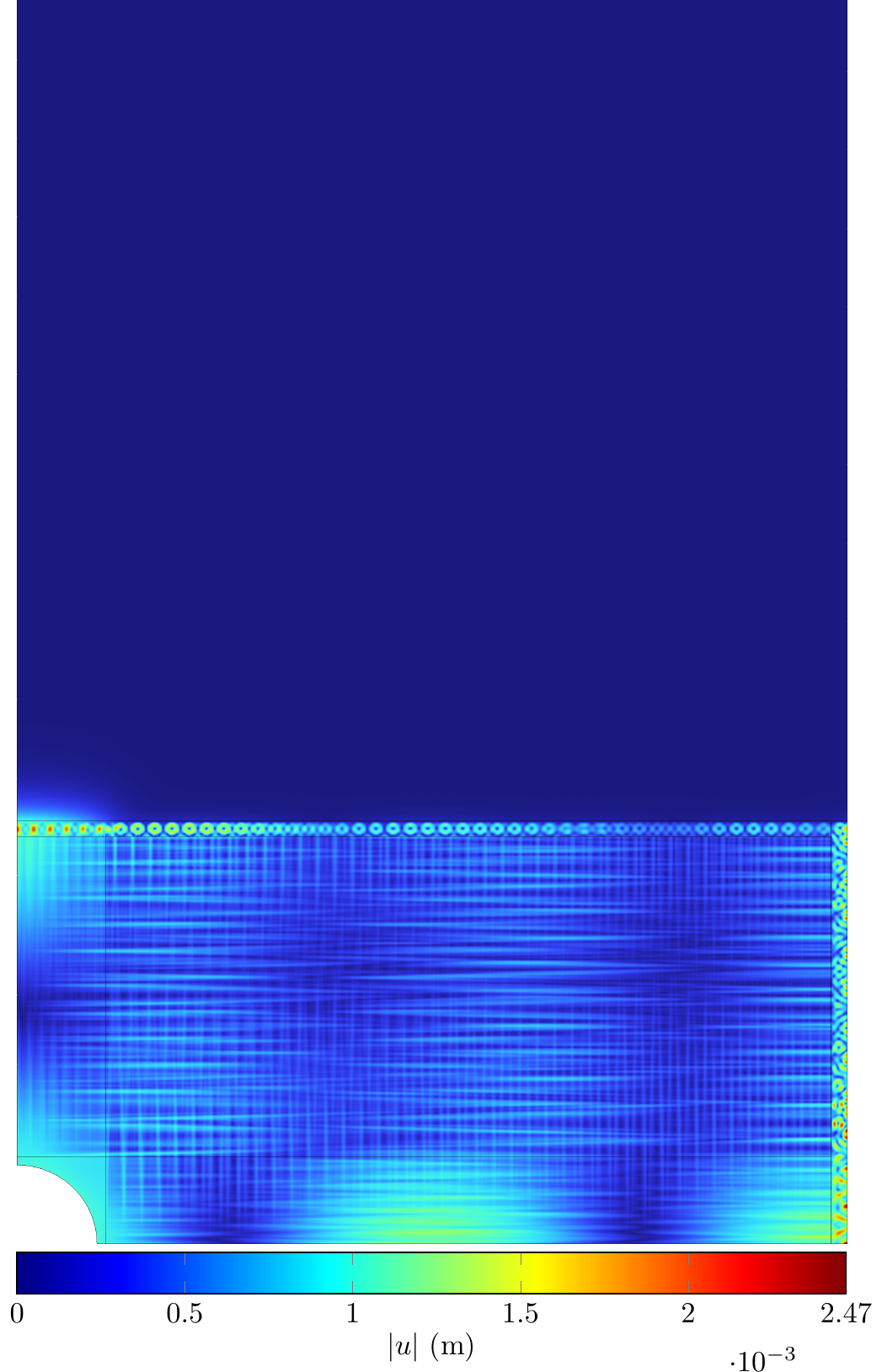}
    \caption{Displacement field for the considered metastructure at 857.5~Hz.}
    \label{fig:newdisp}
\end{figure}

\section{Conclusions and perspectives}
In this paper we showed for the first time how the relaxed micromorphic model can be operatively used to design realistic large-scale structures that can control elastic waves and eventually recover energy. Our model was calibrated on an acoustic metamaterial and the theoretical results were compared to the experimental ones performed on a specimen manufactured via metal etching techniques.
We presented a metastructure combining metamaterials’ bricks of different type/shape and bricks of classical Cauchy materials in such a way that the elastic energy produced by a source of vibrations is concentrated in specific collection points, thus easing eventual subsequent re-use.
The structure conceived in this paper will be optimized in forthcoming works and a prototype will be eventually realized to show the new possibilities that a micromorphic-type modeling of metamaterials opens for practical applications.
\section{Acknowledgments}
Angela Madeo and Gianluca Rizzi acknowledge support from the European Commission through the funding of the ERC Consolidator Grant META-LEGO, N◦ 101001759. Angela Madeo and Gianluca Rizzi acknowledge funding from the French Research Agency ANR, “METASMART” (ANR-17CE08-0006). Patrizio Neff acknowledges support in the framework of the DFG-Priority Programme 2256 “Variational Methods for Predicting Complex Phenomena in Engineering Structures and Materials”, Neff 902/10-1, Project-No. 440935806.
%%%%%%%%%%%%%%%%%%%%%%%%%%%%%%%%%%%%%%%%%%%%%%%%%%%%%%%%%%%%%%%%%%%%%%%%%%%%%%%%
%%%%%%%%%%%%%%%%%%%%%%%%%%%%%%%%%%%%%%%%%%%%%%%%%%%%%%%%%%%%%%%%%%%%%%%%%%%%%%%%
\begingroup
\setstretch{0.8}
\setlength\bibitemsep{3pt}
\printbibliography

@article{neff_unifying_2014,
title = {A unifying perspective: the relaxed linear micromorphic continuum},
volume = {26},
issn = {0935-1175, 1432-0959},
shorttitle = {A unifying perspective},
url = {http://link.springer.com/10.1007/s00161-013-0322-9},
doi = {10.1007/s00161-013-0322-9},
language = {en},
number = {5},
urldate = {2021-04-23},
journal = {Continuum Mechanics and Thermodynamics},
author = {Neff, Patrizio and Ghiba, Ionel-Dumitrel and Madeo, Angela and Placidi, Luca and Rosi, Giuseppe},
month = sep,
year = {2014},
pages = {639--681},
}

@article{lakes1987foam,
title={Foam structures with a negative Poisson's ratio},
author={Lakes, Roderic},
journal={Science},
volume={235},
number={4792},
pages={1038--1040},
year={1987},
publisher={American Association for the Advancement of Science}
}

@article{frenzel2017three,
title={Three-dimensional mechanical metamaterials with a twist},
author={Frenzel, Tobias and Kadic, Muamer and Wegener, Martin},
journal={Science},
volume={358},
number={6366},
pages={1072--1074},
year={2017},
publisher={American Association for the Advancement of Science}
}

@article{bilal,
author={O.R. Bilal, D. Ballagi, C. Daraio},
title={Architected lattices for simultaneous broadband attenuation of
airborne sound and mechanical vibrations in all directions.},
journal={Physical Review Applied},
year={2018},
pages={},
volume={10.5:054060},
}

@article{tallarico,
author={Domenico, Tallarico and A. Trevisan, N.V. Movchan, A.B. Movchan},
title={Edge waves and localization in lattices containing tilted resonators.},
journal={Frontiers in Materials},
volume={4(16)},
url={DOI: 10.3389/fmats.2017.00016}
}

@article{chen2001dispersive,
title={A dispersive model for wave propagation in periodic heterogeneous media based on homogenization with multiple spatial and temporal scales},
author={Chen, Wen and Fish, Jacob},
journal={J. Appl. Mech.},
volume={68},
number={2},
pages={153--161},
year={2001}
}

@article{willis2011effective,
title={Effective constitutive relations for waves in composites and metamaterials},
author={Willis, John R},
journal={Proceedings of the Royal Society A: Mathematical, Physical and Engineering Sciences},
volume={467},
number={2131},
pages={1865--1879},
year={2011},
publisher={The Royal Society Publishing}
}

@article{willis2012construction,
title={The construction of effective relations for waves in a composite},
author={Willis, John R},
journal={Comptes Rendus M{\'e}canique},
volume={340},
number={4-5},
pages={181--192},
year={2012},
publisher={Elsevier}
}

@article{boutin2014large,
title={Large scale modulation of high frequency waves in periodic elastic composites},
author={Boutin, Claude and Rallu, Antoine and Hans, St{\'e}phane},
journal={Journal of the Mechanics and Physics of Solids},
volume={70},
pages={362--381},
year={2014},
publisher={Elsevier}
}

@article{sridhar2018general,
title={A general multiscale framework for the emergent effective elastodynamics of metamaterials},
author={Sridhar, Ashwin and Kouznetsova, Varvara G and Geers, Marc GD},
journal={Journal of the Mechanics and Physics of Solids},
volume={111},
pages={414--433},
year={2018},
publisher={Elsevier}
}

@article{sridhar2016homogenization,
title={Homogenization of locally resonant acoustic metamaterials towards an emergent enriched continuum},
author={Sridhar, Ashwin and Kouznetsova, Varvara G and Geers, Marc GD},
journal={Computational mechanics},
volume={57},
number={3},
pages={423--435},
year={2016},
publisher={Springer}
}

@article{srivastava2017evanescent,
title={Evanescent wave boundary layers in metamaterials and sidestepping them through a variational approach},
author={Srivastava, Ankit and Willis, John R},
journal={Proceedings of the Royal Society A: Mathematical, Physical and Engineering Sciences},
volume={473},
number={2200},
pages={20160765},
year={2017},
publisher={The Royal Society Publishing}
}

@article{forest,
author={S. Forest, F. Barbe, G. Cailletaud},
title={Cosserat modelling of size effects in the mechanical behaviour of polycrystals and multi-phase materials.},
journal={Int. J. Solids Struct},
year={2000},
pages={7105-7126},
volume={37},
}

@article{haberman,
author={M.R Haberman, A.N. Norris},
title={Acoustic metamaterials.},
journal={Acoustics Today},
year={2016},
pages={31-39},
volume={12:3},
}

@article{li2004double,
title={Double-negative acoustic metamaterial},
author={Li, Jensen and Chan, Che Ting},
journal={Physical Review E},
volume={70},
number={5},
pages={055602},
year={2004},
publisher={APS}
}

@article{miniaci2016spider,
title={Spider web-inspired acoustic metamaterials},
author={Miniaci, Marco and Krushynska, Anastasiia and Movchan, Alexander B and Bosia, Federico and Pugno, Nicola M},
journal={Applied Physics Letters},
volume={109},
number={7},
pages={071905},
year={2016},
publisher={AIP Publishing LLC}
}

@article{bordiga2019prestress,
title={Prestress tuning of negative refraction and wave channeling from flexural sources},
author={Bordiga, Giovanni and Cabras, Luigi and Piccolroaz, Andrea and Bigoni, Davide},
journal={Applied Physics Letters},
volume={114},
number={4},
pages={041901},
year={2019},
publisher={AIP Publishing LLC}
}

@article{dagostino_effective_2020,
title = {Effective {description} of {anisotropic} {wave} {dispersion} in {mechanical} {band}-{gap} {metamaterials} via the {relaxed} {micromorphic} {model}},
volume = {139},
issn = {0374-3535, 1573-2681},
url = {http://link.springer.com/10.1007/s10659-019-09753-9},
doi = {10.1007/s10659-019-09753-9},
language = {en},
number = {2},
urldate = {2021-04-23},
journal = {Journal of Elasticity},
author = {d’Agostino, Marco Valerio and Barbagallo, Gabriele and Ghiba, Ionel-Dumitrel and Eidel, Bernhard and Neff, Patrizio and Madeo, Angela},
month = may,
year = {2020},
pages = {299--329},
}

@article{rizzi_exploring_2021,
title = {Exploring {metamaterials}’ {structures} {through} the {relaxed} {micromorphic} {model}: {switching} an {acoustic} {screen} {into} an {acoustic} {absorber}},
volume = {7},
issn = {2296-8016},
shorttitle = {Exploring {Metamaterials}’ {Structures} {Through} the {Relaxed} {Micromorphic} {Model}},
url = {https://www.frontiersin.org/articles/10.3389/fmats.2020.589701/full},
doi = {10.3389/fmats.2020.589701},
urldate = {2021-04-23},
journal = {Frontiers in Materials},
author = {Rizzi, Gianluca and Collet, Manuel and Demore, Félix and Eidel, Bernhard and Neff, Patrizio and Madeo, Angela},
month = mar,
year = {2021},
pages = {589701},
}

@article{rizzi2020towards,
title={Towards the conception of complex engineering meta-structures: relaxed-micromorphic modelling of mechanical diodes},
author={Rizzi, Gianluca and Tallarico, Domenico and Neff, Patrizio and Madeo, Angela},
journal={(arXiv:2012.11192) Accepted in Wave Motion},
year={2021}
}

@article{rizzi2021boundary,
  title={Boundary and interface conditions in the relaxed micromorphic model: Exploring finite-size metastructures for elastic wave control},
  author={Rizzi, Gianluca and d’Agostino, Marco Valerio and Neff, Patrizio and Madeo, Angela},
  journal={Mathematics and Mechanics of Solids},
  pages={10812865211048923},
  year={2021},
  publisher={SAGE Publications Sage UK: London, England}
}

@article{aivaliotis_frequency-_2020,
title = {Frequency- and angle-dependent scattering of a finite-sized meta-structure via the relaxed micromorphic model},
volume = {90},
issn = {0939-1533, 1432-0681},
url = {http://link.springer.com/10.1007/s00419-019-01651-9},
doi = {10.1007/s00419-019-01651-9},
language = {en},
number = {5},
urldate = {2021-04-23},
journal = {Archive of Applied Mechanics},
author = {Aivaliotis, Alexios and Tallarico, Domenico and d’Agostino, Marco-Valerio and Daouadji, Ali and Neff, Patrizio and Madeo, Angela},
month = may,
year = {2020},
pages = {1073--1096},
}

@article{krushynska2017coupling,
title={Coupling local resonance with Bragg band gaps in single-phase mechanical metamaterials},
author={Krushynska, Anastasiia O and Miniaci, Marco and Bosia, Federico and Pugno, Nicola M},
journal={Extreme Mechanics Letters},
volume={12},
pages={30--36},
year={2017},
publisher={Elsevier}
}

@article{willis_effective_2011,
title = {Effective constitutive relations for waves in composites and metamaterials},
volume = {467},
issn = {1364-5021, 1471-2946},
url = {https://royalsocietypublishing.org/doi/10.1098/rspa.2010.0620},
doi = {10.1098/rspa.2010.0620},
language = {en},
number = {2131},
urldate = {2021-04-23},
journal = {Proceedings of the Royal Society A: Mathematical, Physical and Engineering Sciences},
author = {Willis, J. R.},
month = jul,
year = {2011},
pages = {1865--1879},
}

@article{craster_high-frequency_2010,
title = {High-frequency homogenization for periodic media},
volume = {466},
issn = {1364-5021, 1471-2946},
url = {https://royalsocietypublishing.org/doi/10.1098/rspa.2009.0612},
doi = {10.1098/rspa.2009.0612},
language = {en},
number = {2120},
urldate = {2021-04-23},
journal = {Proceedings of the Royal Society A: Mathematical, Physical and Engineering Sciences},
author = {Craster, R. V. and Kaplunov, J. and Pichugin, A. V.},
month = aug,
year = {2010},
pages = {2341--2362},
}

@article{nolde_high_2011,
title = {High frequency homogenization for structural mechanics},
volume = {59},
number = {3},
journal = {Journal of the Mechanics and Physics of Solids},
author = {Nolde, E. and Craster, R. V. and Kaplunov, J.},
year = {2011},
pages = {651--671},
}

@article{sridhar_general_2018,
title = {A general multiscale framework for the emergent effective elastodynamics of metamaterials},
volume = {111},
issn = {00225096},
url = {https://linkinghub.elsevier.com/retrieve/pii/S0022509617306245},
doi = {10.1016/j.jmps.2017.11.017},
language = {en},
urldate = {2021-04-23},
journal = {Journal of the Mechanics and Physics of Solids},
author = {Sridhar, A. and Kouznetsova, V.G. and Geers, M.G.D.},
month = feb,
year = {2018},
pages = {414--433},
}

@article{pham_transient_2013,
title = {Transient computational homogenization for heterogeneous materials under dynamic excitation},
volume = {61},
issn = {00225096},
url = {https://linkinghub.elsevier.com/retrieve/pii/S0022509613001269},
doi = {10.1016/j.jmps.2013.07.005},
language = {en},
number = {11},
urldate = {2021-04-23},
journal = {Journal of the Mechanics and Physics of Solids},
author = {Pham, K. and Kouznetsova, V.G. and Geers, M.G.D.},
month = nov,
year = {2013},
pages = {2125--2146},
}

@article{sridhar_homogenization_2016,
title = {Homogenization of locally resonant acoustic metamaterials towards an emergent enriched continuum},
volume = {57},
issn = {0178-7675, 1432-0924},
url = {http://link.springer.com/10.1007/s00466-015-1254-y},
doi = {10.1007/s00466-015-1254-y},
language = {en},
number = {3},
urldate = {2021-04-23},
journal = {Computational Mechanics},
author = {Sridhar, A. and Kouznetsova, V. G. and Geers, M. G. D.},
month = mar,
year = {2016},
pages = {423--435},
}

@article{sridhar_frequency_2020,
title = {Frequency domain boundary value problem analyses of acoustic metamaterials described by an emergent generalized continuum},
volume = {65},
issn = {0178-7675, 1432-0924},
url = {http://link.springer.com/10.1007/s00466-019-01795-z},
doi = {10.1007/s00466-019-01795-z},
language = {en},
number = {3},
urldate = {2021-04-23},
journal = {Computational Mechanics},
author = {Sridhar, A. and Kouznetsova, V. G. and Geers, M. G. D.},
month = mar,
year = {2020},
pages = {789--805},
}

@article{celli_bandgap_2019,
title = {Bandgap widening by disorder in rainbow metamaterials},
volume = {114},
issn = {0003-6951, 1077-3118},
url = {http://aip.scitation.org/doi/10.1063/1.5081916},
doi = {10.1063/1.5081916},
language = {en},
number = {9},
urldate = {2021-04-23},
journal = {Applied Physics Letters},
author = {Celli, Paolo and Yousefzadeh, Behrooz and Daraio, Chiara and Gonella, Stefano},
month = mar,
year = {2019},
pages = {091903},
}

@article{wang_harnessing_2014,
title = {Harnessing {buckling} to {design} {tunable} {locally} {resonant} {acoustic} {metamaterials}},
volume = {113},
issn = {0031-9007, 1079-7114},
url = {https://link.aps.org/doi/10.1103/PhysRevLett.113.014301},
doi = {10.1103/PhysRevLett.113.014301},
language = {en},
number = {1},
urldate = {2021-04-23},
journal = {Physical Review Letters},
author = {Wang, Pai and Casadei, Filippo and Shan, Sicong and Weaver, James C. and Bertoldi, Katia},
month = jul,
year = {2014},
pages = {014301},
}

@article{buckmann_mechanical_2015,
title = {Mechanical cloak design by direct lattice transformation},
volume = {112},
issn = {0027-8424, 1091-6490},
url = {http://www.pnas.org/lookup/doi/10.1073/pnas.1501240112},
doi = {10.1073/pnas.1501240112},
language = {en},
number = {16},
urldate = {2021-04-23},
journal = {Proceedings of the National Academy of Sciences},
author = {Bückmann, Tiemo and Kadic, Muamer and Schittny, Robert and Wegener, Martin},
month = apr,
year = {2015},
pages = {4930--4934},
}

@article{misseroni_cymatics_2016,
title = {Cymatics for the cloaking of flexural vibrations in a structured plate},
volume = {6},
issn = {2045-2322},
url = {http://www.nature.com/articles/srep23929},
doi = {10.1038/srep23929},
language = {en},
number = {1},
urldate = {2021-04-23},
journal = {Scientific Reports},
author = {Misseroni, D. and Colquitt, D. J. and Movchan, A. B. and Movchan, N. V. and Jones, I. S.},
month = apr,
year = {2016},
pages = {23929},
}

@article{guenneau_acoustic_2007,
title = {Acoustic metamaterials for sound focusing and confinement},
volume = {9},
issn = {1367-2630},
url = {https://iopscience.iop.org/article/10.1088/1367-2630/9/11/399},
doi = {10.1088/1367-2630/9/11/399},
number = {11},
urldate = {2021-04-23},
journal = {New Journal of Physics},
author = {Guenneau, Sébastien and Movchan, Alexander and Pétursson, Gunnar and Anantha Ramakrishna, S},
month = nov,
year = {2007},
pages = {399--399},
}

@article{kaina_slow_2017,
title = {Slow waves in locally resonant metamaterials line defect waveguides},
volume = {7},
issn = {2045-2322},
url = {http://www.nature.com/articles/s41598-017-15403-8},
doi = {10.1038/s41598-017-15403-8},
language = {en},
number = {1},
urldate = {2021-04-23},
journal = {Scientific Reports},
author = {Kaina, Nadège and Causier, Alexandre and Bourlier, Yoan and Fink, Mathias and Berthelot, Thomas and Lerosey, Geoffroy},
month = dec,
year = {2017},
pages = {15105},
}

@article{willis_negative_2016,
title = {Negative refraction in a laminate},
volume = {97},
issn = {00225096},
url = {https://linkinghub.elsevier.com/retrieve/pii/S0022509615302623},
doi = {10.1016/j.jmps.2015.11.004},
language = {en},
urldate = {2021-04-23},
journal = {Journal of the Mechanics and Physics of Solids},
author = {Willis, J. R.},
month = dec,
year = {2016},
pages = {10--18},
}

@article{zhu_negative_2014,
title = {Negative refraction of elastic waves at the deep-subwavelength scale in a single-phase metamaterial},
volume = {5},
issn = {2041-1723},
url = {http://www.nature.com/articles/ncomms6510},
doi = {10.1038/ncomms6510},
language = {en},
number = {1},
urldate = {2021-04-23},
journal = {Nature Communications},
author = {Zhu, R. and Liu, X. N. and Hu, G. K. and Sun, C. T. and Huang, G. L.},
month = dec,
year = {2014},
pages = {5510},
}

@article{kaina_negative_2015,
title = {Negative refractive index and acoustic superlens from multiple scattering in single negative metamaterials},
volume = {525},
issn = {0028-0836, 1476-4687},
url = {http://www.nature.com/articles/nature14678},
doi = {10.1038/nature14678},
language = {en},
number = {7567},
urldate = {2021-04-23},
journal = {Nature},
author = {Kaina, Nadège and Lemoult, Fabrice and Fink, Mathias and Lerosey, Geoffroy},
month = sep,
year = {2015},
pages = {77--81},
}

@article{liu_broadband_2018,
title = {Broadband locally resonant metamaterials with graded hierarchical architecture},
volume = {123},
issn = {0021-8979, 1089-7550},
url = {http://aip.scitation.org/doi/10.1063/1.5003264},
doi = {10.1063/1.5003264},
language = {en},
number = {9},
urldate = {2021-04-23},
journal = {Journal of Applied Physics},
author = {Liu, Chenchen and Reina, Celia},
month = mar,
year = {2018},
pages = {095108},
}

@article{nassar_willis_2015,
title = {Willis elastodynamic homogenization theory revisited for periodic media},
volume = {77},
issn = {00225096},
url = {https://linkinghub.elsevier.com/retrieve/pii/S0022509614002579},
doi = {10.1016/j.jmps.2014.12.011},
language = {en},
urldate = {2021-04-23},
journal = {Journal of the Mechanics and Physics of Solids},
author = {Nassar, H. and He, Q.-C. and Auffray, N.},
month = apr,
year = {2015},
pages = {158--178},
}

@article{bacigalupo_second-gradient_2014,
title = {Second-gradient homogenized model for wave propagation in heterogeneous periodic media},
volume = {51},
issn = {00207683},
url = {https://linkinghub.elsevier.com/retrieve/pii/S0020768313004721},
doi = {10.1016/j.ijsolstr.2013.12.001},
language = {en},
number = {5},
urldate = {2021-04-23},
journal = {International Journal of Solids and Structures},
author = {Bacigalupo, A. and Gambarotta, L.},
month = mar,
year = {2014},
pages = {1052--1065},
}

@article{neff_relaxed_2015,
title = {The relaxed linear micromorphic continuum: well-posedness of the static problem and relations to the gauge theory of dislocations},
volume = {68},
issn = {0033-5614, 1464-3855},
shorttitle = {The relaxed linear micromorphic continuum},
url = {https://academic.oup.com/qjmam/article-lookup/doi/10.1093/qjmam/hbu027},
doi = {10.1093/qjmam/hbu027},
language = {en},
number = {1},
urldate = {2021-04-25},
journal = {The Quarterly Journal of Mechanics and Applied Mathematics},
author = {Neff, P. and Ghiba, I. D. and Lazar, M. and Madeo, A.},
month = feb,
year = {2015},
pages = {53--84},
}

@article{misseroni2019omnidirectional,
title={Omnidirectional flexural invisibility of multiple interacting voids in vibrating elastic plates},
author={Misseroni, D and Movchan, AB and Bigoni, D},
journal={Proceedings of the Royal Society A},
volume={475},
number={2229},
pages={20190283},
year={2019},
publisher={The Royal Society Publishing}
}

@article{miniaci2016large,
title={Large scale mechanical metamaterials as seismic shields},
author={Miniaci, Marco and Krushynska, Anastasiia and Bosia, Federico and Pugno, Nicola M},
journal={New Journal of Physics},
volume={18},
number={8},
pages={083041},
year={2016},
publisher={IOP Publishing}
}

@article{krushynska2018accordion,
title={Accordion-like metamaterials with tunable ultra-wide low-frequency band gaps},
author={Krushynska, Anastasiia O and Amendola, Ada and Bosia, Federico and Daraio, Chiara and Pugno, Nicola M and Fraternali, Fernando},
journal={New Journal of Physics},
volume={20},
number={7},
pages={073051},
year={2018},
publisher={IOP Publishing}
}

@article{rizzi2019identification,
title={Identification of second-gradient elastic materials from planar hexagonal lattices. Part II: Mechanical characteristics and model validation},
author={Rizzi, G and Dal Corso, F and Veber, D and Bigoni, D},
journal={International Journal of Solids and Structures},
volume={176},
pages={19--35},
year={2019},
publisher={Elsevier}
}
\endgroup
%%%%%%%%%%%%%%%%%%%%%%%%%%%%%%%%%%%%%%%%%%%%%%%%%%%%%%%%%%%%%%%%%%%%%%%%%%%%%%%%
%%%%%%%%%%%%%%%%%%%%%%%%%%%%%%%%%%%%%%%%%%%%%%%%%%%%%%%%%%%%%%%%%%%%%%%%%%%%%%%%
\newpage

\appendix

\section{Boundary conditions on a symmetry plane for a relaxed micromorphic medium using Curie's Symmetry Principle}

$u$ and $P$ not depending on the orientation of space, supposing that our problem to have a symmetry with respect to the plane $\mathcal {N}$ of normal $n$, one can apply Curie's Symmetry Principle, which gives
\begin{align}
\begin{cases}
 u(x^{\star}) = u^{\star}(x) \\
    P(x^{\star}) = P^{\star}(x)
    \label{curie}
\end{cases}
\end{align}
$\chi^{\star}$ being the symmetric of $\chi$ with respect to $\mathcal {N}$. Let's define $t_1$ and $t_2$ so that $(t_1,t_2,n)$ forms an orthonormal basis, thus
\begin{equation*}
    t_1^\star = t_1 \text{ , } t_2^\star = t_2 \text{ and } n^\star = -n
\end{equation*}And
\begin{equation*}
    (t_1\otimes t_1)^\star = t_1\otimes t_1 \text{ , } (t_1\otimes t_2)^\star = t_1\otimes t_2 \text{ , } (t_1\otimes n)^\star = -t_1\otimes n
\end{equation*}
\begin{equation*}
    (t_2\otimes t_1)^\star = t_2\otimes t_1 \text{ , } (t_2\otimes t_2)^\star = t_2\otimes t_2 \text{ , } (t_2\otimes n)^\star = -t_2\otimes n
\end{equation*}
\begin{equation*}
    (n\otimes t_1)^\star = -n\otimes t_1 \text{ , } (n\otimes t_2)^\star = -n\otimes t_2 \text{ and } (n\otimes n)^\star = n\otimes n
\end{equation*}
Let's write $u$ and $P$ in this base :
\begin{equation*}
    \exists ! (u_1,u_2,u_3) \in {\mathcal {C}}^{0}(\mathbb{R}^3,\mathbb{R})^3 \text{ : } u = u_1t_1+u_2t_2+u_3n
\end{equation*}
\begin{equation*}
    \exists ! (P_{11},P_{22},P_{33},P_{12},P_{13},P_{21},P_{31},P_{32},P_{23}) \in {\mathcal {C}}^{0}(\mathbb{R}^3,\mathbb{R})^9 \text{ : }
\end{equation*}
\begin{equation*}
    P = P_{11} t_1 \otimes t_1+
    P_{12} t_1 \otimes t_2+
    P_{13} t_1 \otimes n+\\
    P_{21} t_2 \otimes t_1+
    P_{22} t_2 \otimes t_2+
    P_{23} t_2 \otimes n+\\
    P_{31} n \otimes t_1+
    P_{32} n \otimes t_2+
    P_{33} n \otimes n
\end{equation*}
Let's also write $x$ as
\begin{equation*}
    \exists ! (x_0,\epsilon) \in \mathcal {N}\times \mathbb{R} : x = x_0 + \epsilon  n 
\end{equation*}Then
\begin{equation*}
    x^{\star} = x_0 - \epsilon  n 
\end{equation*}
For $u$, substituting in \ref{curie}, one can get
\begin{equation*}
    u_1(x_0-\epsilon n) t_1 + u_2(x_0-\epsilon n) t_2 + u_3(x_0-\epsilon n) n = u_1(x_0+\epsilon n) t_1 + u_2(x_0+\epsilon n) t_2 -u_3(x_0+\epsilon n) n
\end{equation*}By identification
\begin{displaymath}
\forall x_0\in \mathcal {N}, \forall \epsilon \in \mathbb{R} 
\left \{
\begin{array}{rcl}
u_1(x_0-\epsilon n) & = & u_1(x_0+\epsilon n) \\
u_2(x_0-\epsilon n) & = & u_2(x_0+\epsilon n) \\
u_3(x_0-\epsilon n) & = & -u_3(x_0+\epsilon n)
\end{array}
\right .
\end{displaymath}
In the same way, we have for $P$
\begin{displaymath}
\forall x_0\in \mathcal {N}, \forall \epsilon \in \mathbb{R} \left \{
\begin{array}{rcl}
P_{11}(x_0-\epsilon n) & = & P_{11}(x_0+\epsilon n) \\
P_{12}(x_0-\epsilon n) & = & P_{12}(x_0+\epsilon n) \\
P_{13}(x_0-\epsilon n) & = & -P_{13}(x_0+\epsilon n) \\
P_{21}(x_0-\epsilon n) & = & P_{21}(x_0+\epsilon n) \\
P_{22}(x_0-\epsilon n) & = & P_{22}(x_0+\epsilon n) \\
P_{23}(x_0-\epsilon n) & = & -P_{23}(x_0+\epsilon n) \\
P_{31}(x_0-\epsilon n) & = & -P_{31}(x_0+\epsilon n) \\
P_{32}(x_0-\epsilon n) & = & -P_{32}(x_0+\epsilon n) \\
P_{33}(x_0-\epsilon n) & = & P_{33}(x_0+\epsilon n)
\end{array}
\right .
\end{displaymath}
These conditions allow to reconstruct the displacement and microdistorsion fields in the whole plate, when knowing them in one fourth of the plate. For consistency reasons, when considering the symmetry planes, these conditions imply :
\begin{displaymath}
\forall x \in \mathcal {N} \text{, } \left \{
\begin{array}{ccl}
\langle u , n \rangle &= & 0 \\
\langle P , n \otimes t_1 \rangle & = & 0 \\
\langle P , n \otimes t_2 \rangle & = & 0 \\
\langle P , t_1 \otimes n \rangle & = & 0 \\
\langle P , t_2 \otimes n \rangle & = & 0
\end{array} \right . \textit{ i.e. } \forall x \in \mathcal {N},
u_{[t_1,t_2,n]}(x) = \begin{pmatrix} \star \\ \star \\ 0 \end{pmatrix}
    \text{ and }
    P_{[t_1,t_2,n]}(x)  =
    \begin{pmatrix}
    \star & \star & 0 \\
    \star & \star & 0 \\
    0 & 0 & \star
    \end{pmatrix}
\end{displaymath}
This can be written, using Einstein's convention, as
\begin{align*}
\forall x \in \mathcal {N},
\begin{cases}
 u_in_i = 0 \\
    (\delta_{ki}-n_kn_i)(P_{ij}n_j) = 0 \text{ } \forall k =\{1,2,3\}
    \label{usym}
\end{cases}
\end{align*}
\section{2D Stress-free surface boundary conditions}
We have, in the frequency domain
\begin{equation*}
    (\Tilde{\sigma}+\hat{\sigma})_{\star 1} =
    \begin{pmatrix}
    2\mu_e(-P_{11}+u_{1,1})+\lambda_e(-P_{11}-P_{22}+u_x+v_y)-\omega^2(2\bar{\eta}_1 u_{1,1}+\bar{\eta}_3(u_{1,1}+u_{2,2})) \\
    \mu_{\rm c}(-p_{12}+p_{21}+u_{1,2}-u_{2,1})+\mu_e^*(-P_{12}-P_{21}+u_{1,2}+u_{2,1})-\omega^2(\bar{\eta}_2(u_{1,2} - u_{2,1}) + \bar{\eta}^*_1(u_{1,2}+u_{2,1})) \\
    \star
    \end{pmatrix}
\end{equation*}
\begin{equation*}
    (\Tilde{\sigma}+\hat{\sigma})_{\star 2} = \begin{pmatrix}
    \mu_{\rm c}(P_{12}-P_{21}-u_{1,2}+u_{2,1})+\mu_e^*(-P_{12}-P_{21}+u_{1,2}+u_{2,1})+\omega^2((\bar{\eta}_2-\bar{\eta}_1^*)u_{1,2}-(\bar{\eta}_2+\bar{\eta}_1^*)u_{2,1}) \\
    2\mu_e(-P_{22}+u_{2,2})+\lambda_e(-P_{11}-P_{22}+u_{1,1}+u_{2,2}) - \omega^2 (2\bar{\eta}_1 u_{2,2}+ \bar{\eta}_3(u_{1,1}+u_{2,2})) \\
    \star
    \end{pmatrix}
\end{equation*}
For a vertical stress-free border, the relaxed micromorphic medium verifies
\begin{align*}
\begin{cases}
 2\mu_e(-P_{11}+u_{1,1})+\lambda_e(-P_{11}-P_{22}+u_x+v_y)-\omega^2(2\bar{\eta}_1 u_{1,1}+\bar{\eta}_3(u_{1,1}+u_{2,2})) & = 0 \\
    \mu_{\rm c}(-p_{12}+p_{21}+u_{1,2}-u_{2,1})+\mu_e^*(-P_{12}-P_{21}+u_{1,2}+u_{2,1})-\omega^2(\bar{\eta}_2(u_{1,2} - u_{2,1}) + \bar{\eta}^*_1(u_{1,2}+u_{2,1})) & = 0
\end{cases}
\end{align*}
For a horizontal stress-free border, the relaxed micromorphic medium verifies
\begin{align*}
    \begin{cases}
     \mu_{\rm c}(P_{12}-P_{21}-u_{1,2}+u_{2,1})+\mu_e^*(-P_{12}-P_{21}+u_{1,2}+u_{2,1})+\omega^2((\bar{\eta}_2-\bar{\eta}_1^*)u_{1,2}-(\bar{\eta}_2+\bar{\eta}_1^*)u_{2,1}) & = 0 \\
    2\mu_e(-P_{22}+u_{2,2})+\lambda_e(-P_{11}-P_{22}+u_{1,1}+u_{2,2}) - \omega^2 (2\bar{\eta}_1 u_{2,2}+ \bar{\eta}_3(u_{1,1}+u_{2,2})) & = 0
    \end{cases}
\end{align*}
The continuity of the traction forces at a vertical border is
\begin{align*}
\begin{cases}
 2\mu_e(-P_{11}+u_{1,1})+\lambda_e(-P_{11}-P_{22}+u_x+v_y)-\omega^2(2\bar{\eta}_1 u_{1,1}+\bar{\eta}_3(u_{1,1}+u_{2,2})) = 2\mu u_{1,1}+\lambda (u_{1,1}+u_{2,2}) \\
    \mu_{\rm c}(-p_{12}+p_{21}+u_{1,2}-u_{2,1})+\mu_e^*(-P_{12}-P_{21}+u_{1,2}+u_{2,1})-\omega^2(\bar{\eta}_2(u_{1,2} - u_{2,1}) + \bar{\eta}^*_1(u_{1,2}+u_{2,1})) = \mu (u_{1,2}+u_{2,1})
\end{cases}
\end{align*}
The continuity of the traction forces at a horizontal border is
\begin{align*}
    \begin{cases}
     \mu_{\rm c}(P_{12}-P_{21}-u_{1,2}+u_{2,1})+\mu_e^*(-P_{12}-P_{21}+u_{1,2}+u_{2,1})+\omega^2((\bar{\eta}_2-\bar{\eta}_1^*)u_{1,2}-(\bar{\eta}_2+\bar{\eta}_1^*)u_{2,1}) = \mu (u_{1,2}+u_{2,1}) \\
    2\mu_e(-P_{22}+u_{2,2})+\lambda_e(-P_{11}-P_{22}+u_{1,1}+u_{2,2}) - \omega^2 (2\bar{\eta}_1 u_{2,2}+ \bar{\eta}_3(u_{1,1}+u_{2,2})) = 2\mu u_{2,2}+\lambda (u_{1,1}+u_{2,2})
    \end{cases}
\end{align*}

\newpage

\section{Characteristics of the metamaterials used in the design of the structure in Fig. 26}

We present here the fitting (Fig. 24) for the larger unit cell used in the metastructure's design, as well as the relative position of the dispersion curves for the smaller and larger unit cells (Fig. 25).

\begin{figure}[H]
    \centering
    \includegraphics{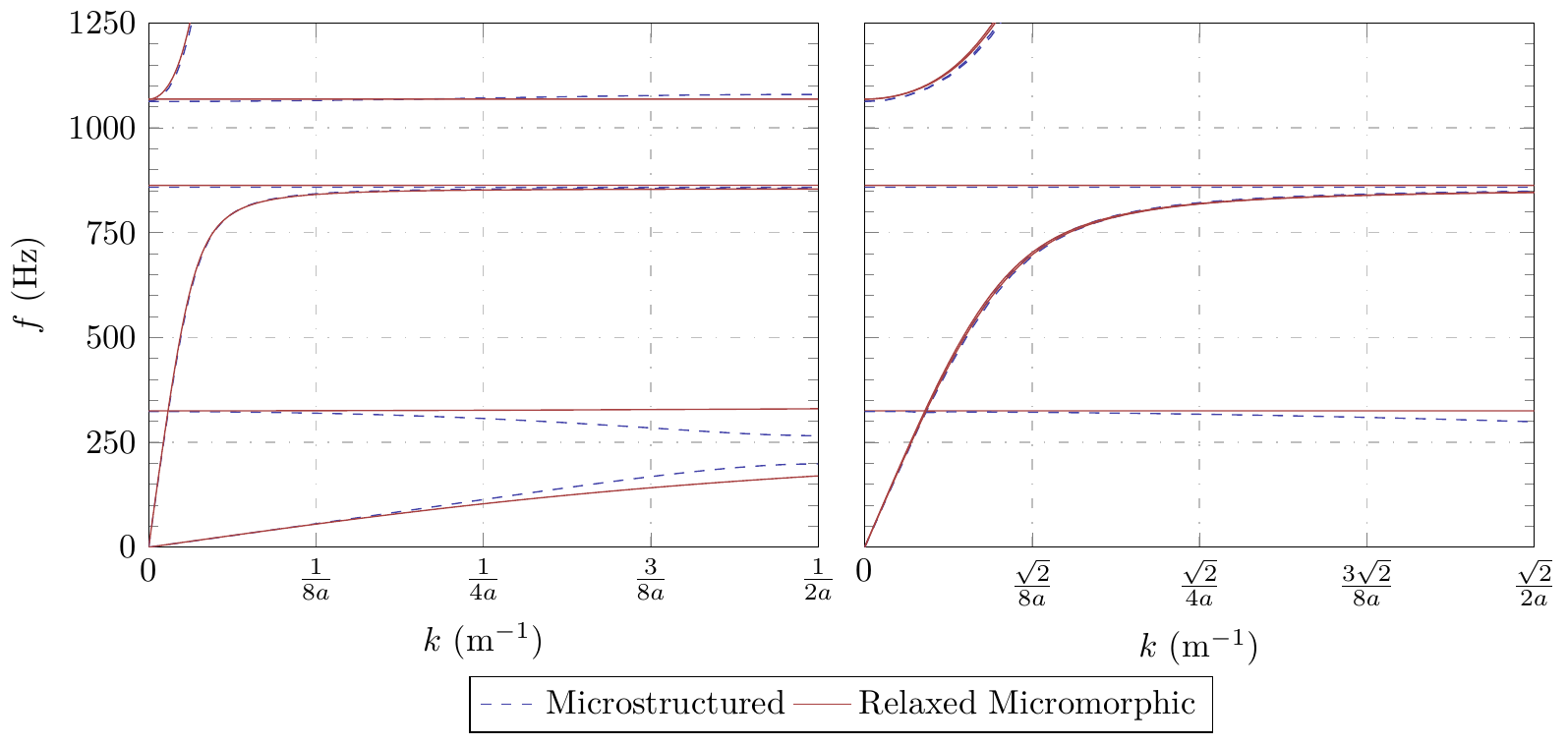}
    \caption{(\textit{left}) Dispersion curves of the microstructured and the relaxed micromorphic ``double cell'' along $\Gamma$X (propagation at 0°). (\textit{right}) Dispersion curves of the microstructured and the relaxed micromorphic ``double cell'' along $\Gamma$M (propagation at 45°).}
    \label{fig:dispersiondouble}
\end{figure}
\begin{figure}[H]
    \centering
    \includegraphics{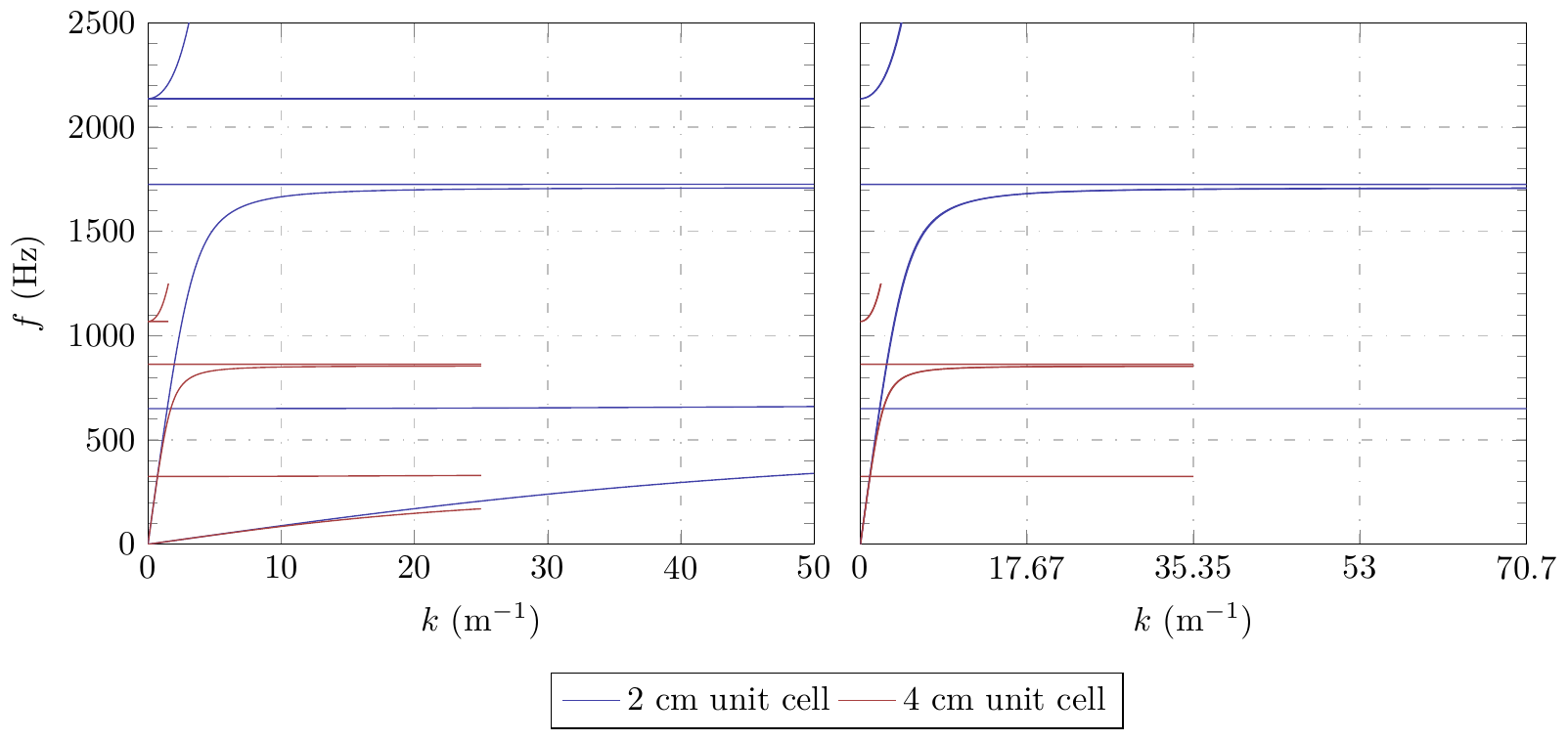}
    \caption{(\textit{left}) Dispersion curves for the 4cm unit cell (red) and the 2 cm unit cell (blue) along $\Gamma$X (propagation at 0°) and (\textit{right}) along $\Gamma$M (propagation at 45°).}
    \label{fig:dispersionboth}
\end{figure}

\end{document}